\documentclass[12pt]{article}
\usepackage{a4wide,epsfig,psfrag,amsmath,amssymb,cite,scalefnt}
\usepackage{color}

\newcommand{\gsim}{\;\rlap{\lower 3.5 pt \hbox{$\mathchar \sim$}} \raise 1pt
 \hbox {$>$}\;}
\newcommand{\lsim}{\;\rlap{\lower 3.5 pt \hbox{$\mathchar \sim$}} \raise 1pt
 \hbox {$<$}\;}

\allowdisplaybreaks

\newcommand{\lp}{\left(}
\newcommand{\rp}{\right)}
\newcommand{\lbk}{\left \lbrack}
\newcommand{\rbk}{\right \rbrack}
\newcommand{\lbc}{\left \lbrace}
\newcommand{\rbc}{\right \rbrace}
\newcommand{\la}{\left.}
\newcommand{\ra}{\right.}
\newcommand{\forcetag}{\addtocounter{equation}{1} \tag{\theequation}}

\newcommand{\LiB}[2]{\ensuremath{\textrm{Li}_{ #1 } \left( #2 \right)}}

\begin{document}

\title{\vskip-3cm{\baselineskip14pt
    \begin{flushleft}
      \normalsize SFB/CPP-11-05\\
      \normalsize TTP11-02
  \end{flushleft}}
  \vskip1.5cm
  Moments of heavy-light current correlators\\ up to three loops
}

\author{\small 
  Jens Hoff and
  Matthias Steinhauser
  \\
  {\small\it Institut f{\"u}r Theoretische Teilchenphysik, }
  {\small\it Karlsruhe Institute of Technology (KIT)}\\
  {\small\it 76128 Karlsruhe, Germany}
}

\date{}

\maketitle

\thispagestyle{empty}

\begin{abstract}
  We consider moments of the non-diagonal vector, axial-vector, scalar
  and pseudo-scalar current correlators involving 
  two different massive quark flavours up to three-loop accuracy.
  Expansions around the limits where one mass is zero and the 
  equal-mass case are computed. These results are used to construct
  approximations valid for arbitrary mass values.

\medskip


\end{abstract}

\thispagestyle{empty}


\newpage


\section{Introduction}

Current correlators are building blocks for a number of physical
quantities. Among them is the total cross section of
hadrons in electron positron annihilation which is obtained from the
imaginary part of the vector correlator. The axial-vector correlator
leads in an analogue way to contributions to the $Z$ boson decay rate.
The scalar and pseudo-scalar correlators govern the decay rate of
Higgs bosons with the respective CP property.

An important role in the context of current correlators is taken over
by moments obtained from an expansion of the two-point functions 
for small external momenta. As far as the correlators are concerned
where the current couples to the same quark flavour
they have been used to extract precise values for the charm and bottom quark
masses by comparing the results for the vector
correlator up to four-loop
order~\cite{Kallen:1955fb,Chetyrkin:1997mb,Maier:2007yn,Chetyrkin:2006xg,Boughezal:2006px,Sturm:2008eb,Maier:2008he,Kiyo:2009gb} 
with the moments obtained from the experimentally measured cross section
(see, e.g.,
Refs.~\cite{Kuhn:2007vp,Chetyrkin:2009fv,Chetyrkin:2010ic}).
Also the moments for the other
correlators~\cite{Sturm:2008eb,Maier:2008he,Kiyo:2009gb} can be
employed once 
the experimental data is replaced by precise lattice simulations as
has been done in Refs.~\cite{Allison:2008xk,McNeile:2010ji}. 
Besides the mass values also the strong coupling constant can be
extracted from the comparison of moments of the pseudo-scalar currents
obtained both in perturbation theory and with the help of
lattice gauge theory calculations~\cite{Allison:2008xk,McNeile:2010ji}.

The focus of the paper lies on moments of heavy-light current
correlators, i.e., two-point functions involving two quark flavours with
different masses. Results for the case where one of the masses is zero
have been obtained in Refs.~\cite{Djouadi:1993ss,Djouadi:1994gf}
and~\cite{Chetyrkin:2000mq,Chetyrkin:2001je} to  
two- and three-loop order, respectively.
These results have been used
in~\cite{Chetyrkin:2000mq,Chetyrkin:2001je} to reconstruct, in 
combination with high-energy expansions and information about the
threshold behaviour, approximations valid for
all external momenta and quark masses. Applications of these results are 
corrections for single-top production or decay rates of
charged Higgs bosons in extensions of the Standard
Model~\cite{Chetyrkin:2000mq,Chetyrkin:2001je}.
The three-loop corrections have furthermore been used to
obtain precise results for the $B$ and $D$ meson decay constants,
$f_B$ and $f_D$, based on sum rules~\cite{Penin:2001ux,Jamin:2001fw}.
Also the extraction of $f_B$ and $f_D$ from lattice gauge simulations
requires perturbative input. As discussed in Ref.~\cite{:2010jy}
it is particularly desirable to have moments of the heavy-light
currents for general values of the two quark masses, not only for
the physical masses of the involved quarks in order to perform simulations for
a variety of different masses, which then can be extrapolated to the mass
values of physical interest. We therefore extend
the known three-loop results for the moments of the 
vector, axial-vector, scalar and pseudo-scalar current correlators
where one of the quark masses is zero to 
the case of two different masses. This is done by computing
analytic expansions around two limiting cases, the one for equal masses,
the other for one of the masses equal to zero.
Subsequently, approximations are constructed which cover the whole range.

In the next Section we provide details on the calculation,
the results for the pseudo-scalar correlator
are presented in Section~\ref{sec::results}. We conclude with a summary
in Section~\ref{sec::sum}. Long analytic formulae are relegated to the
Appendix where also numerical results for the vector, axial-vector and scalar
correlator can be found. All results for the moments can be downloaded
from~\cite{progdata}. 


\section{\label{sec::calc}Calculation}

\begin{figure}[t]
  \centering
  \epsfxsize=.8\textwidth
  \leavevmode
  \epsffile[80 300 520 550]{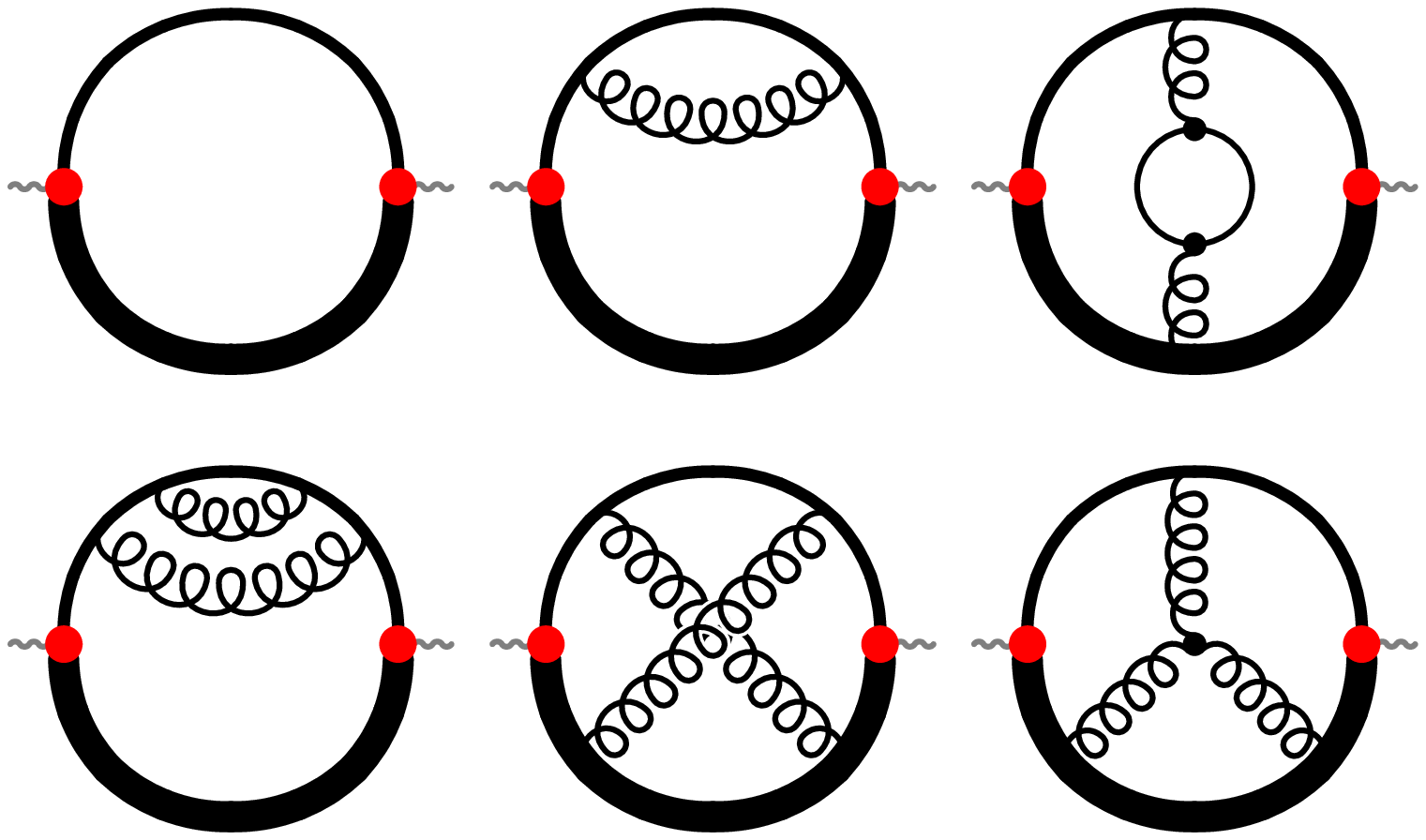}
  \caption[]{\label{fig::diags}Sample diagrams contributing to 
    $\Pi^\delta$ at one, two and three loops.
    The thick (lower) and thin (upper) lines correspond to quarks with mass
    $m_1$ and $m_2$, respectively, and the curly lines represent gluons.
    }
\end{figure}

Let us in a first step present our notation. We define the vector (v),
axial-vector (a), scalar (s) and pseudo-scalar (p) currents via
\begin{eqnarray}
  j_\mu^v &=& \bar{\psi}_1\gamma_\mu\psi_2\,,\nonumber\\
  j_\mu^a &=& \bar{\psi}_1\gamma_\mu\gamma^5\psi_2\,,\nonumber\\
  j^s     &=& \bar{\psi}_1\psi_2\,,\nonumber\\
  j^p     &=& \bar{\psi}_1 i\gamma^5 \psi_2
  \,,
  \label{eq::currents}
\end{eqnarray}
where $\psi_1$ and $\psi_2$ denote the two quark flavours.
Using these currents the vector and axial-vector correlator is defined through
($\delta=v,a$)
\begin{eqnarray}
  \left(-q^2g_{\mu\nu}+q_\mu q_\nu\right)\,\Pi^\delta(q^2)
  +q_\mu q_\nu\,\Pi^\delta_L(q^2)
  &=&
  i\int {\rm d}x\,e^{iqx}\langle 0|Tj^\delta_\mu(x) j^{\delta\dagger}_\nu(0)|0 \rangle\,,
  \label{eq::pivadef}
\end{eqnarray}
where $\Pi^\delta(q^2)$ and $\Pi_L^\delta(q^2)$ are the transverse and
longitudinal contributions, respectively.
The scalar and pseudo-scalar polarization function reads
($\delta=s,p$)
\begin{eqnarray}
  q^2\,\Pi^\delta(q^2)
  &=&
  i\int {\rm d}x\,e^{iqx} \langle 0|Tj^\delta(x)j^{\delta\dagger}(0)|0 \rangle
  \,.
  \label{eq::pispdef}
\end{eqnarray}
Throughout this paper we consider anti-commuting $\gamma_5$
which is justified as for $\psi_1\not=\psi_2$ only non-singlet diagrams
contribute.
As a consequence the axial-vector (pseudo-scalar) correlator coincides
with the vector (scalar) one if either $m_1$ or $m_2$ is zero.
Sample Feynman diagrams occurring at one-, two- and three-loop order
are shown in Fig.~\ref{fig::diags}.

It is convenient to introduce the dimensionless variables
\begin{eqnarray}
  z &=& \frac{q^2}{m_1^2}\,,\\
  x &=& \frac{m_2}{m_1}\,,
\end{eqnarray}
which enables us to cast $\Pi^\delta(q^2)$ for $q^2\to0$ in the form
\begin{eqnarray}
  \bar{\Pi}^\delta(q^2) &=& \frac{3}{16\pi^2}
  \sum_{n\ge-1} \bar{C}^\delta_n(x) z^n
  \,,
\end{eqnarray}
where the bar indicates that the overall renormalization of the polarization
function is performed in the $\overline{\rm MS}$ scheme.
Note that we have the symmetry relations 
$\bar{C}^s_n(x)=\bar{C}^p_n(-x)$ and $\bar{C}^v_n(x)=\bar{C}^a_n(-x)$
(see, e.g., Ref.~\cite{Chetyrkin:1996ia}).

We perform an explicit calculation only for the transverse part of the vector
and axial-vector current and reconstruct the coefficients for the longitudinal
part with the help of ($n=-1,\ldots,3$)
\begin{eqnarray}
  \bar{C}_{L,n}^v &=& (1-x)^2 \bar{C}_{n+1}^s\,,\nonumber\\
  \bar{C}_{L,n}^a &=& (1+x)^2 \bar{C}_{n+1}^p\,.
\end{eqnarray}
For $n=-1,0,1$ and $2$ and six expansion terms in $x$ and $(1-x)$ we have checked that
these equations hold.

The expansion of the coefficients in $\alpha_s$ is given by
\begin{eqnarray}
  \bar{C}^\delta_n = \bar{C}^{(0),\delta}_n 
  + \frac{\alpha_s}{\pi} \bar{C}^{(1),\delta}_n 
  + \left(\frac{\alpha_s}{\pi}\right)^2 \bar{C}^{(2),\delta}_n 
  + \ldots
  \,.
\end{eqnarray}
where the arguments $x$ and $\mu$ are suppressed.
Often it is
advantageous to require a QED-like normalization
$\Pi^\delta(0)=0$ where the relation to $\bar{\Pi}^\delta(q^2)$ is
given by
\begin{eqnarray}
  \Pi^\delta(q^2) &=& \bar{\Pi}^\delta(q^2) - \frac{3}{16\pi^2}\left(\bar{C}^\delta_0 +
  \frac{\bar{C}^\delta_{-1}}{z}\right)
  \,.
\end{eqnarray}

For the practical calculation we use a well tested set-up which is
highly automated in order to avoid errors.
All Feynman diagrams are generated with {\tt
  QGRAF}~\cite{Nogueira:1991ex} and the 
various diagram topologies are identified and transformed to {\tt
  FORM}~\cite{Vermaseren:2000nd} with the help of {\tt q2e} and {\tt
  exp}~\cite{Harlander:1997zb,Seidensticker:1999bb}. {\tt exp} also
applies the asymptotic expansion for $x\to0$ on a diagrammatic level.
The expansion around $x=1$ leads to a naive Taylor series.
In a next step the expressions are passed to {\tt FORM} where appropriate
projectors are applied, the expansions in small quantities are
performed and traces are taken. Once the analytical expressions for
each diagram are reduced to scalar integrals the program {\tt
  MATAD}~\cite{Steinhauser:2000ry} is called in order to handle 
vacuum integrals up to three loops.

The three-loop calculation requires the inclusion of counterterms up
to two-loop order (see, e.g., Ref.~\cite{Chetyrkin:2004mf})
which we implement in the $\overline{\rm MS}$ scheme
both for the parameters $m_1$, $m_2$ and $\alpha_s$ and the
renormalization of the current itself. The latter is equal to the mass
renormalization for the scalar and pseudo-scalar case and unity for the
vector and axial-vector current.
At that point all remaining poles are local and can be subtracted in
the $\overline{\rm MS}$ scheme in order to arrive at $\bar{\Pi}^\delta(q^2)$.
(See Ref.~\cite{Chetyrkin:1996ia} for details.) 

Following this procedure we were able to evaluate the
coefficients $\bar{C}_n(x)$ with $n\le4$ where terms up to 
order $x^8$ and $(1-x)^9$ are included in the expansions around $x=0$
and $x=1$, respectively.


\section{\label{sec::results}Results}

In this Section we discuss the quality of our expansions in $x$. At
one and two loops we can compare to the exact result which provides
confidence concerning the validity of the three-loop approximations. 
In the main text we restrict ourselves to the pseudo-scalar case.
Results for the remaining three correlators are presented in the
Appendix and in Ref.~\cite{progdata}.

\subsection{One- and two-loop results for the pseudo-scalar moments}

The exact dependence on $x$ of the 
one-loop results for the pseudo-scalar moments 
$\bar{C}_n^p(x)$ with $n\le4$ is given by
\begin{flalign}
\bar{C}_{-1}^{(0), p} (x) =\;& 
  - 4\lp 1 + l_\mu \rp \lp 1 - x + x^2 \rp + l_x \, \frac{8 \, x^3}{1 + x} \nonumber\,,\\
\bar{C}_0^{(0), p} (x) =\;& 
  \frac{1 + 4 \, x + x^2}{\lp 1 + x \rp^2} - l_x \, \frac{4 \, x^3 \lp 2 + x \rp}{\lp - 1 + x \rp \lp 1 + x \rp^3} + 2 \, l_\mu \nonumber\,,\\
\bar{C}_1^{(0), p} (x) =\;& 
  \frac{2 \lp 1 - x + x^2 \rp \lp 1 + 4 \, x + x^2 \rp}{3 \lp - 1 + x \rp^2 \lp 1 + x \rp^4} 
    - l_x \, \frac{8 \, x^3}{\lp - 1 + x \rp^3 \lp 1 + x \rp^5} \nonumber\,,\\
\bar{C}_2^{(0), p} (x) =\;& 
  \frac{1 + 4 \, x - 7 \, x^2 + 40 \, x^3 - 7 \, x^4 + 4 \, x^5 + x^6}{6 \lp - 1 + x \rp^4 \lp 1 + x \rp^6} 
    - l_x \, \frac{4 \, x^3 \lp 2 - x + 2 \, x^2 \rp}{\lp - 1 + x \rp^5 \lp 1 + x \rp^7} \nonumber\,,\\
\bar{C}_3^{(0), p} (x) =\;& 
  \frac{1 + 5 \, x - 14 \, x^2 + 145 \, x^3 - 94 \, x^4 + 145 \, x^5 - 14 \, x^6 + 5 \, x^7 + x^8}
    {15 \lp - 1 + x \rp^6 \lp 1 + x \rp^8} \notag \\
  & - l_x \, \frac{8 \, x^3 \lp 1 - x + 3 \, x^2 - x^3 + x^4 \rp}{\lp - 1 + x \rp^7 \lp 1 + x \rp^9} \nonumber\,,\\
\bar{C}_4^{(0), p} (x) =\;& 
  \frac{1 + 6 \, x - 23 \, x^2 + 356 \, x^3 - 398 \, x^4 + 956 \, x^5}{30 \lp - 1 + x \rp^8 \lp 1 + x \rp^{10}} \notag \\
  & + \frac{ - 398 \, x^6 + 356 \, x^7 - 23 \, x^8 + 6 \, x^9 + x^{10}}{30 \lp - 1 + x \rp^8 \lp 1 + x \rp^{10}} \notag \\
  & - l_x \, \frac{4 \, x^3 \lp 2 - 3 \, x + 12 \, x^2 - 8 \, x^3 + 12 \, x^4 - 3 \, x^5 + 2 \, x^6 \rp}
    {\lp - 1 + x \rp^9 \lp 1 + x \rp^{11}}
\,,
\label{eq::Cn0p}
\end{flalign}
where $l_\mu=\ln(\mu^2/m_1^2)$ and $l_x=\ln(x^2)/2$.
These results have been extracted from Ref.~\cite{Djouadi:1994gf} 
where $\Pi^p(q^2)$ is given in analytic form up to two loops.
We have checked the results of Eq.~(\ref{eq::Cn0p}) evaluating directly
the moments both exactly in $x$ and restricting ourselves to
expansions in $x^n$ and $(1-x)^n$. Note that the latter method has to
be applied at three loops.

\begin{figure}[t]
  \centering
  \begin{tabular}{cc}
    \includegraphics[width=.45\linewidth]{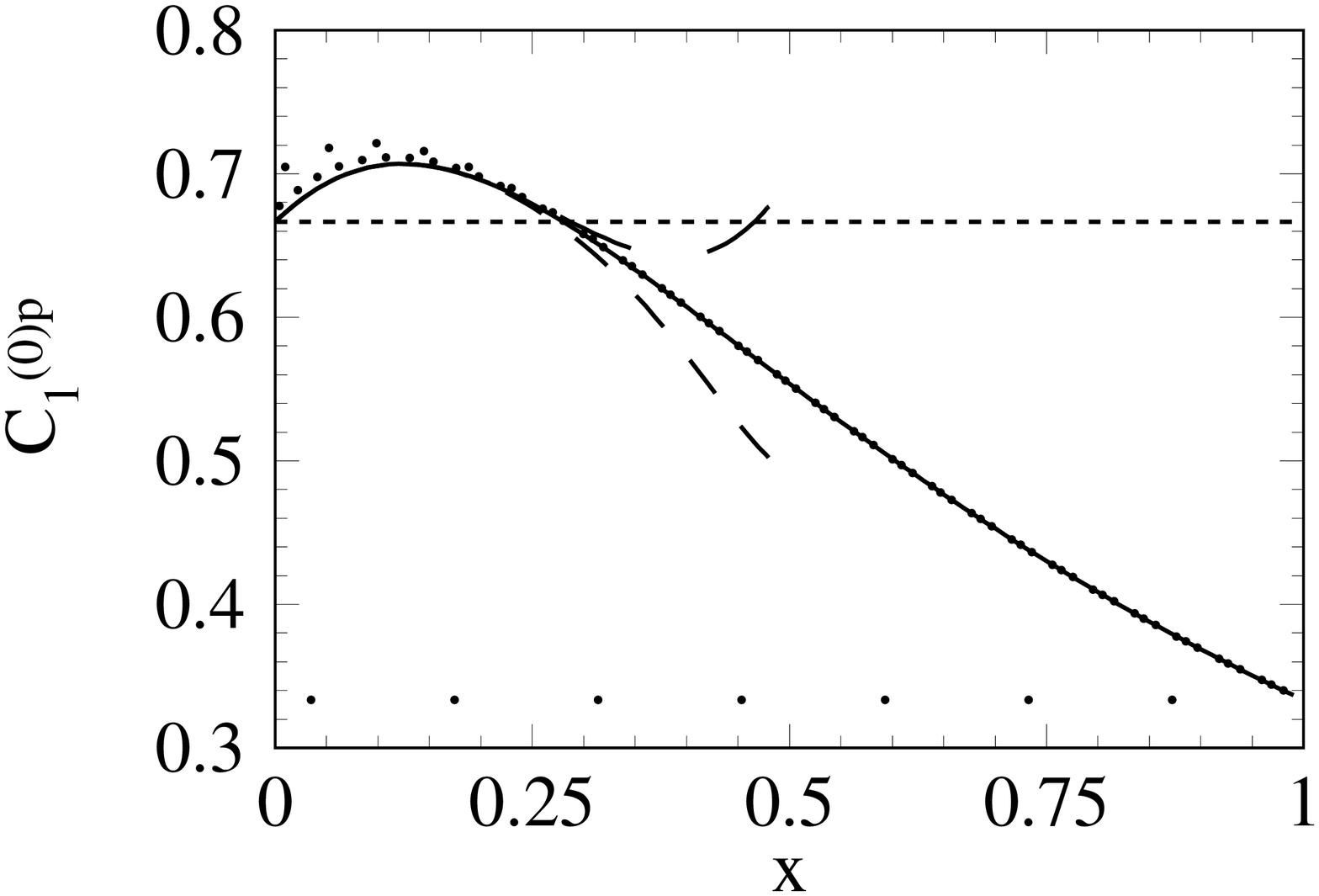}
    &
    \includegraphics[width=.45\linewidth]{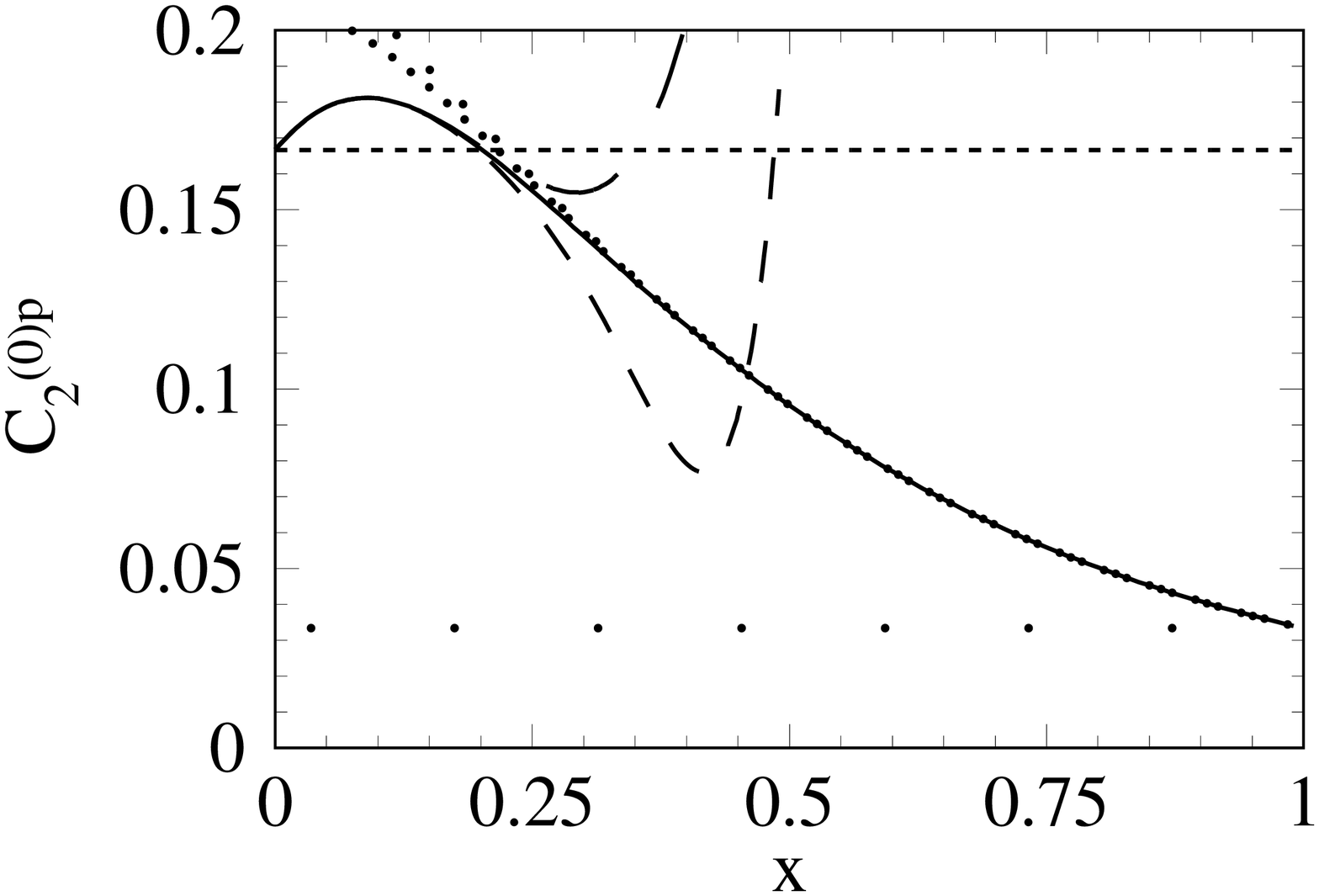}
    \\
    \includegraphics[width=.45\linewidth]{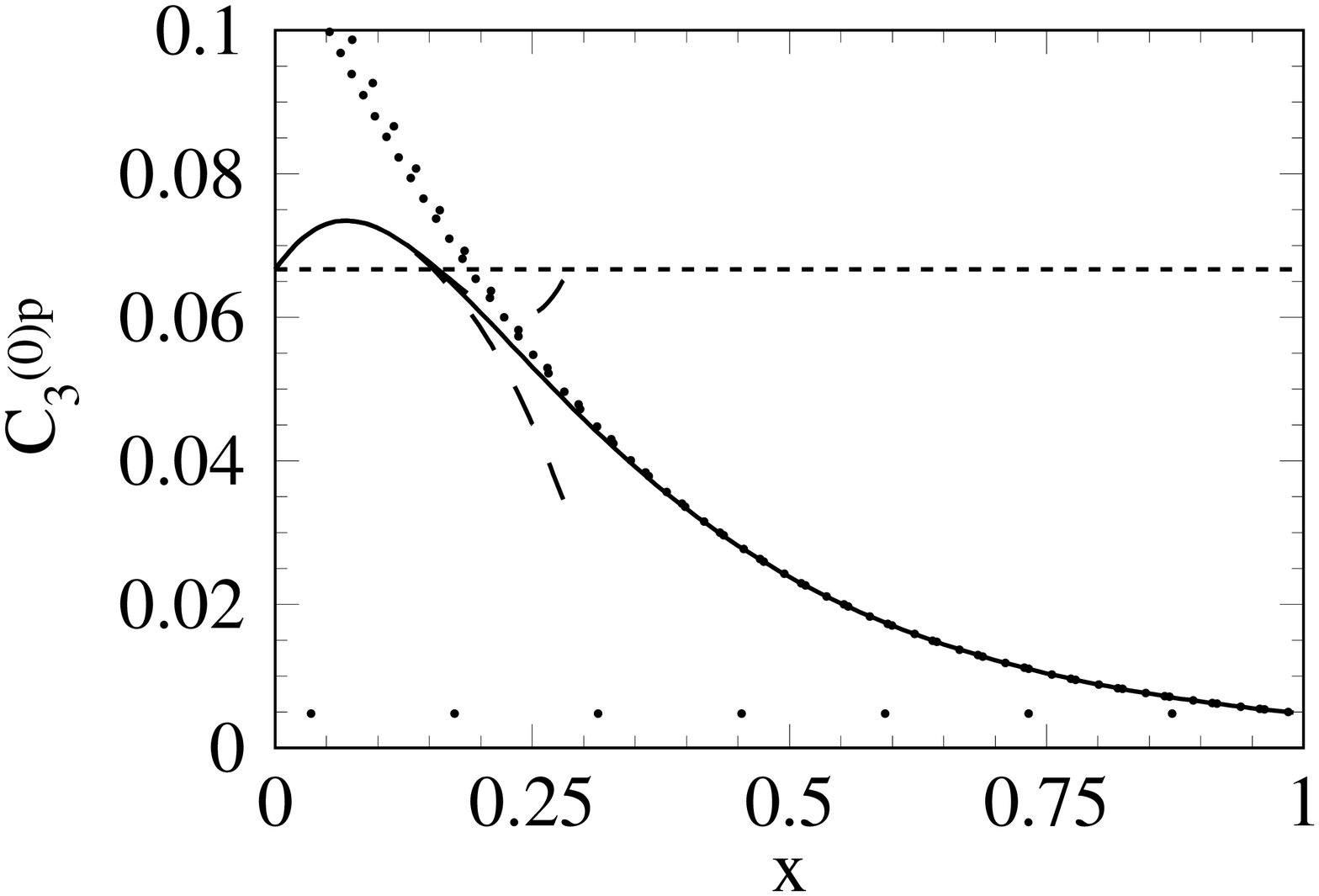}
    &
    \includegraphics[width=.45\linewidth]{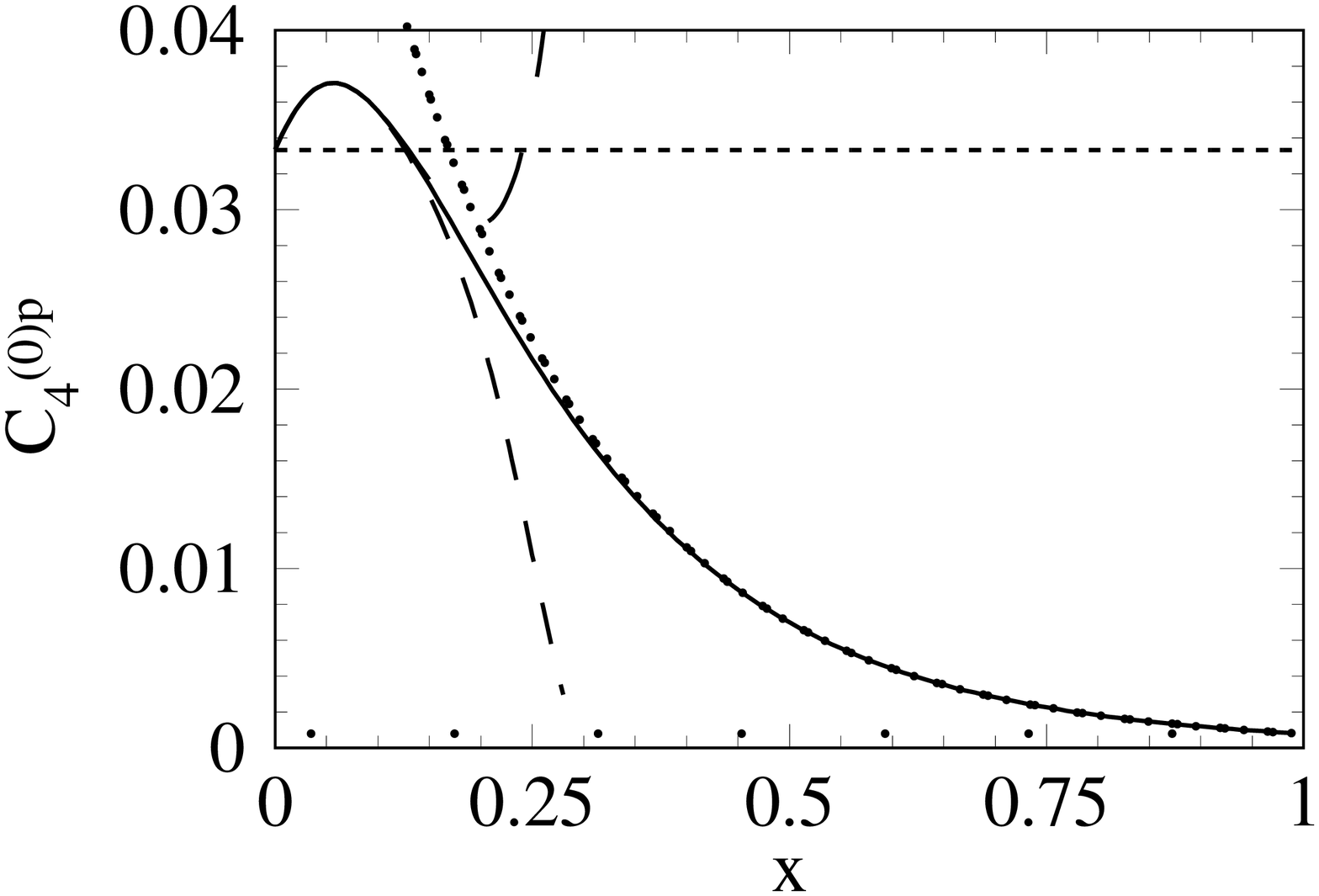}
  \end{tabular}
  \caption[]{\label{fig::c3pas0}One-loop
    contribution to $\bar{C}_n^p$. The dashed and dotted lines
    correspond to the expansions in $x$ and $(1-x)$, respectively, where
    the lowest and the two highest approximations are shown.
    The solid line represents the exact result.
    Note that for $n=4$ the curve including terms up to 
    order $(1-x)^8$ and the one including terms up to 
    order $(1-x)^9$ lie on top of each other in the shown interval.
  }
\end{figure}

In Fig.~\ref{fig::c3pas0} we discuss $\bar{C}_n^{(0),p}(x)$ for
$n=1,2,3$ and $4$
and show the curves including $x^0$, $x^7$ and $x^8$ for the small-$x$
expansion as dashed lines where longer dashes correspond to
approximations including higher order power corrections.
The dotted lines correspond to expansions around the equal-mass
case where curves including $(1-x)^0$, $(1-x)^8$ and $(1-x)^9$
(decreasing distance between dots) are shown. The exact result is
shown as solid curve. One observes that the approximation including
corrections up to order $x^8$ coincides with the exact result up to
$x\approx 0.2$, for $n=1$ even up to $x\approx 0.3$. On the other
hand, the approximation based on the 
${\cal O}((1-x)^9)$ terms is indistinguishable from the solid line
for $x\gsim 0.2-0.3$. 
These limits are close to the ones obtained by considering the
approximations including $x^7$ and $(1-x)^8$ terms only.
It is obvious that the small gap in between can
easily be closed with the help of a polynomial interpolation.

\begin{figure}[t]
  \centering
  \begin{tabular}{cc}
    \includegraphics[width=.45\linewidth]{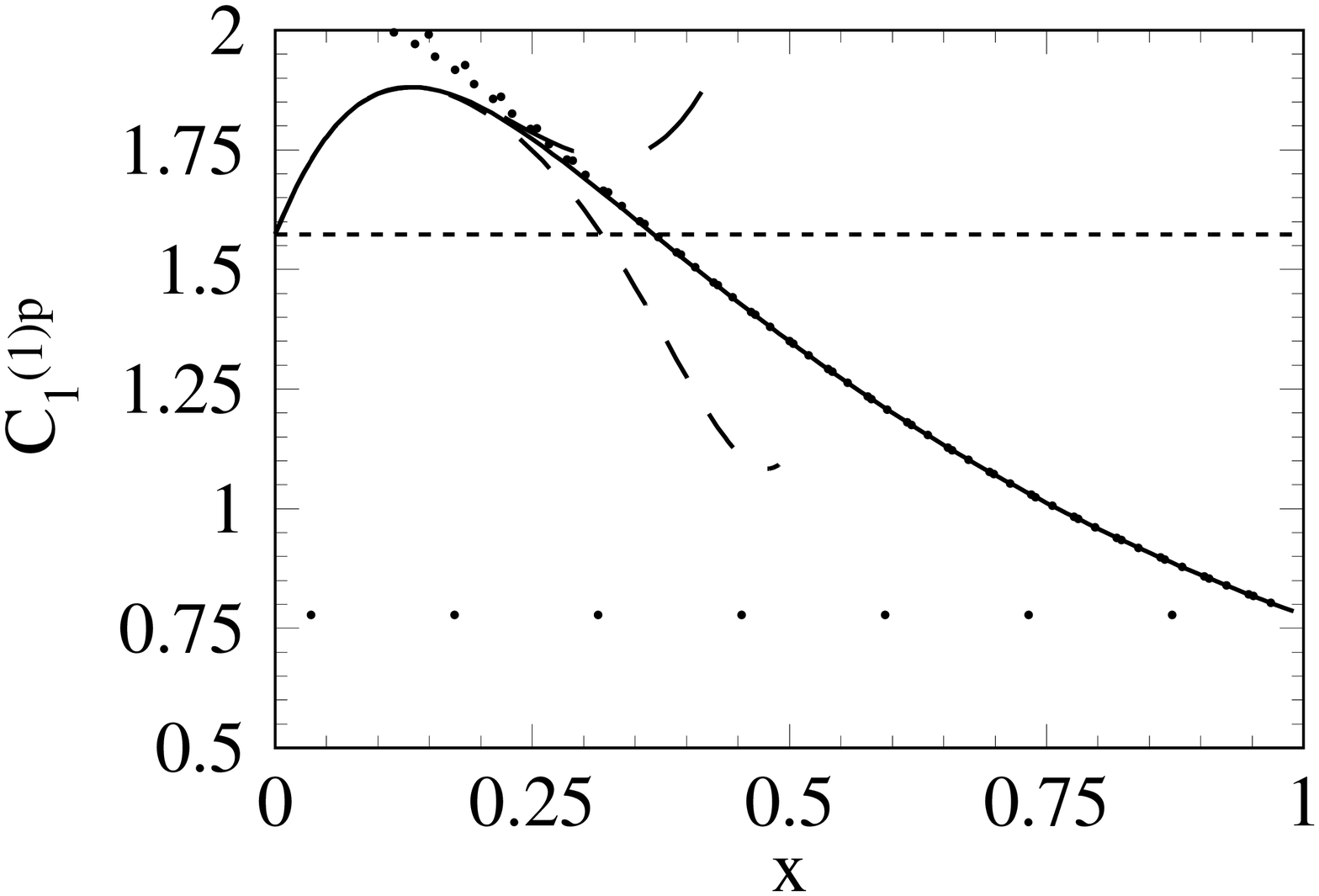}
    &
    \includegraphics[width=.45\linewidth]{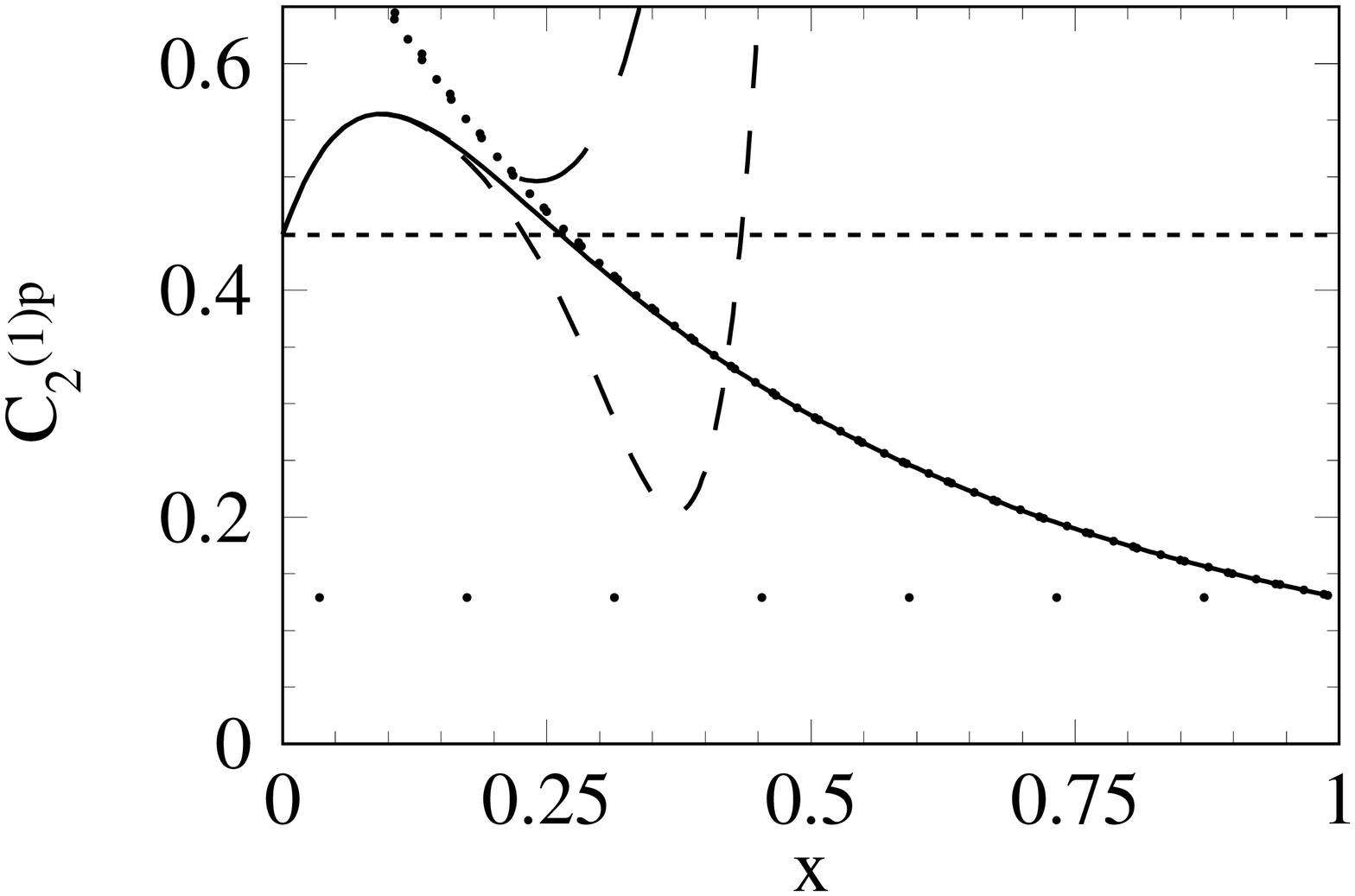}
    \\
    \includegraphics[width=.45\linewidth]{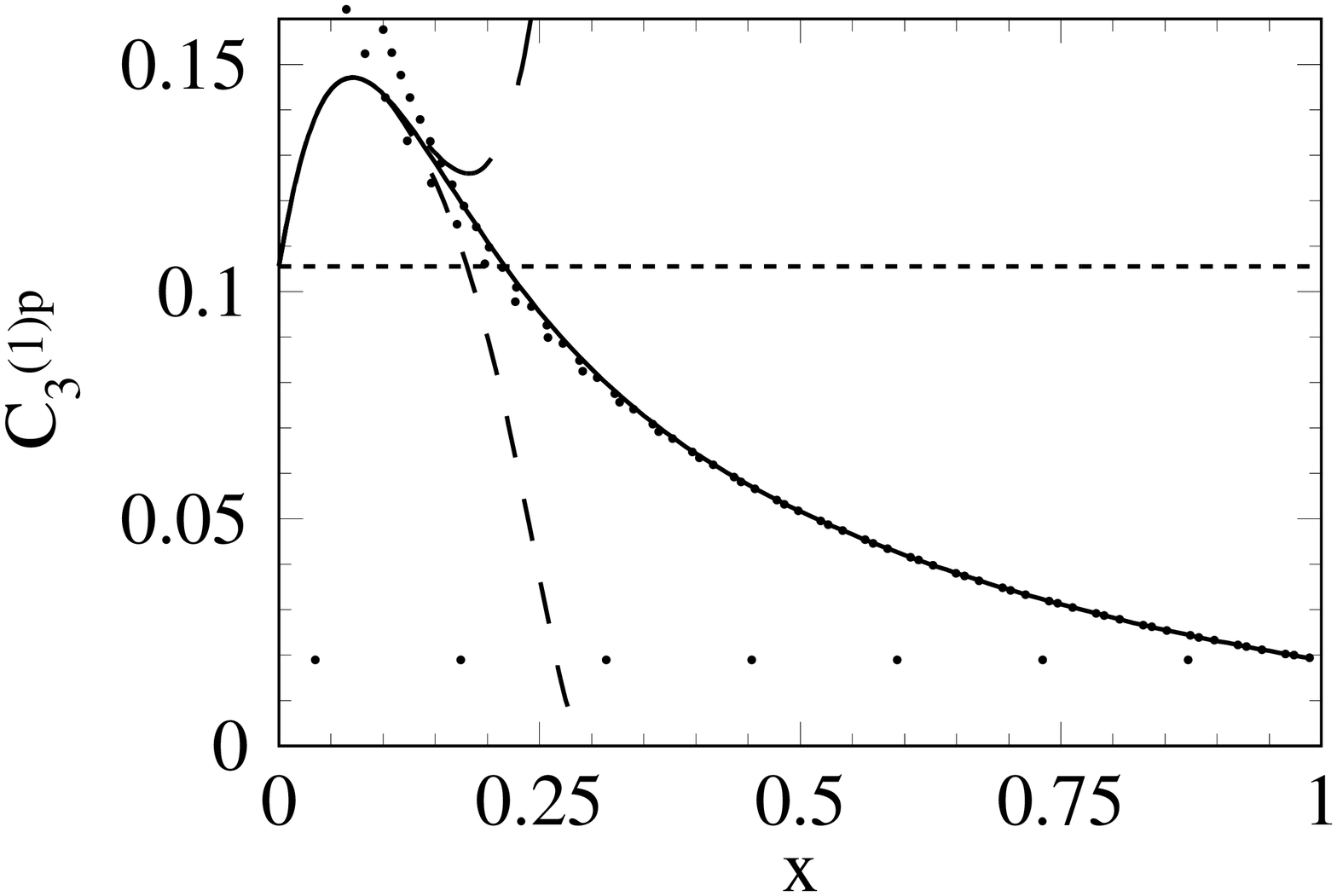}
    &
    \includegraphics[width=.45\linewidth]{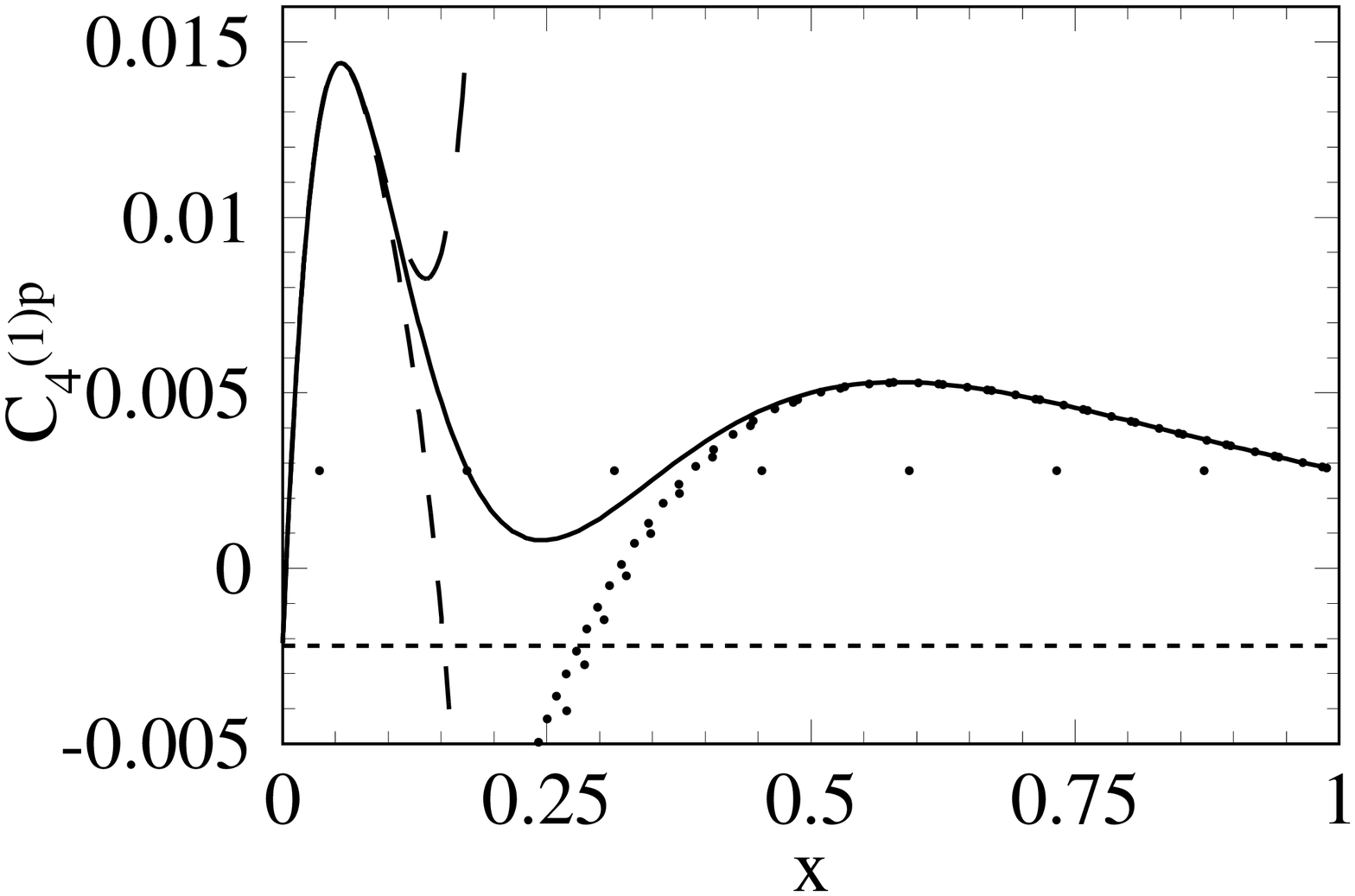}
  \end{tabular}
  \caption[]{\label{fig::c3pas1}Two-loop
    contribution to $\bar{C}_n^p$. The same notation as in
    Fig.~\ref{fig::c3pas0} is adopted.
  }
\end{figure}

The exact analytic results for the two-loop moments of the pseudo-scalar
current are already quite lengthy. Thus we exemplify the structure by
showing $\bar{C}_{1}^{(1),p}(x)$ in the main text. The results for
$\bar{C}_{n}^{(1),p}(x)$ for $n\not=1$ can be found in
Appendix~\ref{app:2lp}.  For $\bar{C}_{1}^{(1),p}(x)$ we have
\begin{flalign*}
\bar{C}_1^{(1), p} (x) =\;& 
  \frac{1 + 36 \, x - 22 \, x^2 + 36 \, x^3 + x^4}{9 \lp -1 + x \rp^2 \lp 1 + x \rp^4} 
    + \lbk \LiB{2}{x^2} + 2 \, l_x \, l_{1- x^2} - \zeta_2 \rbk \frac{8 \lp 1 + 3 \, x + x^2 \rp}
      {9 \lp -1 + x \rp \lp 1 + x \rp^3} \\
  & + \frac{16 \, l_x \, x^2 \lp 1 + 6 \, x + x^2 \rp}{9 \lp -1 + x \rp^3 \lp 1 + x \rp^5} 
    - \frac{16 \, l_x^2 \, x^3 \lp 18 + 9 \, x^2 - 2 \, x^3 + 3 \, x^4 + x^5 \rp}
      {9 \lp -1 + x \rp^4 \lp 1 + x \rp^6} \forcetag 
      \,,
\label{eq::Cn1p}
\end{flalign*}
where $l_{1-x^2}=\ln(1-x^2)$ and $\zeta_2=\pi^2/6$. Note that in
Eq.~(\ref{eq::Cn1p}) the coefficient of $\ln(\mu^2/m_1^2)$ is zero.
Using the same notation as in Fig.~\ref{fig::c3pas0} we show in
Fig.~\ref{fig::c3pas1} the results for $\bar{C}_n^{(1),p}(x)$.
The same conclusions as in the one-loop case can be drawn: The
expansion for $x\to0$ provides reliable results for $x\lsim0.2$
whereas the expansion around $x=1$ is trustworthy for $x\gsim0.2-0.4$. 
This is true both for the approximations containing the ninth and the
ones containing only the eighth order in $x$ and $1-x$, respectively.
Thus it is promising to proceed at three-loop
level, where the computation of the exact dependence on $x$ is quite
complex, in the following way: using the combination of computer
programs described in Section~\ref{sec::calc} it is
straightforward to compute expansions in $x$ and $(1-x)$. Afterwards
we combine the expansions applying a simple interpolation.

The approximations for moments of the vector, axial-vector and
scalar correlators have the same quality as in the pseudo-scalar case,
i.e. they approximate the exact result in the same regions of $x$.
We refrain from an explicit discussion in this paper, however, provide
the figures corresponding Fig.~\ref{fig::c3pas0} and~\ref{fig::c3pas1}
in Appendix~\ref{app::apprvas}.

\subsection{Three-loop results}

Let us in a first step present the expansions around $x=0$ and
$x=1$. Since they are quite lengthy we show in the main part of the paper
only results for the first moment adopting $\mu^2=m_1^2$
to exemplify the structure of the result. Furthermore, we restrict ourselves
to the first three terms in the expansion.
The complete expressions for general $\mu$ and
the results for the moments with $n=-1,0,2,3$ and $4$ 
are provided in the {\tt Mathematica} file~\cite{progdata}. 
For $x\to0$ we have
\begin{flalign*}
\bar{C}_1^{(2), p} (x) =\;& 
  \frac{35}{16} + \frac{1055}{162} \, \zeta_2 - \frac{88}{9} \, \zeta_3 - \frac{134}{27} \, \zeta_4 
    + \frac{2}{27} \, D_3 + \frac{105}{4} \, S_2 \\
  & + x \lp \frac{13465}{972} + \frac{4493}{243} \, \zeta_2 + \frac{409}{27} \, \zeta_3 - \frac{137}{27} \, \zeta_4 
    + \frac{D_3}{9} - \frac{39}{4} \, S_2 \rp \\
  & + x^2 \lp - \frac{176603}{3888} - \frac{12557}{486} \, \zeta_2 - \frac{5564}{81} \, \zeta_3 
    + \frac{94}{27} \, \zeta_4 - \frac{10}{27} \, D_3 + \frac{761}{4} \, S_2 \rp \\
  & + x^3 \lp \frac{163867}{1944} + \frac{401977}{1620} \, \zeta_2 + \frac{221}{81} \, \zeta_3 
    + \frac{5327}{324} \, \zeta_4 \ra \\
  & \la \qquad + \frac{50}{27} \, D_3 - \frac{2228}{243} \, S_2^\epsilon + 600
  \, S_2 + \frac{1114}{81} \, T_1^\epsilon \rp \\
  & + l_x \, x^3 \lp \frac{2368}{27} - \frac{1114}{3} S_2 - \frac{80}{27} \, \zeta_2 - \frac{16}{3} \, \zeta_3 \rp 
    - l_x^2 \, \frac{1024 \, x^3}{9} + l_x^3 \, \frac{1264 \, x^3}{9} \\
  & + n_l \lbk - \frac{5}{72} - \frac{\zeta_2}{9} + \frac{8}{27} \, \zeta_3 
    + x \lp - \frac{55}{36} - \frac{8}{27} \, \zeta_2 + \frac{8}{27} \, \zeta_3 \rp \ra \\
  & \qquad  + x^2 \lp \frac{823}{216} + \frac{\zeta_2}{9} - \frac{8}{27} \, \zeta_3 \rp 
    + x^3 \lp - \frac{571}{54} + \frac{28}{9} \, \zeta_2 + \frac{176}{27} \, \zeta_3 \rp \\
  & \la \qquad + l_x \, x^3 \lp \frac{88}{27} + \frac{16}{3} \, \zeta_2 \rp 
    + l_x^2 \, \frac{80 \, x^3}{9} - l_x^3 \, \frac{32 \, x^3}{9} \rbk \forcetag
  + {\cal O}(x^4)
  \,,
\end{flalign*}
where $D_3\approx-3.0270$, $S_2\approx0.2604$, 
$S_2^\epsilon\approx7.8517$,
$T_1^\epsilon\approx-24.2089$~\cite{Steinhauser:2000ry}  
and $\zeta_3\approx1.2021$.
$n_l$ is the number of massless quarks and the total number of quarks is
$n_f=n_l+2$. For $x\to1$ one obtains
\begin{flalign*}
\bar{C}_1^{(2), p} (x) =\;& 
  - \frac{13139}{2592} + \frac{6977}{1728} \, \zeta_3 + (1 - x)^3 \lp \frac{14229845}{93312} - \frac{13342139}{103680} \, \zeta_3 \rp \\
  & + (1 - x) \lp \frac{8267}{2592} - \frac{6977}{1728} \, \zeta_3 \rp 
    + (1 - x)^2 \lp - \frac{72234997}{933120} + \frac{13760759}{207360} \, \zeta_3 \rp \\
  & + n_l \lbk \frac{25}{162} - \frac{2}{81} (1 - x) - \frac{289}{4860} (1 - x)^2 + \frac{409}{4860} (1 - x)^3 \rbk \forcetag
  + {\cal O}\left((1-x)^4\right)
  \,.
\end{flalign*}

\begin{figure}[t]
  \centering
  \begin{tabular}{cc}
    \includegraphics[width=.45\linewidth]{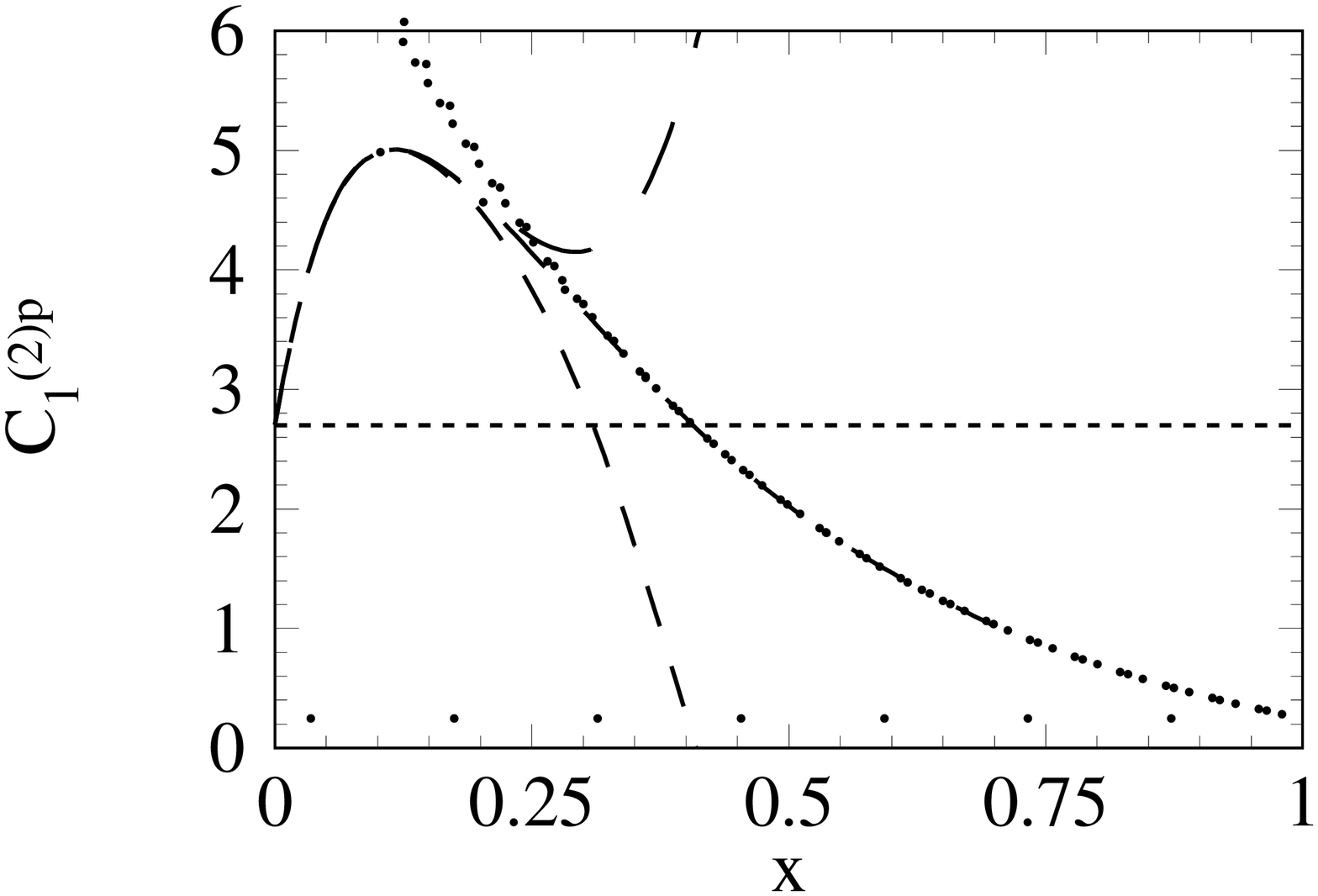}
    &
    \includegraphics[width=.45\linewidth]{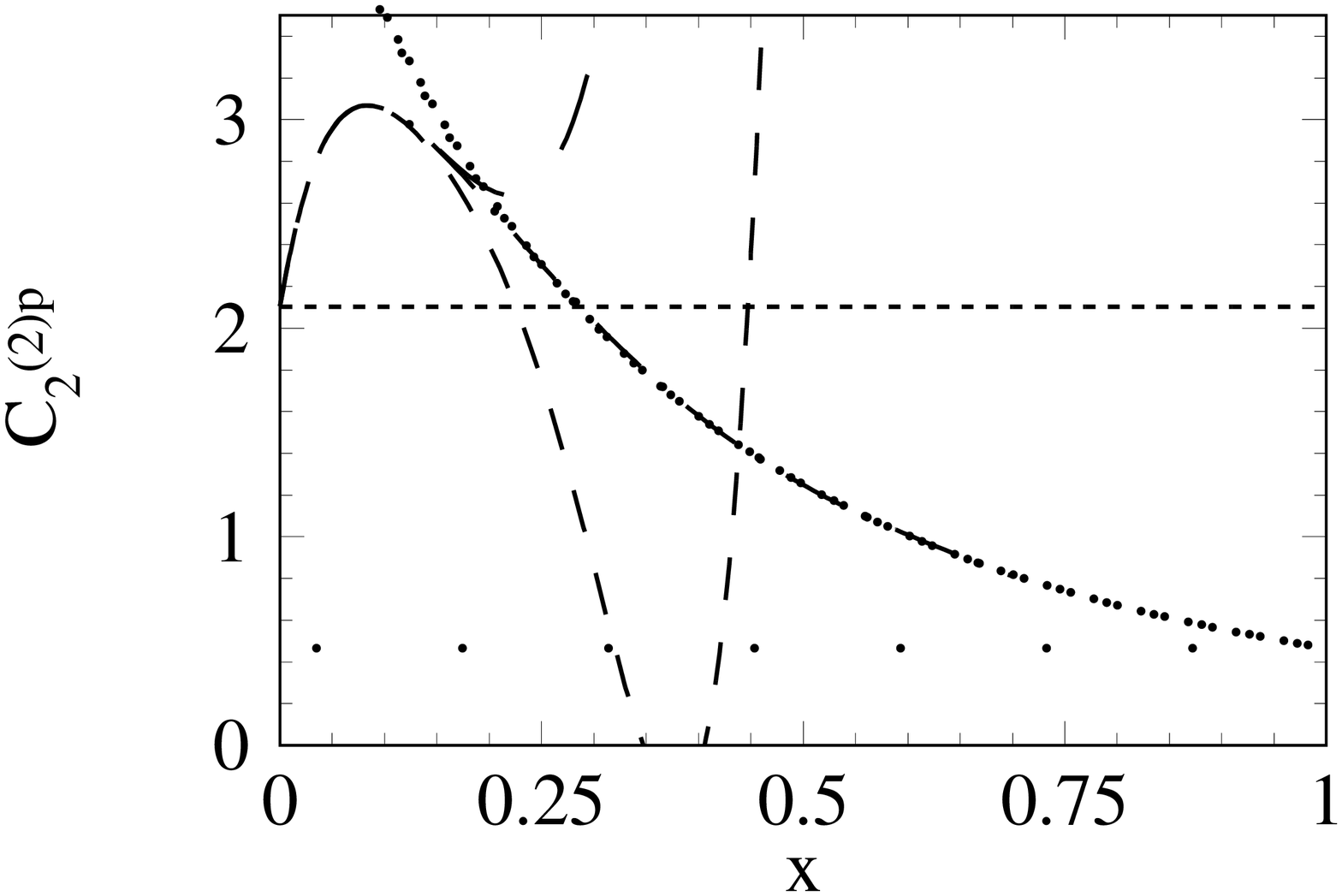}
    \\
    \includegraphics[width=.45\linewidth]{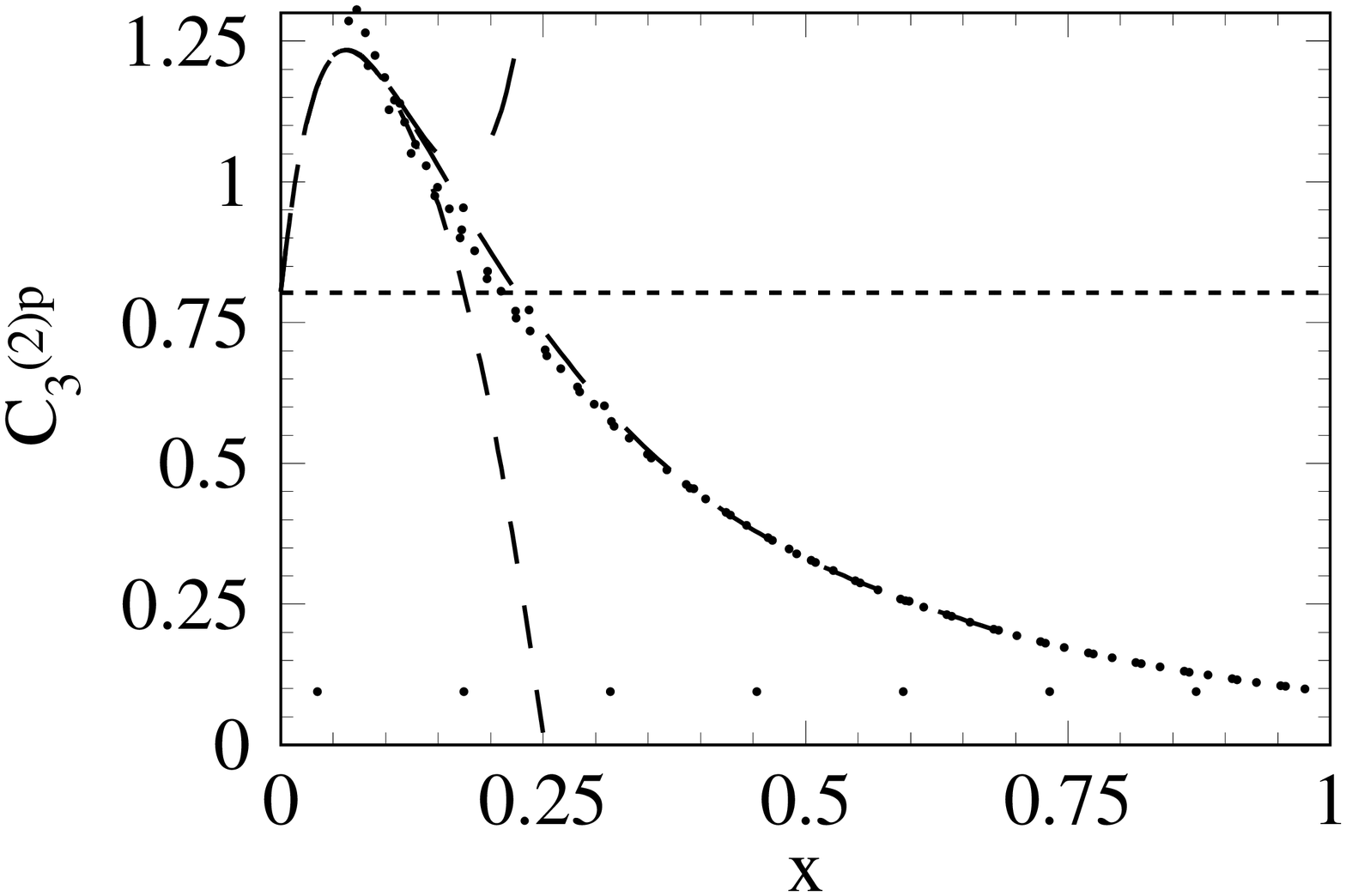}
    &
    \includegraphics[width=.45\linewidth]{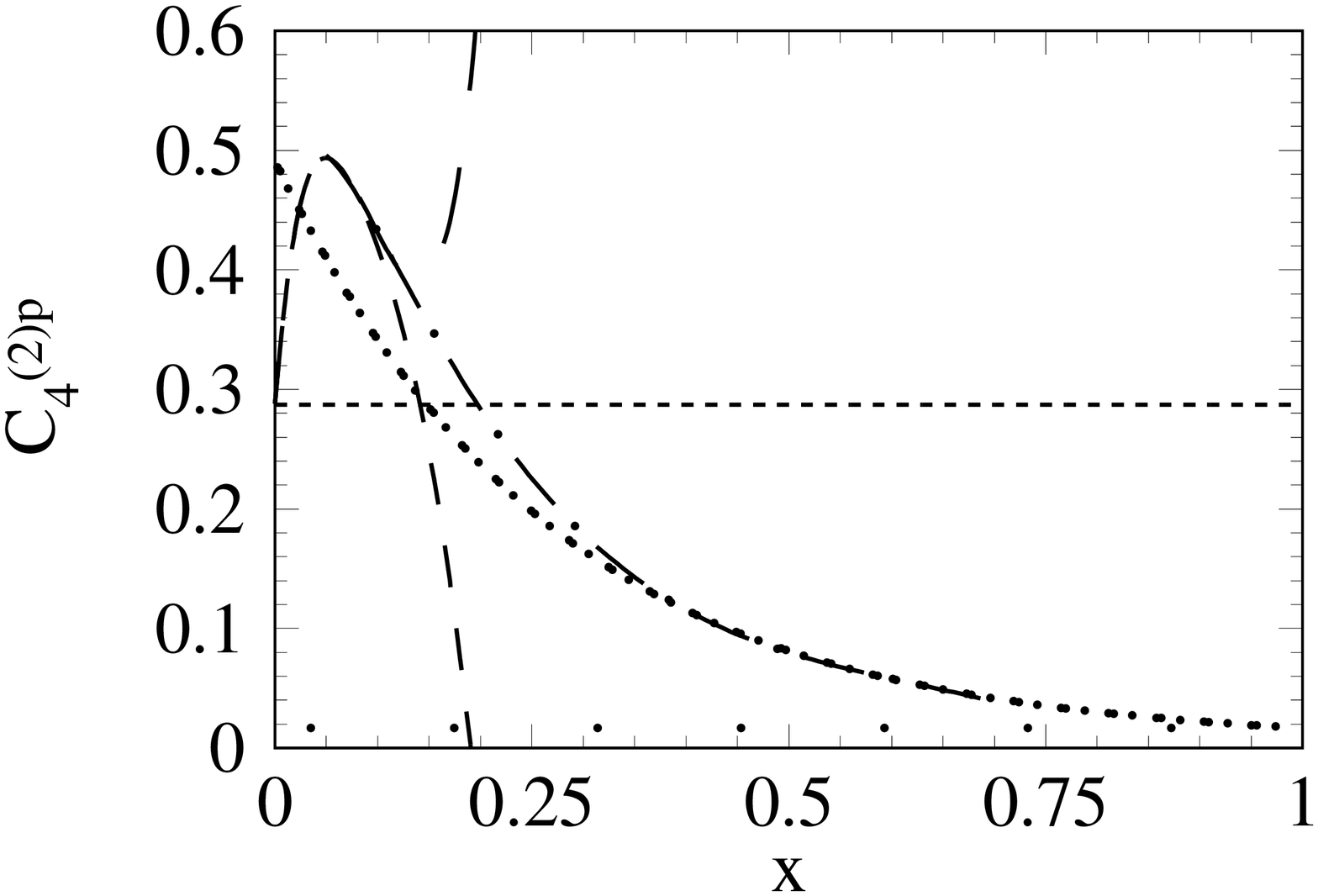}
  \end{tabular}
  \caption[]{\label{fig::c3pas2}Three-loop
    contribution to $\bar{C}_n^p$. 
    The dash-doted line represents the result based on the fit as described
    in the text.
    For the rest the same notation as in Fig.~\ref{fig::c3pas0} is adopted.
  }
\end{figure}

In Fig.~\ref{fig::c3pas2} we show $\bar{C}_n^{(2),p}$ as a function of
$x$ using the same notation as at one and two loops. We have chosen $n_l=3$
which corresponds to a massive charm and bottom quark. We again observe
that the approximations based on the $x^7$ terms and the one including
$x^8$ terms coincide up to $x\approx0.1-0.2$. Furthermore, also for the
expansion around the equal-mass case the two highest approximations
overlap for $x\gsim 0.2-0.3$, depending on $n$. 
From the experience gained from the one- and two-loop considerations
we can conclude that in these regions our expressions constitute
perfect approximations to the exact result. Furthermore, it is again
straightforward to obtain approximations which are valid
for all $x$ values by a simple interpolation.
In order to obtain handy expressions we perform a polynomial
fit which is valid for the intermediate region $x\in [0.1,0.5]$. 
Allowing also for half-integer exponents of $x$ we obtain
\begin{eqnarray}
  \bar{C}_{1,{\rm appr}}^{(2),p}(x)\bigg|_{\mu^2=m_1^2} \!\!\!\!\!\!&=&
  0.382 + 22.191\,x^{1/2} + 5.126\,x - 138.613\,x^{3/2} + 149.094\,x^2 - 39.042\,x^3
  \,,\nonumber\\
  \bar{C}_{2,{\rm appr}}^{(2),p}(x)\bigg|_{\mu^2=m_1^2} \!\!\!\!\!\!&=&
  0.284 + 21.369\,x^{1/2} - 45.897\,x + 11.032\,x^{3/2} +
  26.207\,x^2 - 13.199\,x^3
  \,,\nonumber\\
  \bar{C}_{3,{\rm appr}}^{(2),p}(x)\bigg|_{\mu^2=m_1^2} \!\!\!\!\!\!&=&
  -0.307 + 15.358\,x^{1/2} - 50.366\,x + 61.164\,x^{3/2} - 27.39\,x^2 +
  1.475\,x^3 
  \,,\nonumber\\
  \bar{C}_{4,{\rm appr}}^{(2),p}(x)\bigg|_{\mu^2=m_1^2} \!\!\!\!\!\!&=&
  0.129 + 4.414\,x^{1/2} - 16.562\,x + 20.537\,x^{3/2} - 8.586\,x^2 - 0.031\,x^3
  \,.
  \label{eq::cp3lappr}
\end{eqnarray}
These results are valid for $n_l=3$ and are 
shown as dash-dotted lines in Fig.~\ref{fig::c3pas2}. We applied the same
procedure to the one- and two-loop moments. The comparison with the exact
result provides an estimate for the accuracy of the approximations in
Eqs.~(\ref{eq::cp3lappr}) to better than 5\% for $x\in[0.1,0.5]$.
For $x<0.1$ or $x>0.5$ the corresponding expansions should be used for the
evaluation of the moments since in these regions they provide excellent
approximations to the (unknown) exact result.
Analogue results for the vector, axial-vector and scalar correlators can be
found in~\cite{progdata} where again $n_l=3$ has been chosen. If necessary it
is straightforward to obtain formulae for other values for $n_l$.


\section{\label{sec::sum}Summary}

We have computed three-loop
QCD corrections to the moments of the current correlators
formed by quark fields with different masses.
Our final expressions are based on expansions around the
known limits where either one mass is zero or both
masses are equal. 
We present results which are valid for arbitrary quark masses by combining the
expansions and a simple polynomial fit in the intermediate region.

In the main part of the paper we concentrate on the discussion of the
pseudo-scalar correlator. However, numerical results for the vector,
axial-vector and scalar correlator can be found in the Appendix and analytical
expressions are provided in a {\tt Mathematica} file which comes together with
this paper.

One- and two-loop results with exact quark mass dependence and analytic
results for the expansions at three-loop order are presented in {\tt
  Mathematica} format on the internet page~\cite{progdata}. Also
the exact renormalization scale dependence of the three-loop moments are
provided and numerical approximations valid for arbitrary $m_2/m_1$.

The results obtained in this paper constitute important input for lattice
calculations in the context of semileptonic and leptonic $B$ meson decays. For
such simulations non-perturbative $Z$ factors for heavy-light currents are
needed which can be extracted from the corresponding current-current
correlators. Whereas in the case of the $B_s$ meson expansions for small
strange quark mass might be sufficient this is not true in the case of $B_c$
since $m_c/m_b\approx 0.23$ is in a region where the small-mass expansion
already breaks down.



\vspace*{2em}
{\large\bf Acknowledgments}

We would like to thank Hans K\"uhn for carefully reading the manuscript and
many valuable comments.
M.S. acknowledges discussions with C.~Davies and J.~Koponen which triggered
the calculations performed in this paper.
This work was supported by the BMBF through Grant No. 05H09VKE.


\begin{appendix}


\section{Results for the moments of the pseudo-scalar correlator}

In this Appendix we want to complete the results for the moments of the
pseudo-scalar correlator. The one-loop expressions can be found in
Eq.~(\ref{eq::Cn0p}) and $\bar{C}_{1}^{(1),p}(x)$ is given in
Eq.~(\ref{eq::Cn1p}). In Subsection~\ref{app:2lp} we provide the results for 
$\bar{C}_{n}^{(1),p}(x)$ with $n=-1,0,2,3$ and~$4$.
As far as the three-loop results are concerned we list in
Subsection~\ref{app:3lp} the exact $x$-dependence of the 
$\mu$-dependent terms which complement the results in
Eq.~(\ref{eq::cp3lappr}). 

\subsection{\label{app:2lp}Exact two-loop results}

{\scalefont{0.8}
\begin{flalign*}
\bar{C}_{-1}^{(1), p} (x) =\;& 
  - \frac{40 \lp 1 - x + x^2 \rp}{3} + l_x \, \frac{80 \, x^3}{3 \lp 1 + x \rp} 
    - l_x^2 \, \frac{32 \, x^3}{1 + x} \\
  & + l_\mu \lbk - \frac{40 \lp 1 - x + x^2 \rp}{3} + l_x \, \frac{32 \, x^3}{1 + x} \rbk 
    - 8 \, l_\mu^2 \lp 1 - x + x^2 \rp \forcetag \nonumber\,,\\
\bar{C}_0^{(1), p} (x) =\;& 
  \frac{13 + 66 \, x + 13 \, x^2}{6 \lp 1 + x \rp^2} 
    - \lbk \LiB{2}{x^2} + 2 \, l_x \, \l_{1 - x^2} - \zeta_2 \rbk \frac{4 \lp -1 + x \rp}{3 \lp 1 + x \rp} \\
  & - l_x \, \frac{8 \, x^2 \lp 1 + 5 \, x + x^2 \rp}{3 \lp -1 + x \rp \lp 1 + x \rp^3} 
    + l_x^2 \, \frac{16 \, x^3 \lp -6 - 3 \, x + 3 \, x^2 + 2 \, x^3 \rp}
      {3 \lp -1 + x \rp^2 \lp 1 + x \rp^4} \\
  & + l_\mu \lbk \frac{4 \lp 1 + 5 \, x + x^2 \rp}{3 \lp 1 + x \rp^2} 
    - l_x \, \frac{8 \, x^3 \lp 2 + x \rp}{\lp -1 + x \rp \lp 1 + x \rp^3} \rbk + 2 \, l_\mu^2 \forcetag \nonumber\,,\\
%
%
\bar{C}_2^{(1), p} (x) =\;& 
  \frac{3 + 72 \, x - 127 \, x^2 + 112 \, x^3 - 127 \, x^4 + 72 \, x^5 + 3 \, x^6}{36 \lp -1 + x \rp^4 \lp 1 + x \rp^6} \\
  & + l_x \, \frac{2 \, x^2 \lp 2 + 122 \, x - 27 \, x^2 + 230 \, x^3 - 22 \, x^4 + 12 \, x^5 + 3 \, x^6 \rp}
    {9 \lp -1 + x \rp^5 \lp 1 + x \rp^7} \\
  & - l_x^2 \, \frac{4 \, x^3 \lp 72 - 36 \, x + 216 \, x^2 - 68 \, x^3 + 136 \, x^4 - 3 \, x^5 + 4 \, x^6 + x^7 \rp}
    {9 \lp -1 + x \rp^6 \lp 1 + x \rp^8} \\
  & + l_\mu \lbk - \frac{1 + 4 \, x - 7 \, x^2 + 40 \, x^3 - 7 \, x^4 + 4 \, x^5 + x^6}{3 \lp -1 + x \rp^4 \lp 1 + x \rp^6} 
    + l_x \, \frac{8 \, x^3 \lp 2 - x + 2 \, x^2 \rp}{\lp -1 + x \rp^5 \lp 1 + x \rp^7} \rbk \\
  & + \lbk \LiB{2}{x^2} + 2 \, l_x \, l_{1 - x^2} - \zeta_2 \rbk \frac{2 \lp 1 + 4 \, x + 4 \, x^3 + x^4 \rp}
    {9 \lp -1 + x \rp^3 \lp 1 + x \rp^5} \forcetag \nonumber\,,\\
\bar{C}_3^{(1), p} (x) =\;& 
  - \frac{11 - 219 \, x + 658 \, x^2 + 1479 \, x^3 + 1326 \, x^4 + 1479 \, x^5 + 658 \, x^6 - 219 \, x^7 + 11 \, x^8}
    {270 \lp -1 + x \rp^6 \lp 1 + x \rp^8} \\
  & + l_x \lbk \frac{8 \, x^2 \lp 1 + 263 \, x - 184 \, x^2 + 1109 \, x^3 - 424 \, x^4 \rp}
    {45 \lp -1 + x \rp^7 \lp 1 + x \rp^9} \ra \\
  & \la \qquad + \frac{8 \, x^2 \lp 683 \, x^5 - 44 \, x^6 + 15 \, x^7 + 3 \, x^8 \rp}
    {45 \lp -1 + x \rp^7 \lp 1 + x \rp^9} \rbk \\
  & - l_x^2 \lbk \frac{8 \, x^3 \lp 180 - 180 \, x + 1260 \, x^2 - 800 \, x^3 + 1940 \, x^4 \rp}
    {45 \lp -1 + x \rp^8 \lp 1 + x \rp^{10}} \ra \\
  & \la \qquad + \frac{8 \, x^3 \lp - 535 \, x^5 + 545 \, x^6 - 4 \, x^7 + 5 \, x^8 + x^9 \rp}
    {45 \lp -1 + x \rp^8 \lp 1 + x \rp^{10}} \rbk \\
  & + \lbk \LiB{2}{x^2} + 2 \, l_x \, l_{1 - x^2} - \zeta_2 \rbk \frac{4 \, \lp 1 + 5 \, x - x^2 + 20 \, x^3 - x^4 + 5 \, x^5 + x^6 \rp}
    {45 \lp -1 + x \rp^5 \lp 1 + x \rp^7} \\
  & + l_\mu \lbk l_x \, \frac{32 \, x^3 \lp 1 - x + 3 \, x^2 - x^3 + x^4 \rp}{\lp -1 + x \rp^7 \lp 1 + x \rp^9} \ra \\
  & \la \qquad - \frac{4 \lp 1 + 5 \, x - 14 \, x^2 + 145 \, x^3 - 94 \, x^4 + 145 \, x^5 - 14 \, x^6 + 5 \, x^7 + x^8 \rp}
    {15 \lp -1 + x \rp^6 \lp 1 + x \rp^8} \rbk \forcetag \nonumber\,,\\
%
\bar{C}_4^{(1), p} (x) =\;& 
  - \frac{61 - 201 \, x + 978 \, x^2 + 19344 \, x^3 - 7591 \, x^4 + 51570 \, x^5}{810 \lp -1 + x \rp^8 \lp 1 + x \rp^{10}} \\
  & - \frac{- 7591 \, x^6 + 19344 \, x^7 + 978 \, x^8 - 201 \, x^9 + 61 \, x^{10}}{810 \lp -1 + x \rp^8 \lp 1 + x \rp^{10}} \\
  & + l_x \lbk \frac{2 \, x^2 \lp 2 + 1512 \, x - 1780 \, x^2 + 11358 \, x^3 - 7758 \, x^4 + 16758 \, x^5 \rp}
    {45 \lp -1 + x \rp^9 \lp 1 + x \rp^{11}} \ra \\
  & \la \qquad + \frac{2 \, x^2 \lp - 5155 \, x^6 + 4662 \, x^7 - 214 \, x^8 + 54 \, x^9 + 9 \, x^{10} \rp}
    {45 \lp -1 + x \rp^9 \lp 1 + x \rp^{11}} \rbk \\
  & - l_x^2 \lbk \frac{4 \, x^3 \lp 360 - 540 \, x + 4500 \, x^2 - 4400 \, x^3 + 13200 \, x^4 - 7055 \, x^5 \rp}
    {45 \lp -1 + x \rp^{10} \lp 1 + x \rp^{12}} \ra \\
  & \la \qquad + \frac{4 \, x^3 \lp 9810 \, x^6 - 2169 \, x^7 + 1476 \, x^8 - 5 \, x^9 + 6 \, x^{10} + x^{11} \rp}
    {45 \lp -1 + x \rp^{10} \lp 1 + x \rp^{12}} \rbk \\
  & + \lbk \LiB{2}{x^2} + 2 \, l_x \, l_{1 - x^2} - \zeta_2 \rbk \\
  & \qquad \frac{2 \, \lp 1 + 6 \, x - 2 \, x^2 + 54 \, x^3 - 18 \, x^4 + 54 \, x^5 - 2 \, x^6 + 6 \, x^7 + x^8 \rp}
    {45 \lp -1 + x \rp^7 \lp 1 + x \rp^9} \\
  & + l_\mu \lbk - \frac{1 + 6 \, x - 23 \, x^2 + 356 \, x^3 - 398 \, x^4 + 956 \, x^5}
    {5 \lp -1 + x \rp^8 \lp 1 + x \rp^{10}} \ra \\ 
  & \qquad - \frac{- 398 \, x^6 + 356 \, x^7 - 23 \, x^8 + 6 \, x^9 + x^{10}}{5 \lp -1 + x \rp^8 \lp 1 + x \rp^{10}} \\
  & \la \qquad + l_x \, \frac{24 \, x^3 \lp 2 - 3 \, x + 12 \, x^2 - 8 \, x^3 + 12 \, x^4 - 3 \, x^5 + 2 \, x^6 \rp}
    {\lp -1 + x \rp^9 \lp 1 + x \rp^{11}} \rbk \forcetag
\,.
\end{flalign*}
}

\subsection{\label{app:3lp}Analytic three-loop results}

Let us decompose the the three-loop coefficients as 
\begin{eqnarray}
  \bar{C}_{n}^{(2),p}(x) &=&
  \bar{C}_{n}^{(2),p}(x)\bigg|_{\mu^2=m_1^2}
  + l_\mu \bar{C}_{n,1}^{(2),p}(x)
  + l_\mu^2 \bar{C}_{n,2}^{(2),p}(x)
  \,,
\end{eqnarray}
with $l_\mu=\ln(\mu^2/m_1^2)$. For $n_l=3$ one can find approximations for
$\bar{C}_{n}^{(2),p}(x)\big|_{\mu^2=m_1^2}$ in Eq.~(\ref{eq::cp3lappr}).
The exact $x$-dependence of 
$\bar{C}_{n,1}^{(2),p}(x)$ and $\bar{C}_{n,2}^{(2),p}(x)$, which can be
obtained using renormalization group techniques, reads
{\scalefont{0.8}
\begin{flalign*}
\bar{C}_{1, 1}^{(2), p} (x) =\;& 
  \lp - \frac{29}{2} + n_l \rp \lbc - \frac{1 + 36 \, x - 22 \, x^2 + 36 \, x^3 + x^4}{54 \lp -1 + x \rp^2 \lp 1 + x \rp^4} \ra \\
  & - \lbk \LiB{2}{x^2} + 2 \, l_x \, l_{1 - x^2} - \zeta_2 \rbk \frac{4 \lp 1 + 3 \, x + x^2 \rp}{27 \lp -1 + x \rp \lp 1 + x \rp^3} \\
  & \la - l_x \, \frac{8 \, x^2 \lp 1 + 6 \, x + x^2 \rp}{27 \lp -1 + x \rp^3 \lp 1 + x \rp^5} 
    + l_x^2 \, \frac{8 \, x^3 \lp 18 + 9 \, x^2 - 2 \, x^3 + 3 \, x^4 + x^5 \rp}{27 \lp -1 + x \rp^4 \lp 1 + x \rp^6} \rbc \,,\forcetag \\
\bar{C}_{1, 2}^{(2), p} (x) =\;& 
  0 \,,\\
\bar{C}_{2, 1}^{(2), p} (x) =\;& 
  - \frac{19 \lp 1 + 4 \, x - 7 \, x^2 + 40 \, x^3 - 7 \, x^4 + 4 \, x^5 + x^6 \rp}{36 \lp -1 + x \rp^4 \lp 1 + x \rp^6} 
    + l_x \, \frac{38 \, x^3 \lp 2 - x + 2 \, x^2 \rp}{3 \lp -1 + x \rp^5 \lp 1 + x \rp^7} \\
  & + \lp - \frac{5}{2} + n_l \rp \lbc \frac{7 - 32 \, x + 57 \, x^2 + 288 \, x^3 + 57 \, x^4 - 32 \, x^5 + 7 \, x^6}
    {216 \lp -1 + x \rp^4 \lp 1 + x \rp^6} \ra \\
  & \qquad - \lbk \LiB{2}{x^2} + 2 \, l_x \, l_{1 - x^2} - \zeta_2 \rbk \frac{1 + 4 \, x + 4 \, x^3 + x^4}{27 \lp -1 + x \rp^3 \lp 1 + x \rp^5} \\
  & \qquad - l_x \, \frac{x^2 \lp 2 + 182 \, x - 57 \, x^2 + 290 \, x^3 - 22 \, x^4 + 12 \, x^5 + 3 \, x^6 \rp}
    {27 \lp -1 + x \rp^5 \lp 1 + x \rp^7} \\
  & \la \qquad + l_x^2 \, \frac{2 \, x^3 \lp 72 - 36 \, x + 216 \, x^2 - 68 \, x^3 + 136 \, x^4 - 3 \, x^5 + 4 \, x^6 + x^7 \rp}
  	{27 \lp -1 + x \rp^6 \lp 1 + x \rp^8} \rbc \,,\forcetag \\
\bar{C}_{2, 2}^{(2), p} (x) =\;& 
  \lp - \frac{5}{2} + n_l \rp \lbk \frac{1 + 4 \, x - 7 \, x^2 + 40 \, x^3 - 7 \, x^4 + 4 \, x^5 + x^6}{36 \lp -1 + x \rp^4 \lp 1 + x \rp^6} \ra \\
  & \la \qquad - l_x \, \frac{2 \, x^3 \lp 2 - x + 2 \, x^2 \rp}{3 \lp -1 + x
    \rp^5 \lp 1 + x \rp^7} \rbk \,,\forcetag \\
%
\bar{C}_{3, 1}^{(2), p} (x) =\;& 
   - \frac{13 \lp 1 + 5 \, x - 14 \, x^2 + 145 \, x^3 - 94 \, x^4 + 145 \, x^5 - 14 \, x^6 + 5 \, x^7 + x^8 \rp}
     {15 \lp -1 + x \rp^6 \lp 1 + x \rp^8} \\
   & + l_x \, \frac{104 \, x^3 \lp 1 - x + 3 \, x^2 - x^3 + x^4 \rp}{\lp -1 + x \rp^7 \lp 1 + x \rp^9} + \lp \frac{19}{2} + n_l \rp \\
   & \quad  \lbc \frac{71 + 81 \, x - 182 \, x^2 + 10179 \, x^3 - 4314 \, x^4 + 10179 \, x^5 - 182 \, x^6 + 81 \, x^7 + 71 \, x^8}
     {1620 \lp -1 + x \rp^6 \lp 1 + x \rp^8} \ra \\
   & \qquad - \lbk \LiB{2}{x^2} + 2 \, l_x \, l_{1 - x^2} - \zeta_2 \rbk \frac{2 \lp 1 + 5 \, x - x^2 + 20 \, x^3 - x^4 + 5 \, x^5 + x^6 \rp}
     {135 \lp -1 + x \rp^5 \lp 1 + x \rp^7} \\
   & \qquad - l_x \, \frac{4 \, x^2 \lp 1 + 413 \, x - 334 \, x^2 + 1559 \, x^3 - 574 \, x^4 + 833 \, x^5 - 44 \, x^6 + 15 \, x^7 + 3 \, x^8 \rp}
     {135 \lp -1 + x \rp^7 \lp 1 + x \rp^9} \\
   & \la \qquad + l_x^2 \, 
     \frac{4 \, x^3 \lp 180 - 180 \, x + 1260 \, x^2 - 800 \, x^3 + 1940 \, x^4 - 535 \, x^5 + 545 \, x^6 - 4 \, x^7 + 5 \, x^8 + x^9 \rp}
       {135 \lp -1 + x \rp^8 \lp 1 + x \rp^{10}} \rbc \,,\forcetag \\
\bar{C}_{3, 2}^{(2), p} (x) =\;& 
  \lp \frac{19}{2} + n_l \rp \lbk \frac{1 + 5 \, x - 14 \, x^2 + 145 \, x^3 - 94 \, x^4 + 145 \, x^5 - 14 \, x^6 + 5 \, x^7 + x^8}
    {45 \lp -1 + x \rp^6 \lp 1 + x \rp^8} \ra \\
  & \la \qquad - l_x \, \frac{8 \, x^3 \lp 1 - x + 3 \, x^2 - x^3 + x^4 \rp}{3 \lp -1 + x \rp^7 \lp 1 + x \rp^9} \rbk \,,\forcetag\\
%
\bar{C}_{4, 1}^{(2), p} (x) =\;& 
  - \frac{59 \lp 1 + 6 \, x - 23 \, x^2 + 356 \, x^3 - 398 \, x^4 + 956 \, x^5 - 398 \, x^6 + 356 \, x^7 - 23 \, x^8 + 6 \, x^9 + x^{10} \rp}
    {60 \lp -1 + x \rp^8 \lp 1 + x \rp^{10}} \\
  & + l_x \, \frac{118 \, x^3 \lp 2 - 3 \, x + 12 \, x^2 - 8 \, x^3 + 12 \, x^4 - 3 \, x^5 + 2 \, x^6 \rp}{\lp -1 + x \rp^9 \lp 1 + x \rp^{11}} 
    + \lp \frac{43}{2} + n_l \rp \\
  & \quad \lbc \frac{196 + 609 \, x - 2127 \, x^2 + 67404 \, x^3 - 61321 \, x^4 + 180630 \, x^5}{4860 \lp -1 + x \rp^8 \lp 1 + x \rp^{10}} \ra \\
  & \qquad + \frac{- 61321 \, x^6 + 67404 \, x^7 - 2127 \, x^8 + 609 \, x^9 + 196 \, x^{10}}{4860 \lp -1 + x \rp^8 \lp 1 + x \rp^{10}} \\
  & \qquad - \lbk \LiB{2}{x^2} + 2 \, l_x \, l_{1 - x^2} - \zeta_2 \rbk \frac{1 + 6 \, x - 2 \, x^2 + 54 \, x^3 - 18 \, x^4 + 54 \, x^5 - 2 \, x^6 + 6 \, x^7 + x^8}{135 \lp -1 + x \rp^7 \lp 1 + x \rp^9} \\
  & \qquad - l_x \, x^2 \lbk \frac{2 + 2412 \, x - 3130 \, x^2 + 16758 \, x^3 - 11358 \, x^4 + 22158 \, x^5}{135 \lp -1 + x \rp^9 \lp 1 + x \rp^{11}} \ra \\
  & \la \qquad \qquad + \frac{- 6505 \, x^6 + 5562 \, x^7 - 214 \, x^8 + 54 \, x^9 + 9 \, x^{10}}{135 \lp -1 + x \rp^9 \lp 1 + x \rp^{11}} \rbk \\
  & \qquad + 2 \, l_x^2 \, x^3 \lbk \frac{360 - 540 \, x + 4500 \, x^2 - 4400 \, x^3 + 13200 \, x^4 - 7055 \, x^5}{135 \lp -1 + x \rp^{10} \lp 1 + x \rp^{12}} \ra \\
  & \la \la \qquad \qquad + \frac{9810 \, x^6 - 2169 \, x^7 + 1476 \, x^8 - 5 \, x^9 + 6 \, x^{10} + x^{11}}{135 \lp -1 + x \rp^{10} \lp 1 + x \rp^{12}} \rbk \rbc \,,\forcetag \\
\bar{C}_{4, 2}^{(2), p} (x) =\;& 
  \lp \frac{43}{2} + n_l \rp \lbk - \frac{2 \, l_x \, x^3 \lp 2 - 3 \, x + 12 \, x^2 - 8 \, x^3 + 12 \, x^4 - 3 \, x^5 + 2 \, x^6 \rp}{\lp -1 + x \rp^9 \lp 1 + x \rp^{11}} \ra \\
  & \la \qquad + \frac{1 + 6 \, x - 23 \, x^2 + 356 \, x^3 - 398 \, x^4 + 956 \, x^5 - 398 \, x^6 + 356 \, x^7 - 23 \, x^8 + 6 \, x^9 + x^{10}}{60 \lp -1 + x \rp^8 \lp 1 + x \rp^{10}} \rbk\,. \forcetag
\end{flalign*}
}

\section{\label{app::apprvas}Results for vector, axial-vector and
  scalar correlator}

In this appendix we present the results for the moments 
$\bar{C}_n^\delta(x)$ for $n=1,\ldots,4$ of the vector, axial-vector and
scalar correlator. For the analytic results and numerical approximations for
the three-loop moments in the intermediate $x$-region we refer to~\cite{progdata}
and show in the following the corresponding numerical 
result in analogy to Figs.~\ref{fig::c3pas0},~\ref{fig::c3pas1} and~\ref{fig::c3pas2}.

\begin{figure}[t]
  \centering
  \begin{tabular}{cc}
    \includegraphics[width=0.4\linewidth]{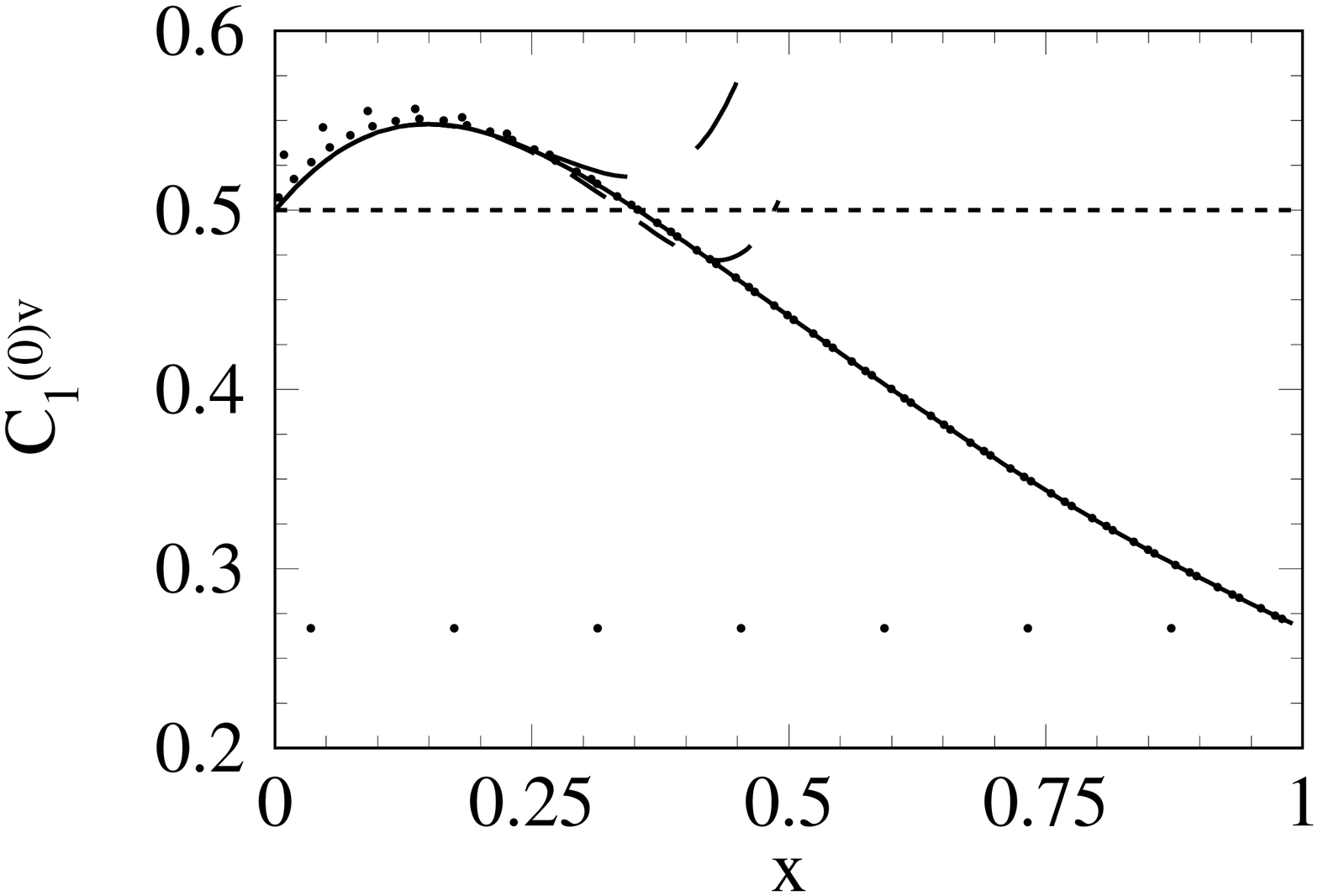}
    &
    \includegraphics[width=0.4\linewidth]{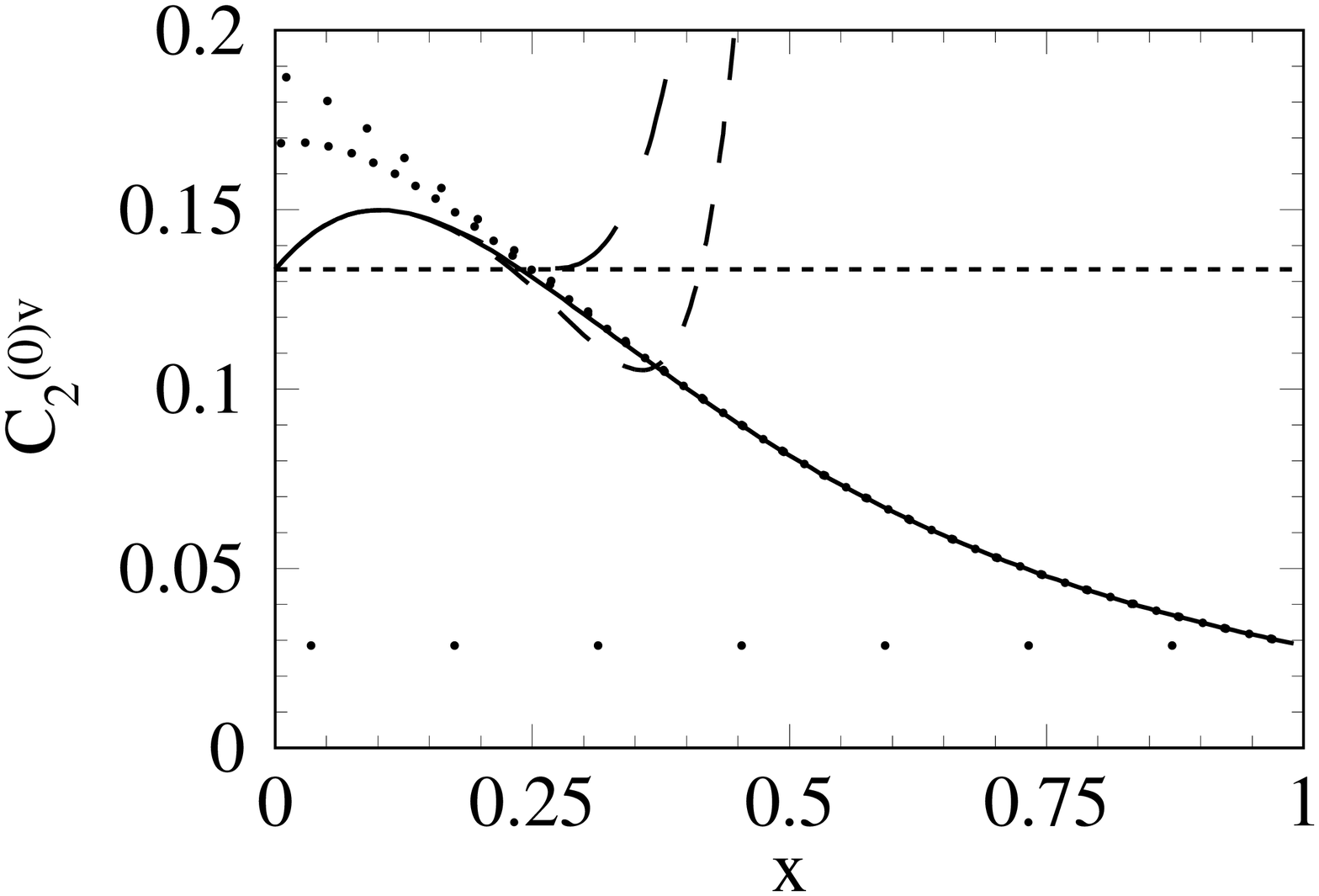}
    \\
    \includegraphics[width=0.4\linewidth]{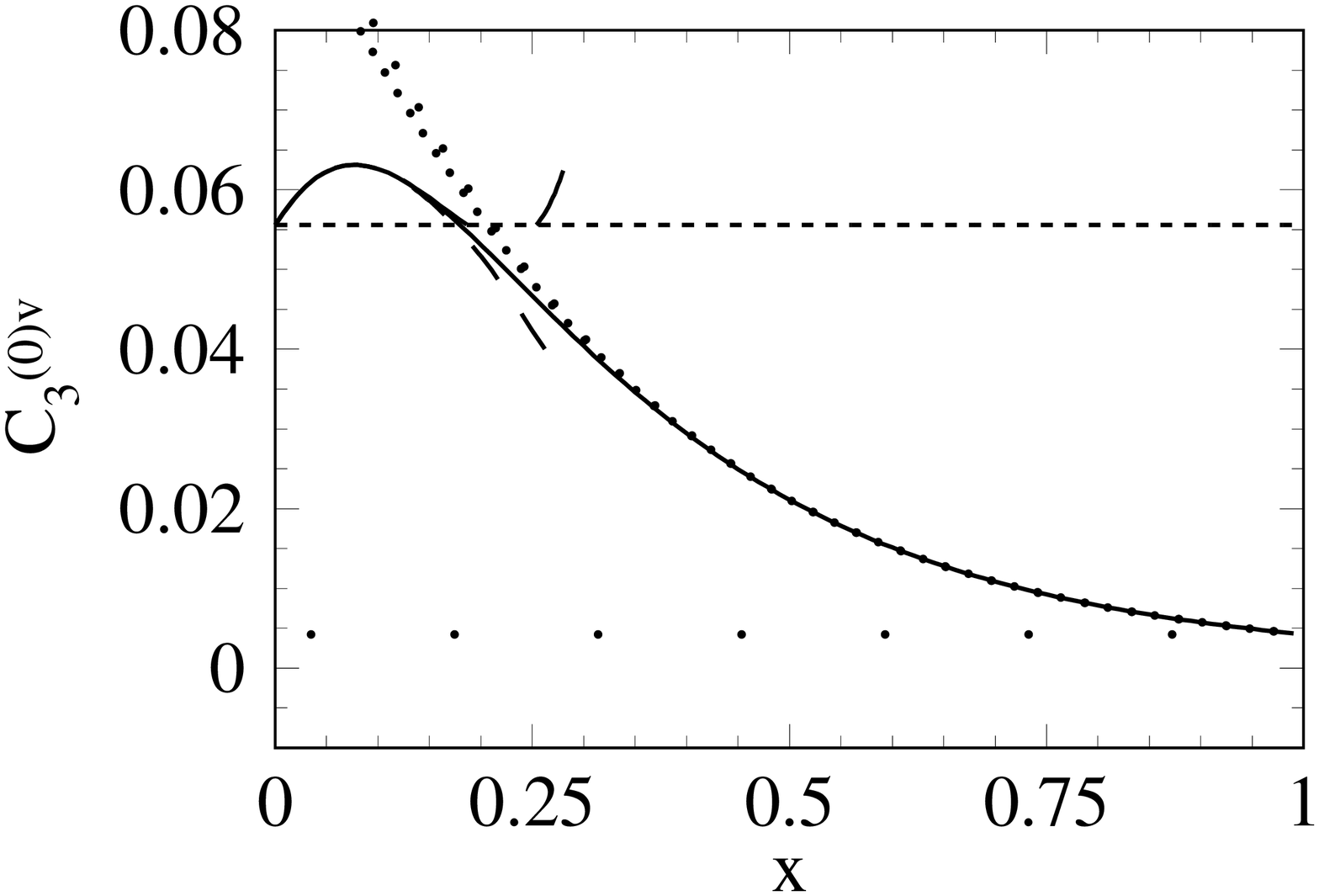}
    &
    \includegraphics[width=0.4\linewidth]{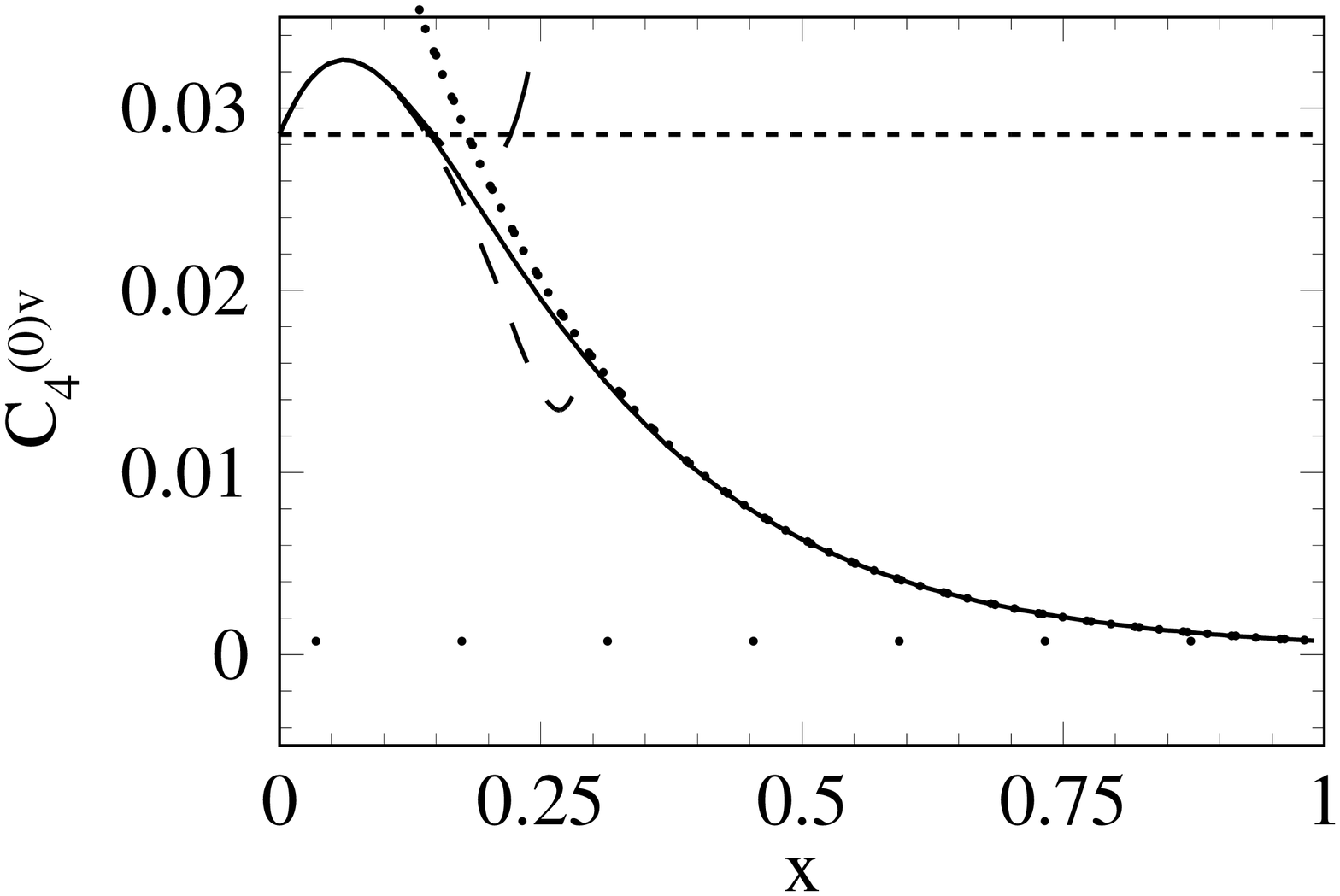}
  \end{tabular}
  \caption[]{\label{fig::c3vas0}One-loop
    contribution to $\bar{C}_n^v$. The same notation as in
    Fig.~\ref{fig::c3pas0} is adopted.
  }
\end{figure}

\begin{figure}[t]
  \centering
  \begin{tabular}{cc}
    \includegraphics[width=0.4\linewidth]{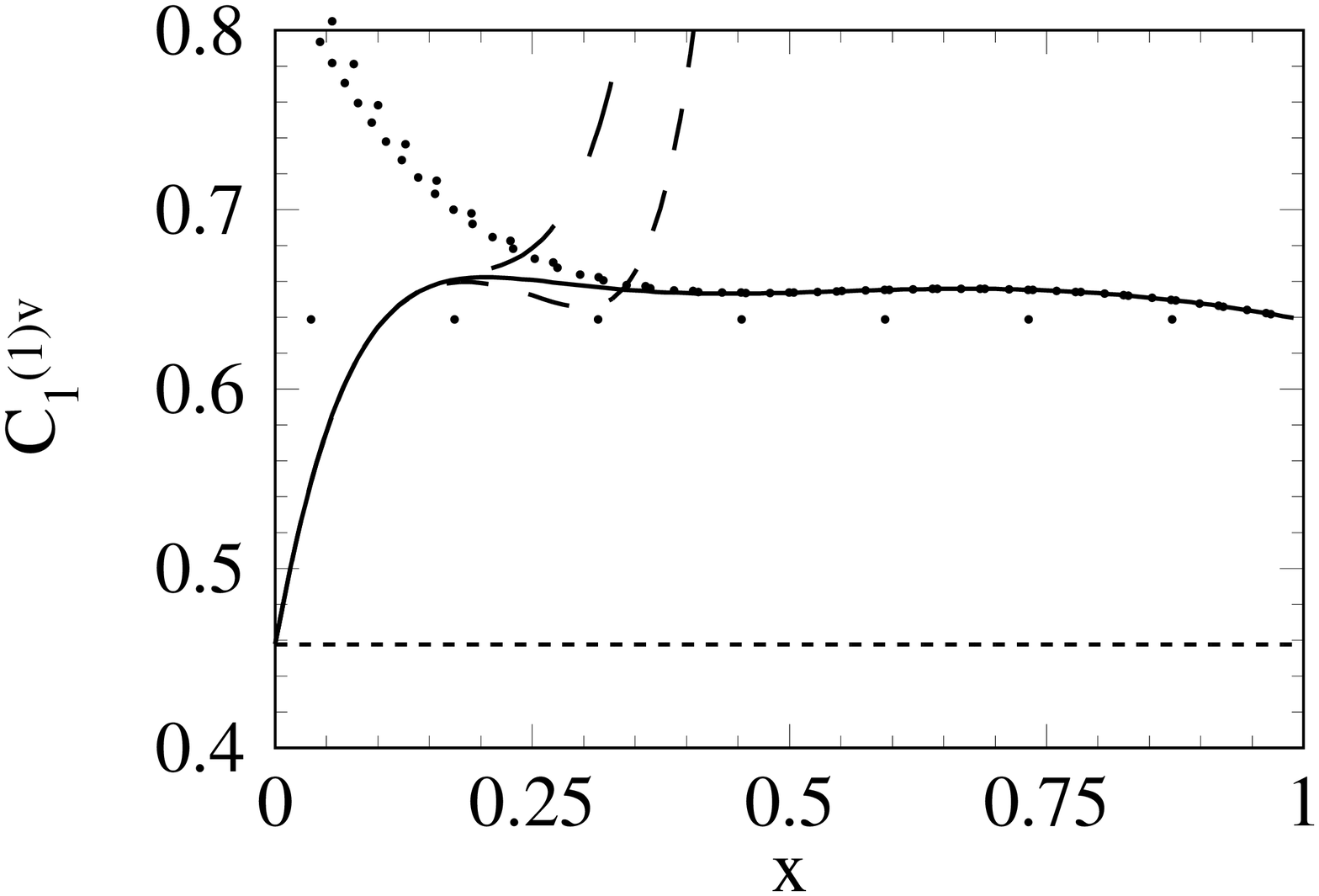}
    &
    \includegraphics[width=0.4\linewidth]{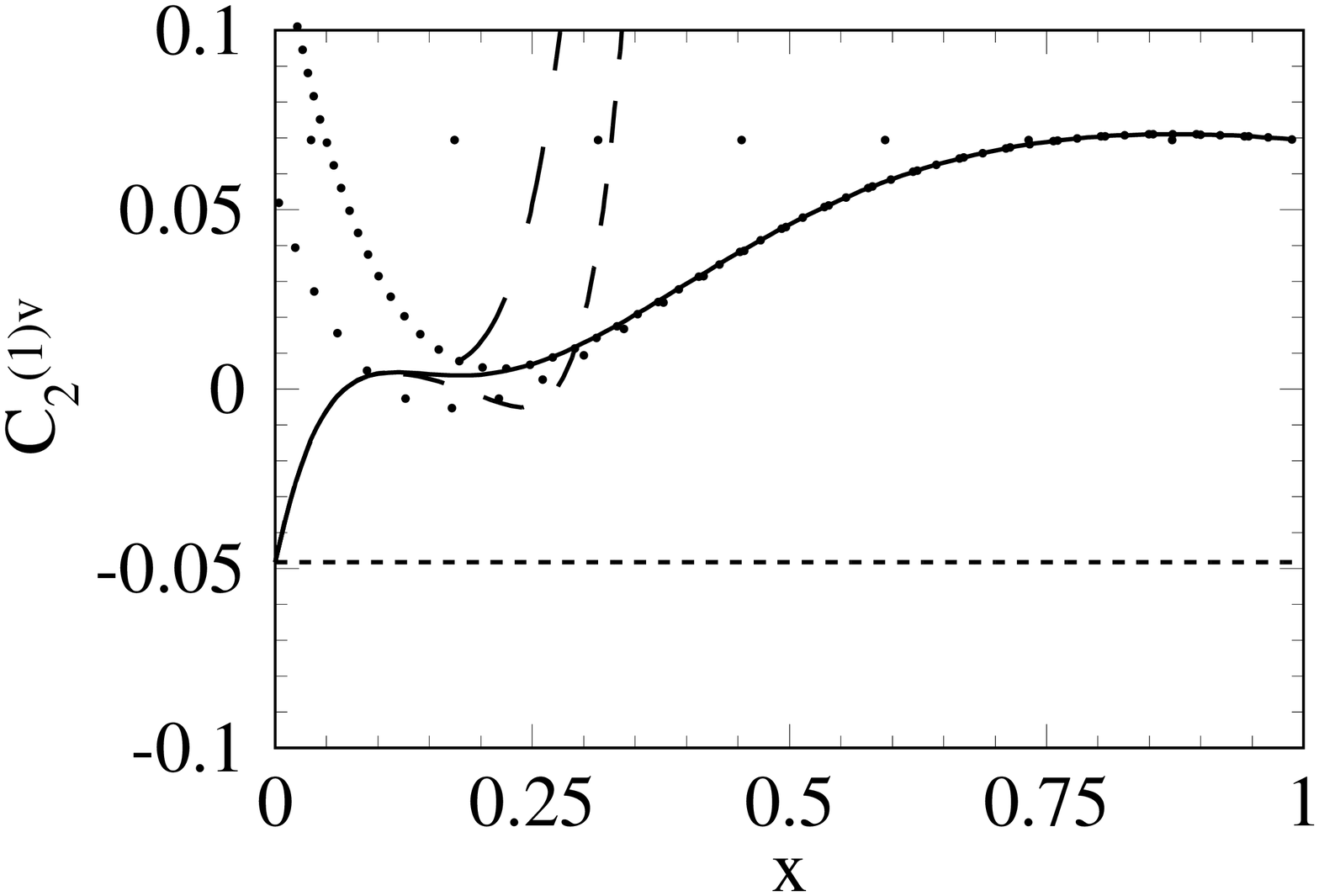}
    \\
    \includegraphics[width=0.4\linewidth]{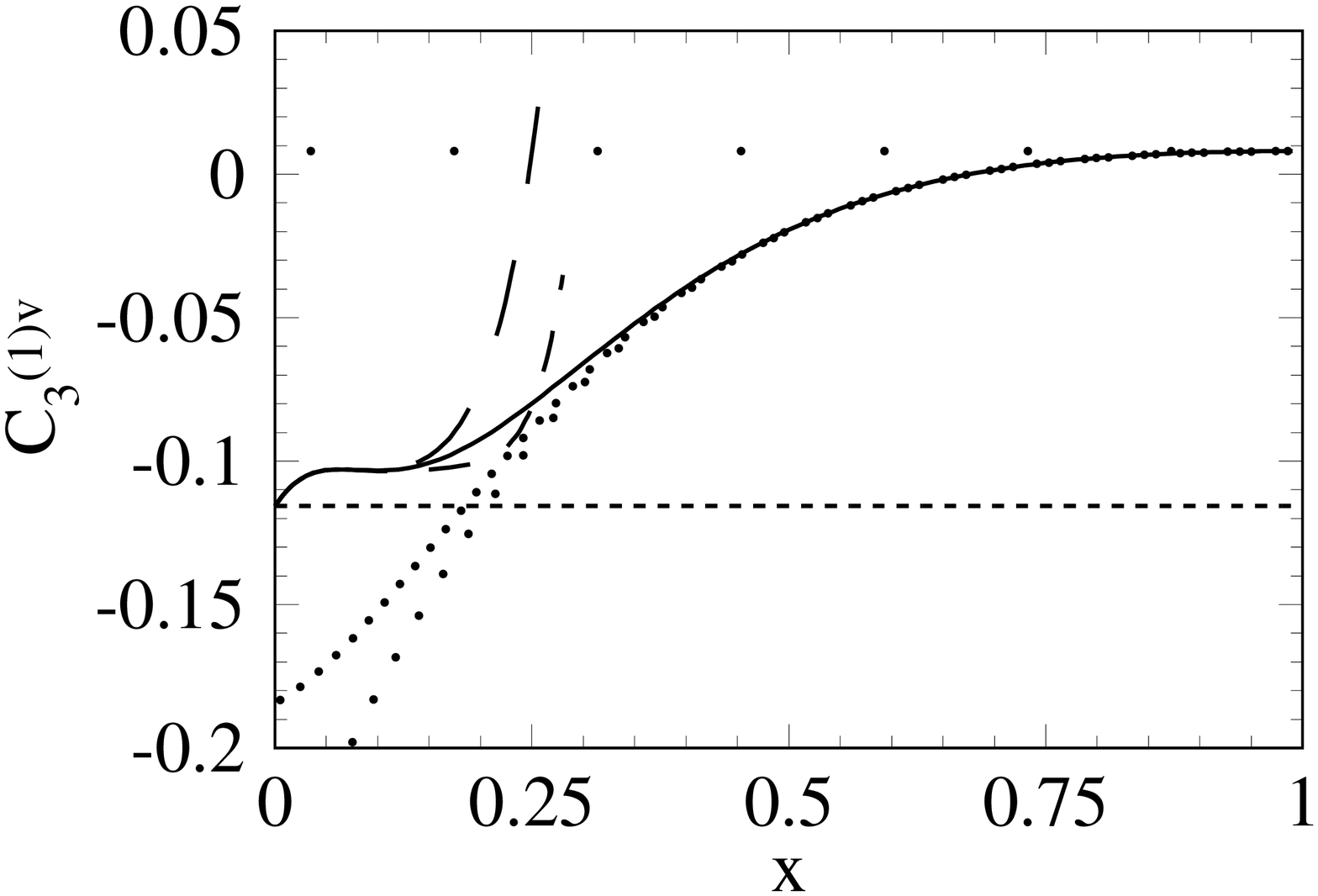}
    &
    \includegraphics[width=0.4\linewidth]{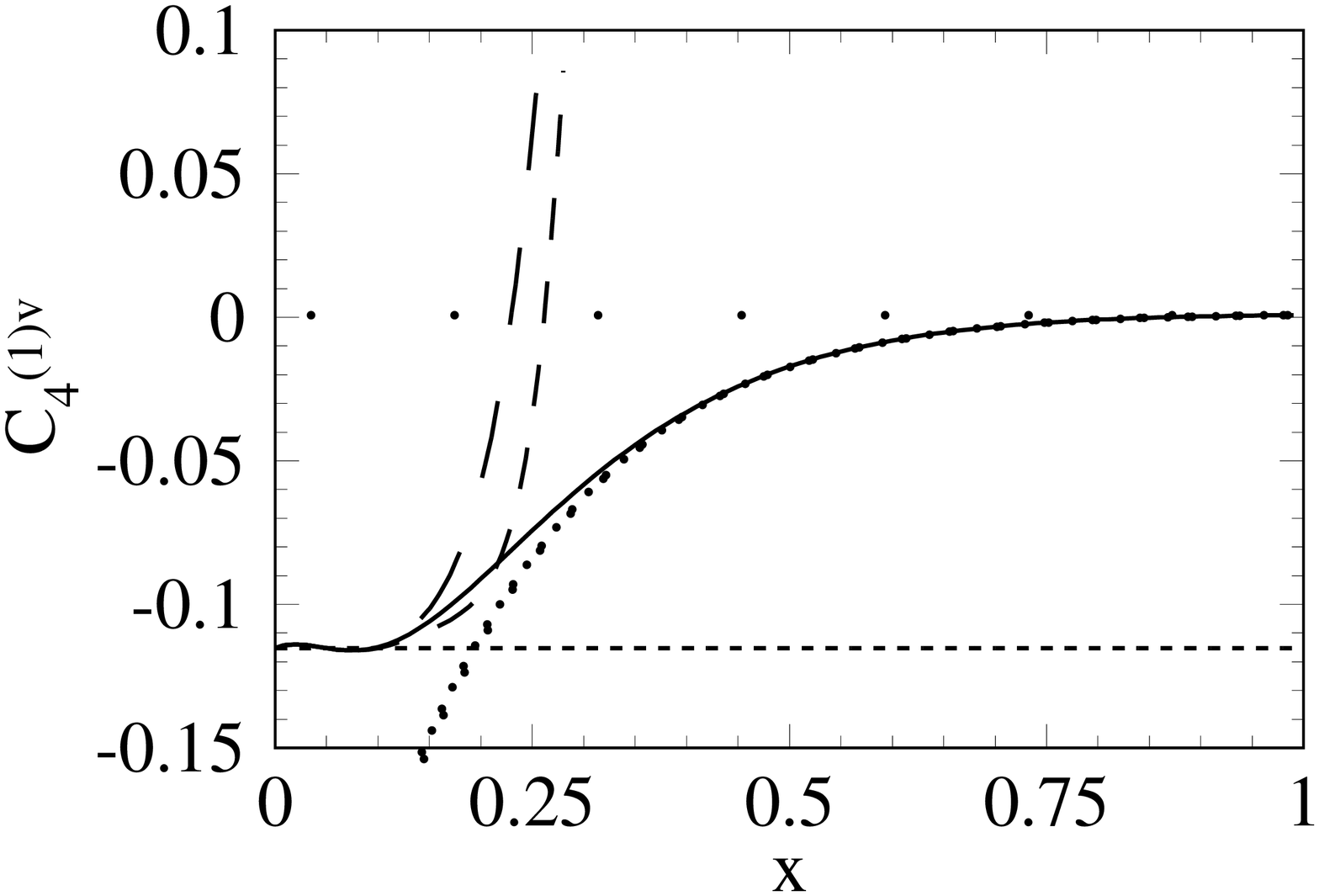}
  \end{tabular}
  \caption[]{\label{fig::c3vas1}Two-loop
    contribution to $\bar{C}_n^v$. The same notation as in
    Fig.~\ref{fig::c3pas0} is adopted.
  }
\end{figure}

\begin{figure}[t]
  \centering
  \begin{tabular}{cc}
    \includegraphics[width=0.4\linewidth]{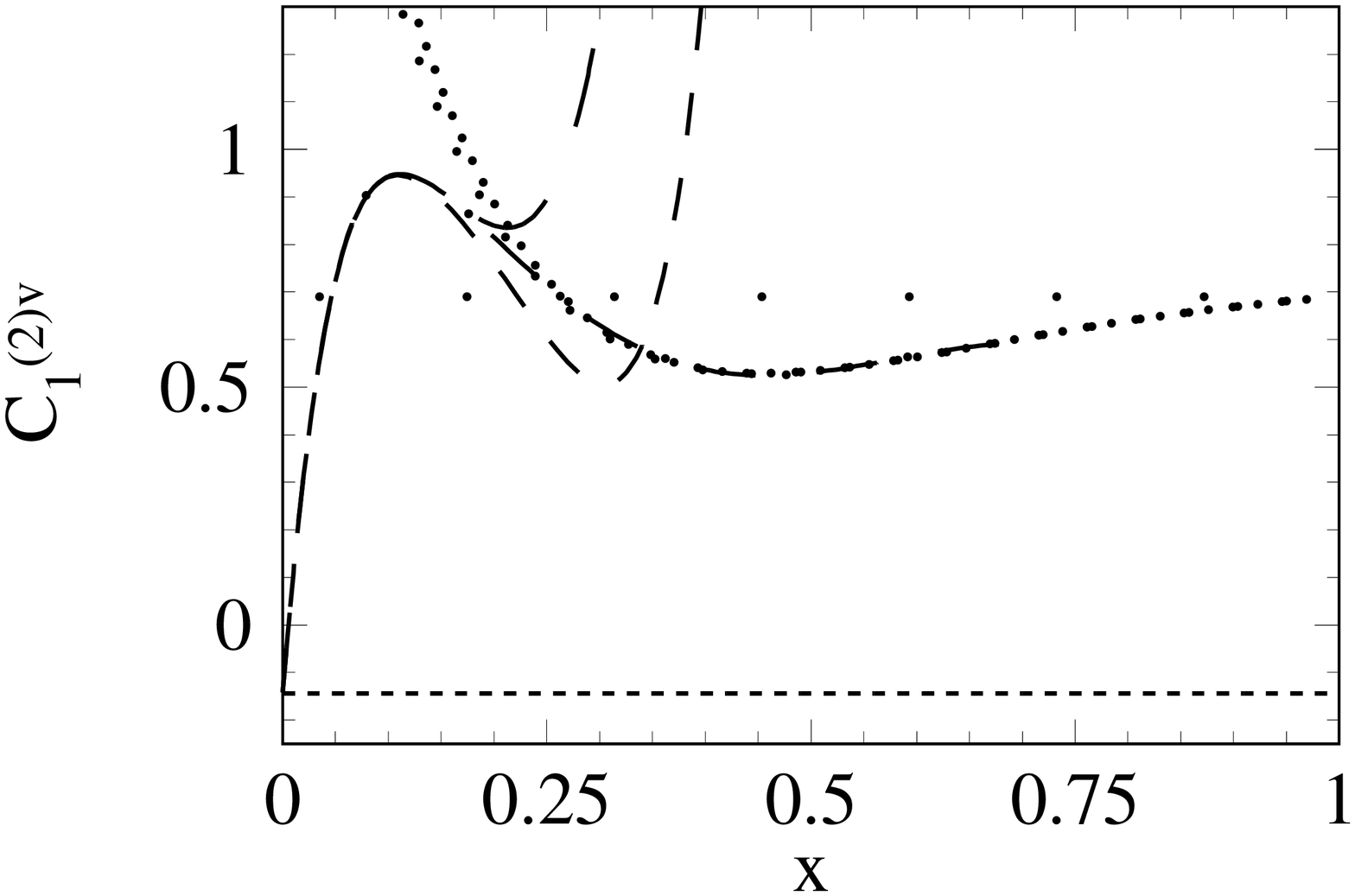}
    &
    \includegraphics[width=0.4\linewidth]{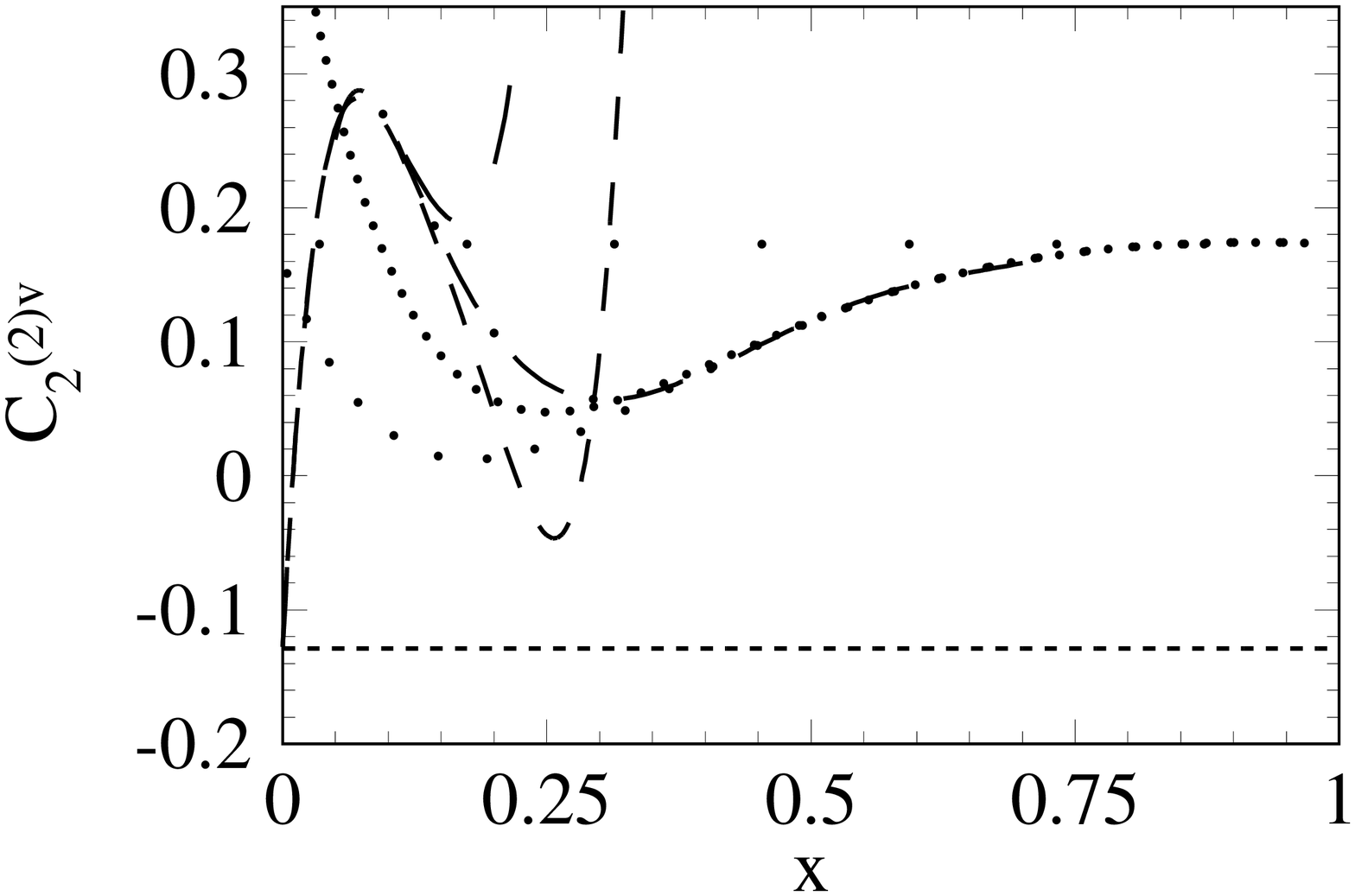}
    \\
    \includegraphics[width=0.4\linewidth]{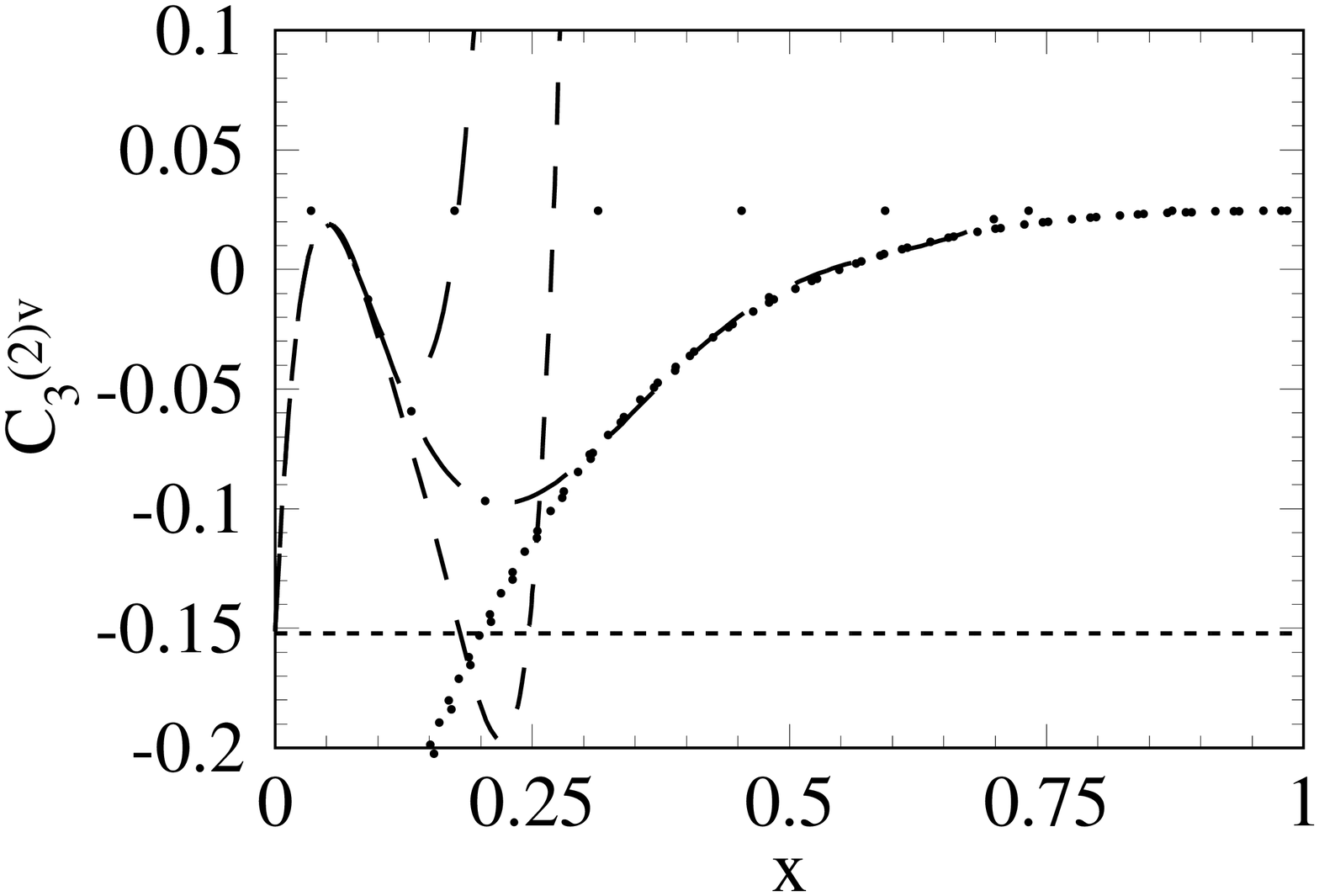}
    &
    \includegraphics[width=0.4\linewidth]{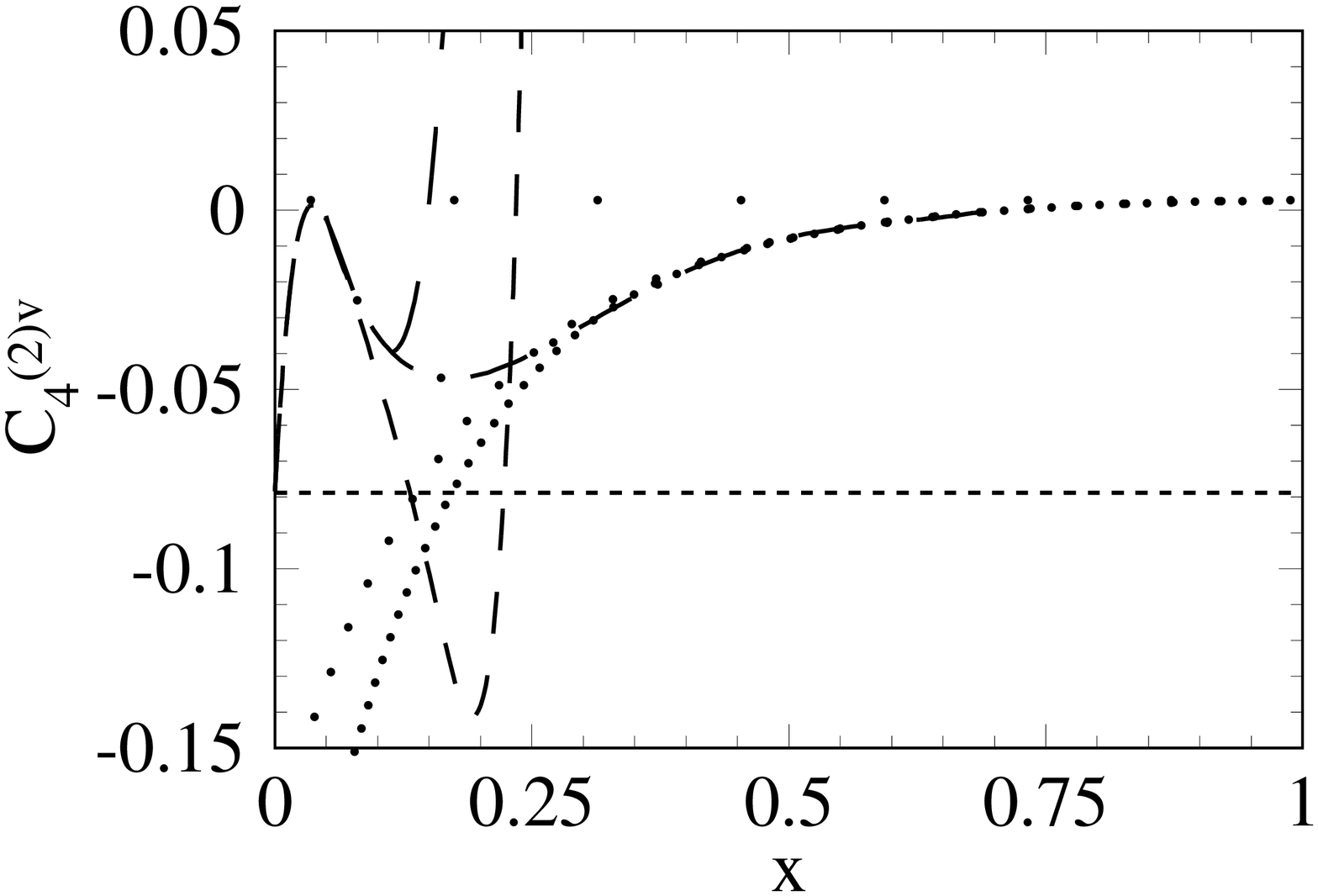}
  \end{tabular}
  \caption[]{\label{fig::c3vas2}Three-loop
    contribution to $\bar{C}_n^v$. The same notation as in
    Fig.~\ref{fig::c3pas2} is adopted.
  }
\end{figure}


\begin{figure}[t]
  \centering
  \begin{tabular}{cc}
    \includegraphics[width=0.4\linewidth]{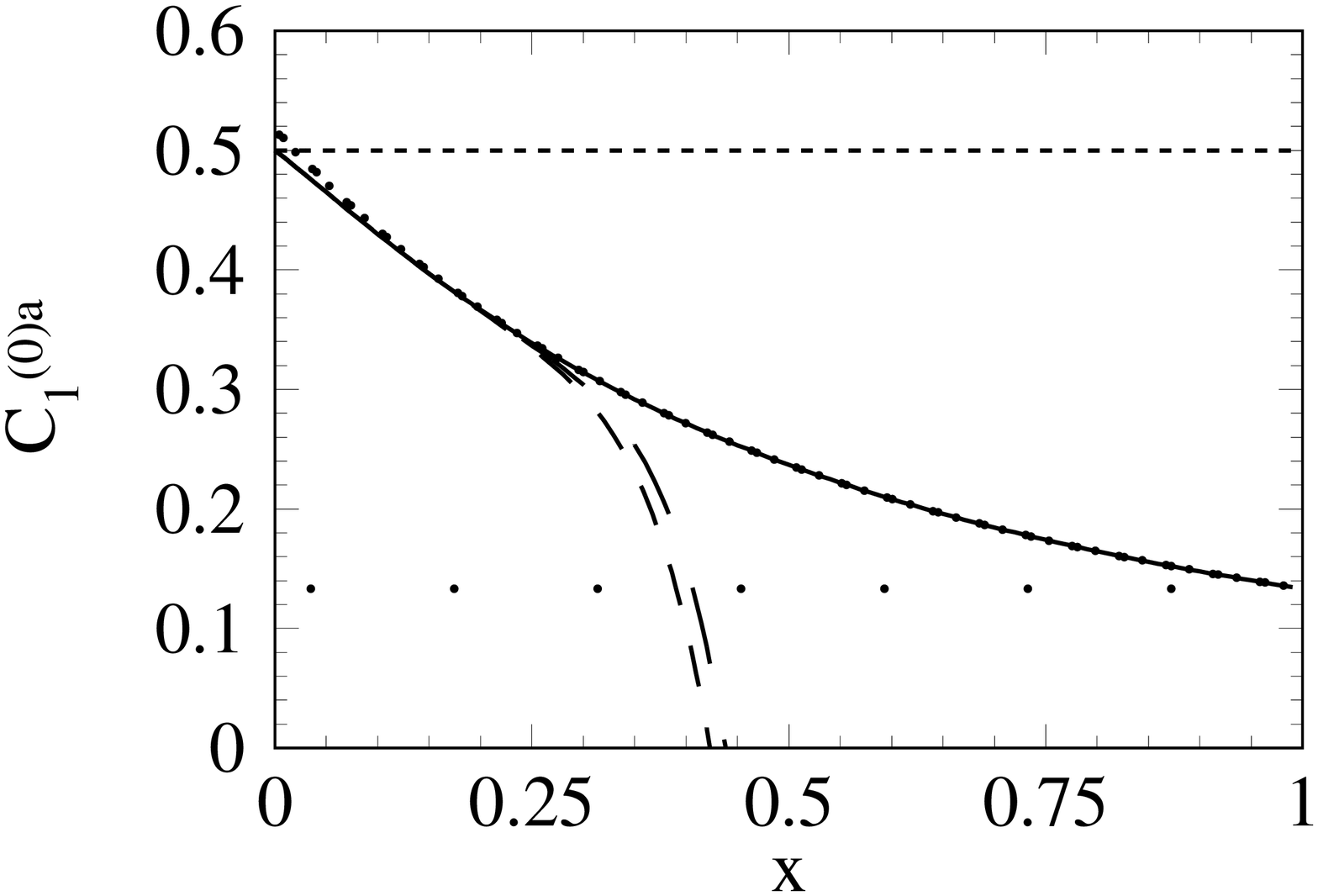}
    &
    \includegraphics[width=0.4\linewidth]{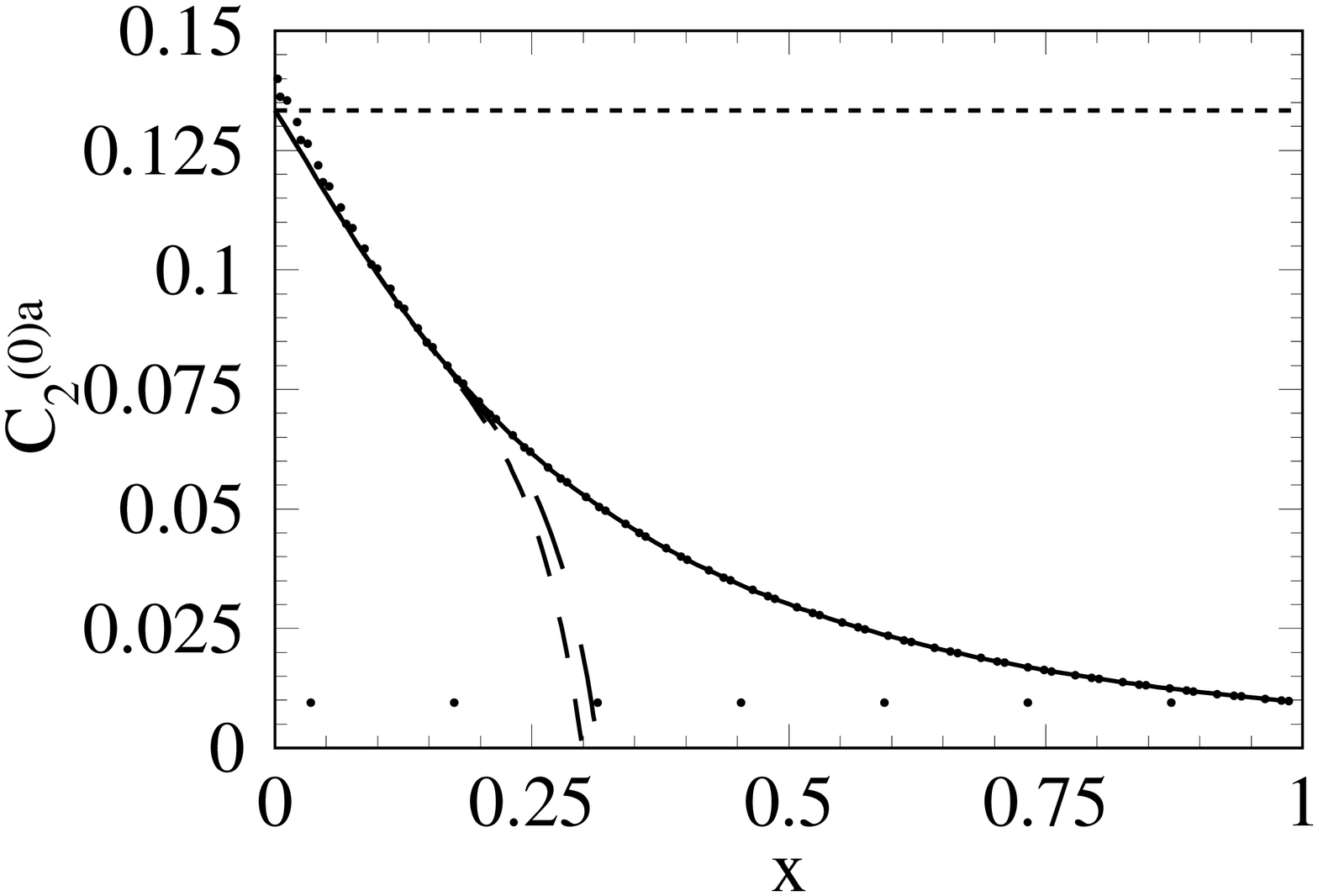}
    \\
    \includegraphics[width=0.4\linewidth]{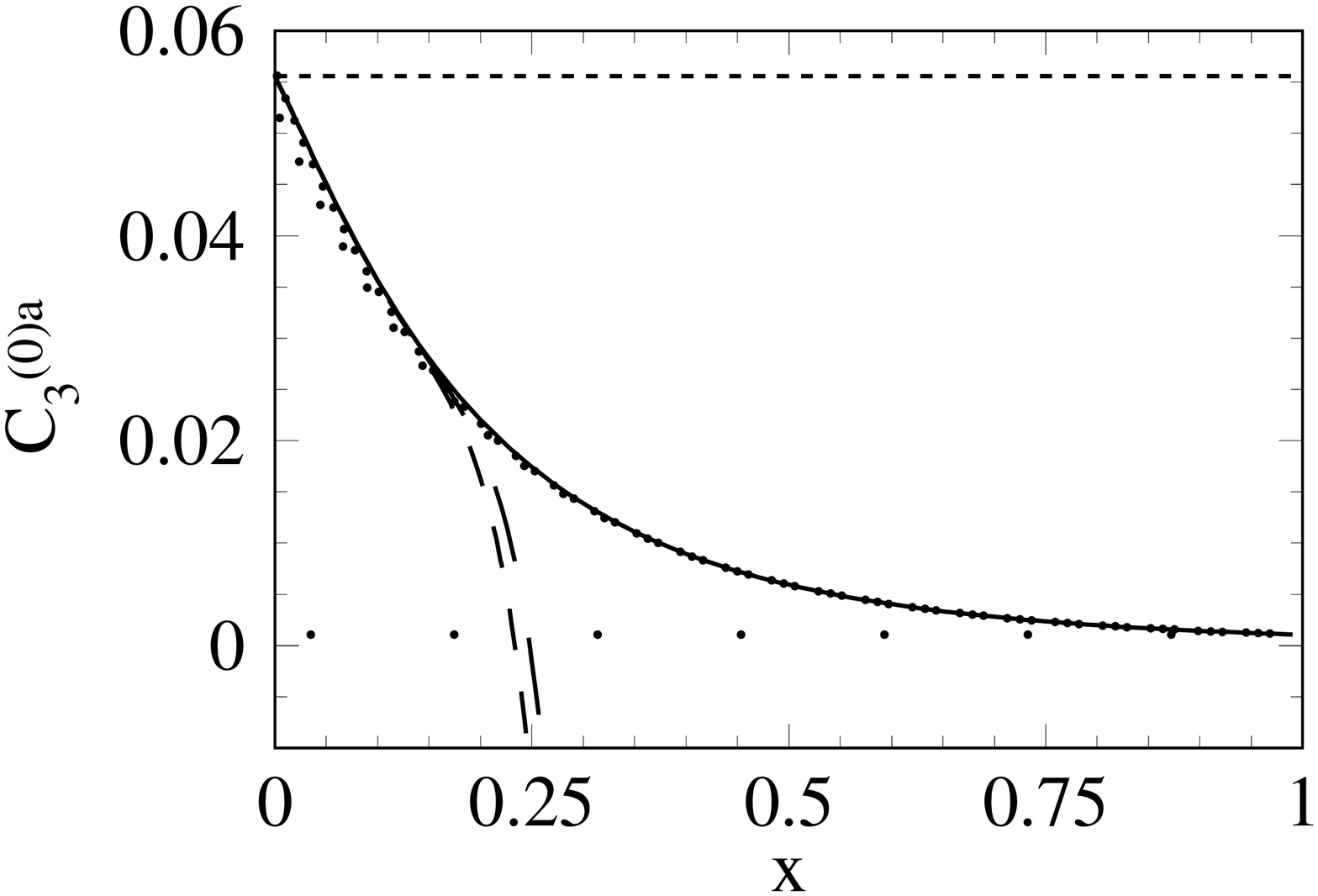}
    &
    \includegraphics[width=0.4\linewidth]{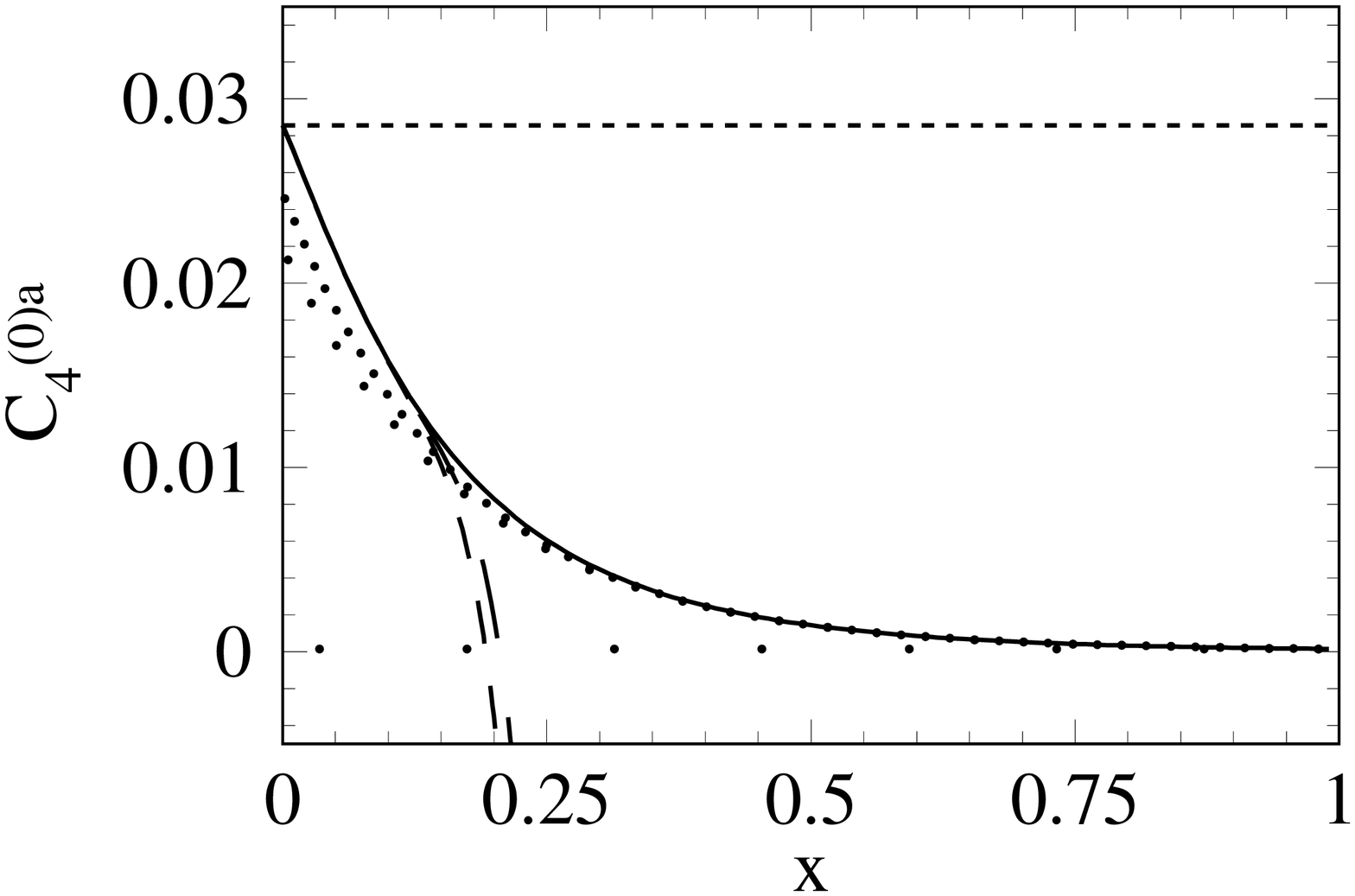}
  \end{tabular}
  \caption[]{\label{fig::c3aas0}One-loop
    contribution to $\bar{C}_n^a$. The same notation as in
    Fig.~\ref{fig::c3pas0} is adopted.
  }
\end{figure}

\begin{figure}[t]
  \centering
  \begin{tabular}{cc}
    \includegraphics[width=0.4\linewidth]{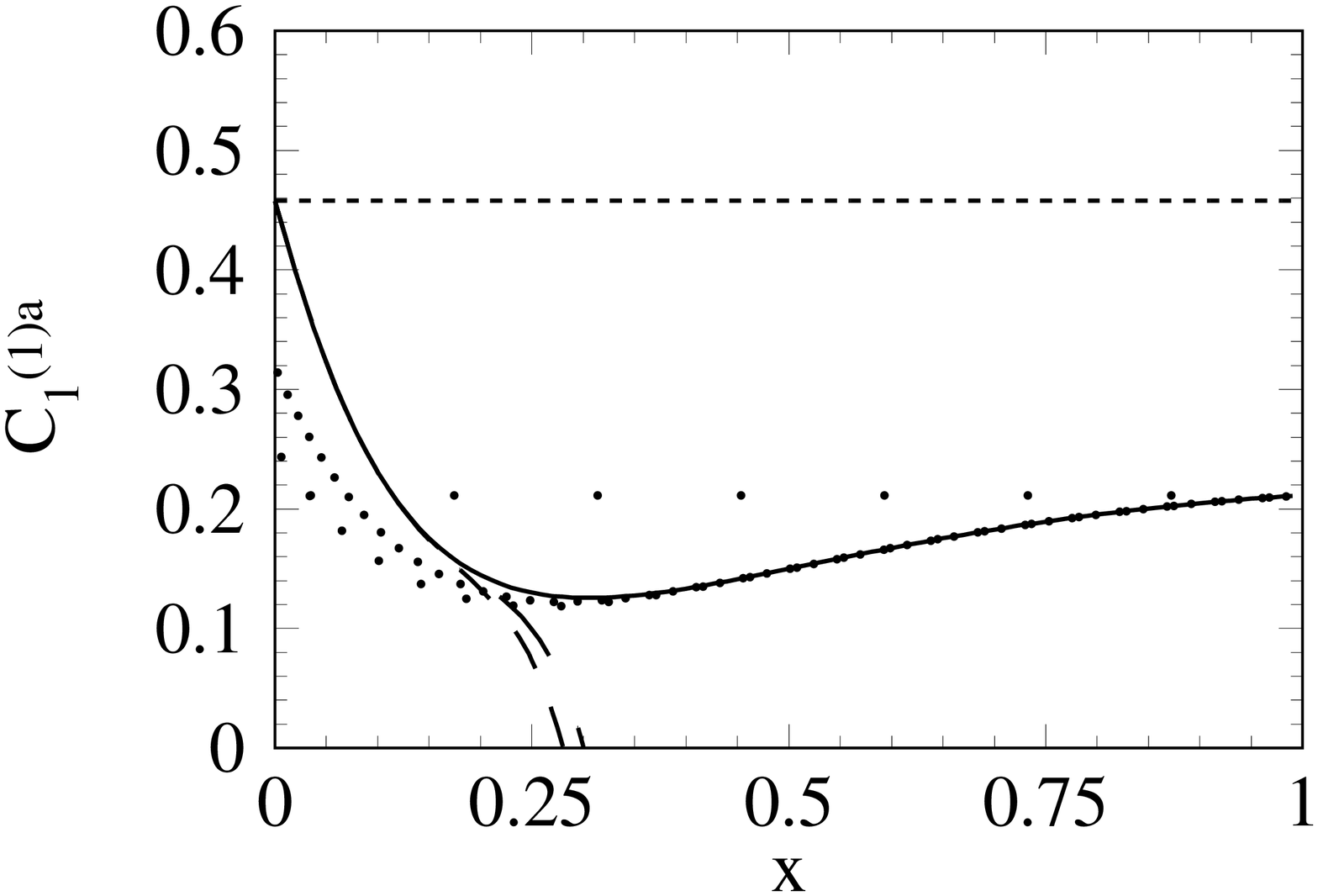}
    &
    \includegraphics[width=0.4\linewidth]{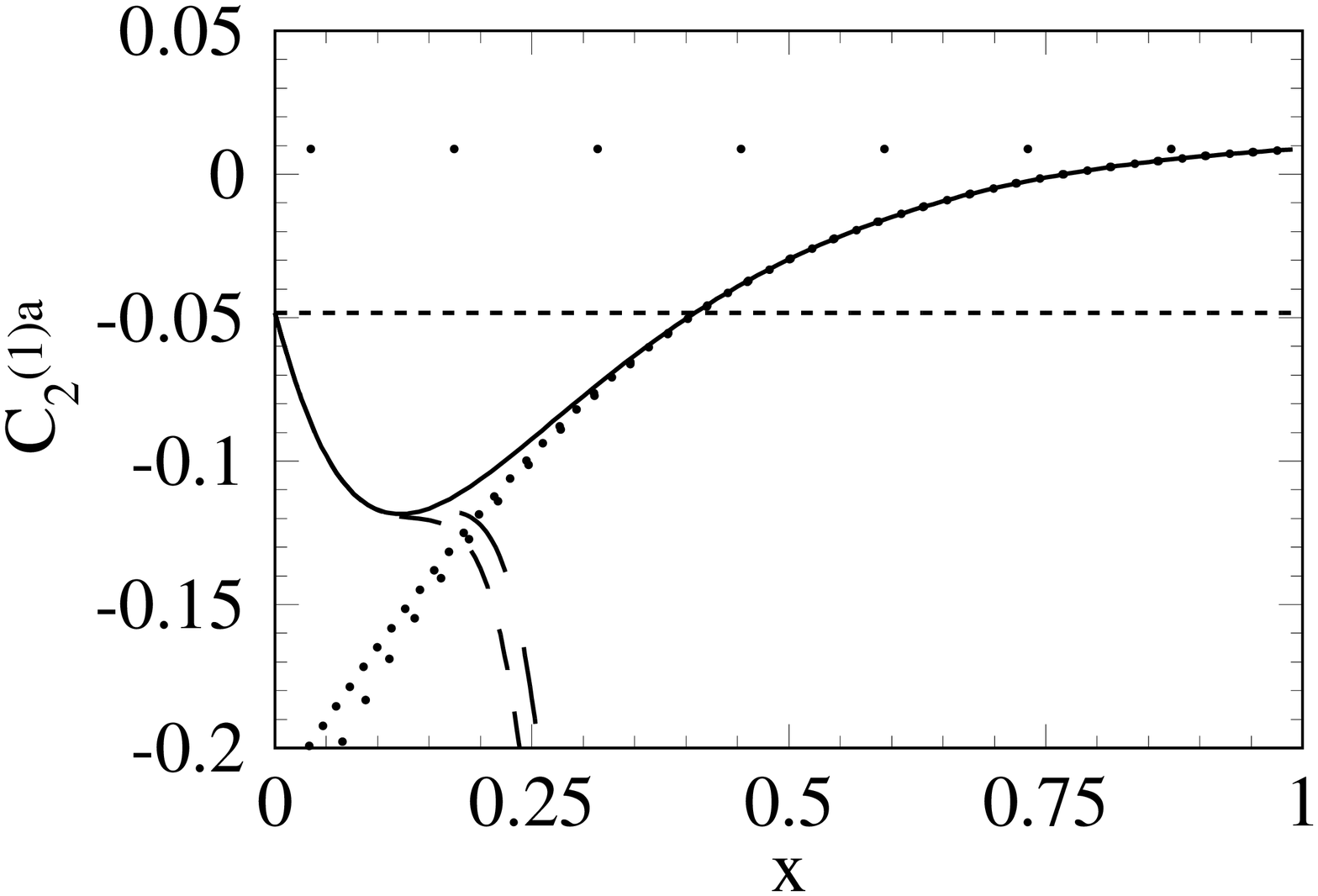}
    \\
    \includegraphics[width=0.4\linewidth]{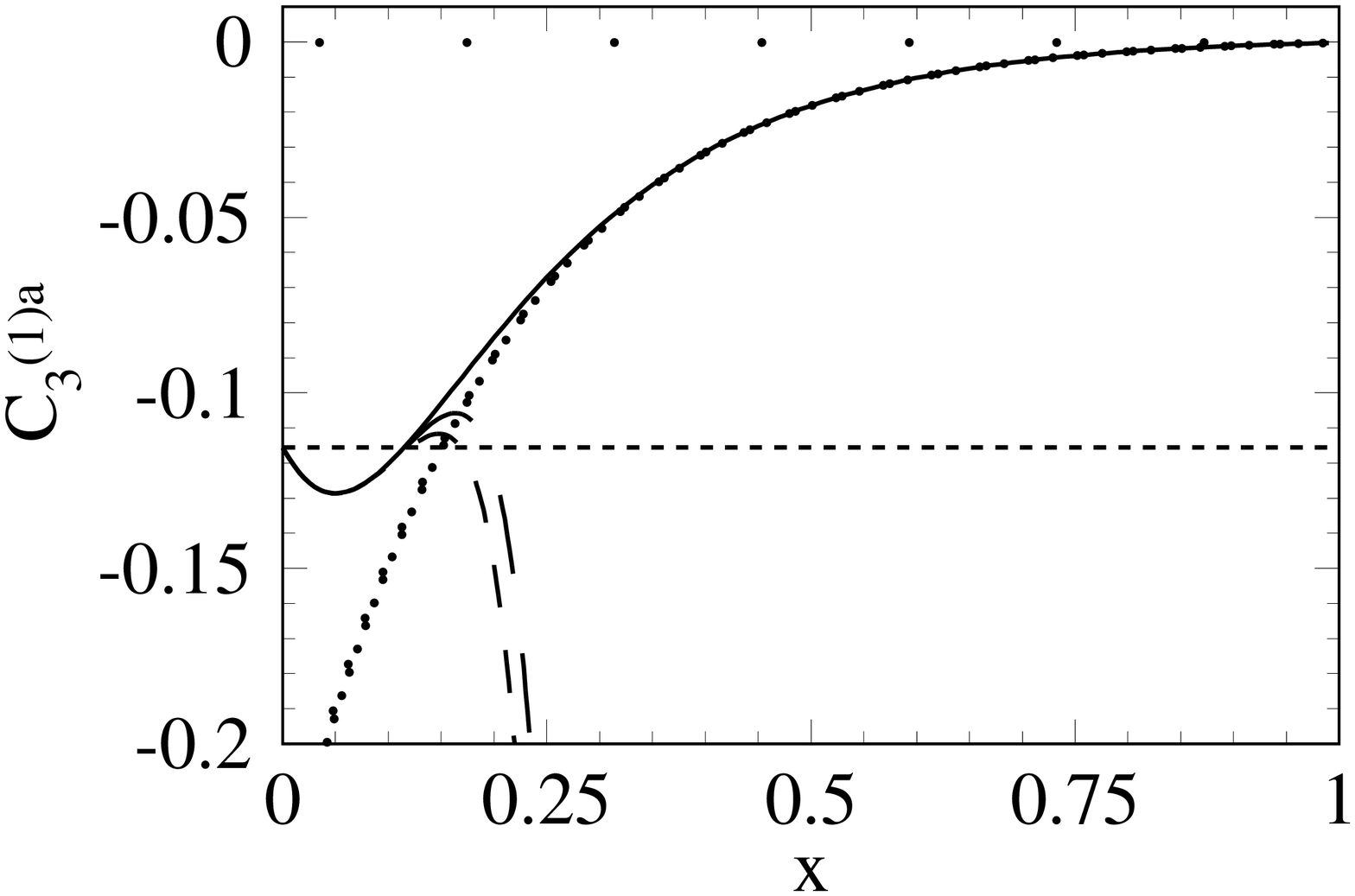}
    &
    \includegraphics[width=0.4\linewidth]{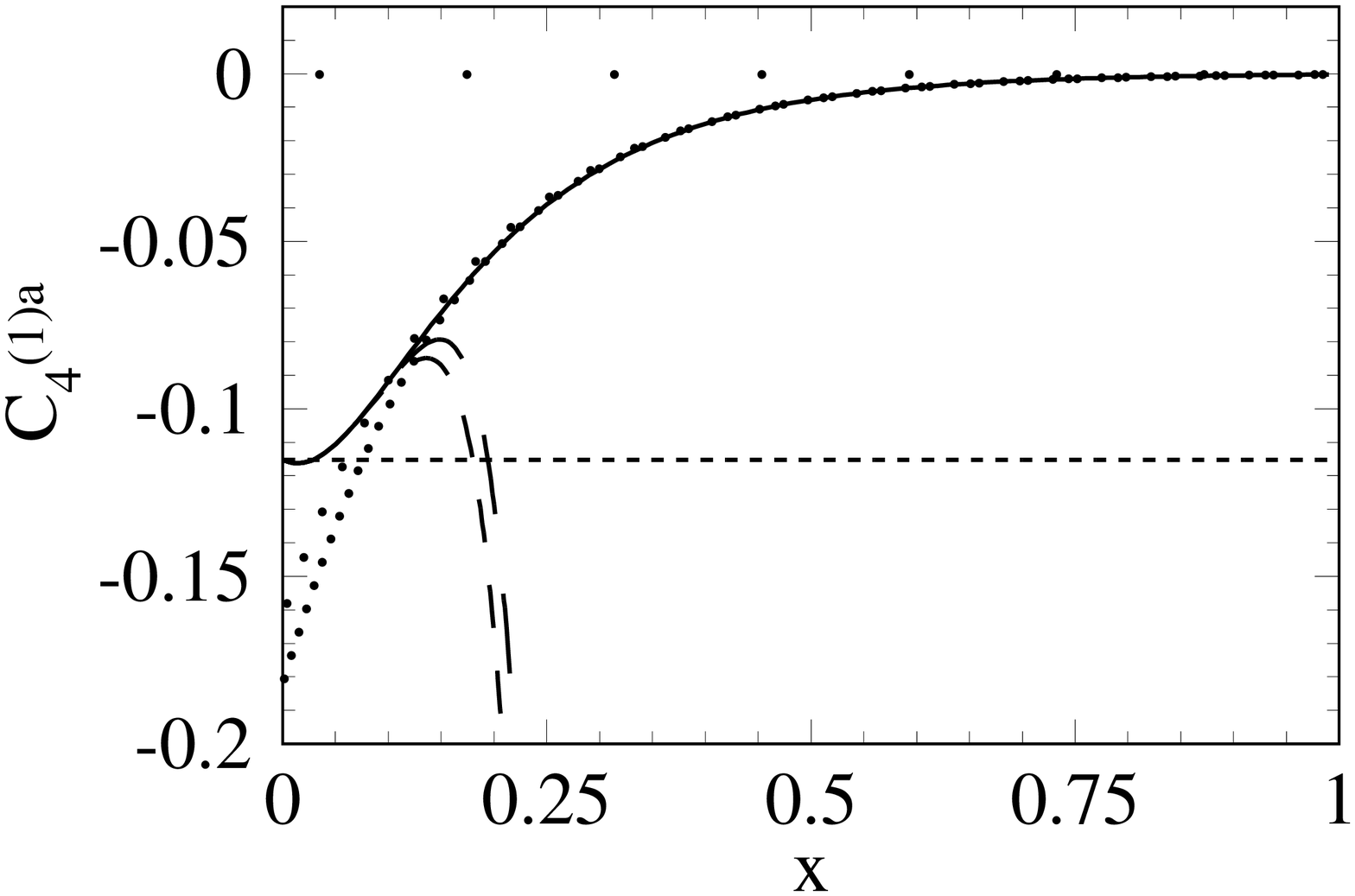}
  \end{tabular}
  \caption[]{\label{fig::c3aas1}Two-loop
    contribution to $\bar{C}_n^a$. The same notation as in
    Fig.~\ref{fig::c3pas0} is adopted.
  }
\end{figure}

\begin{figure}[t]
  \centering
  \begin{tabular}{cc}
    \includegraphics[width=0.4\linewidth]{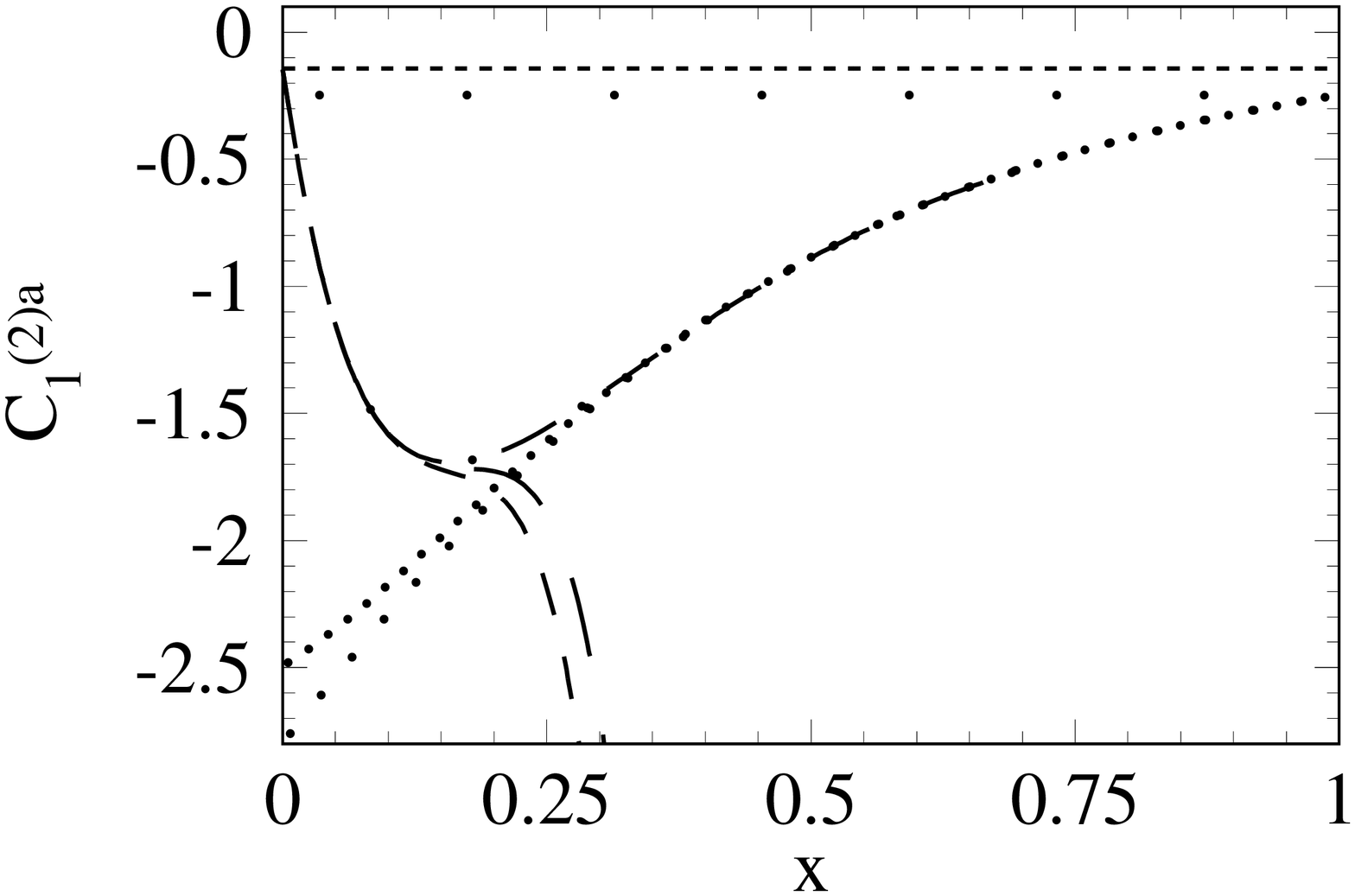}
    &
    \includegraphics[width=0.4\linewidth]{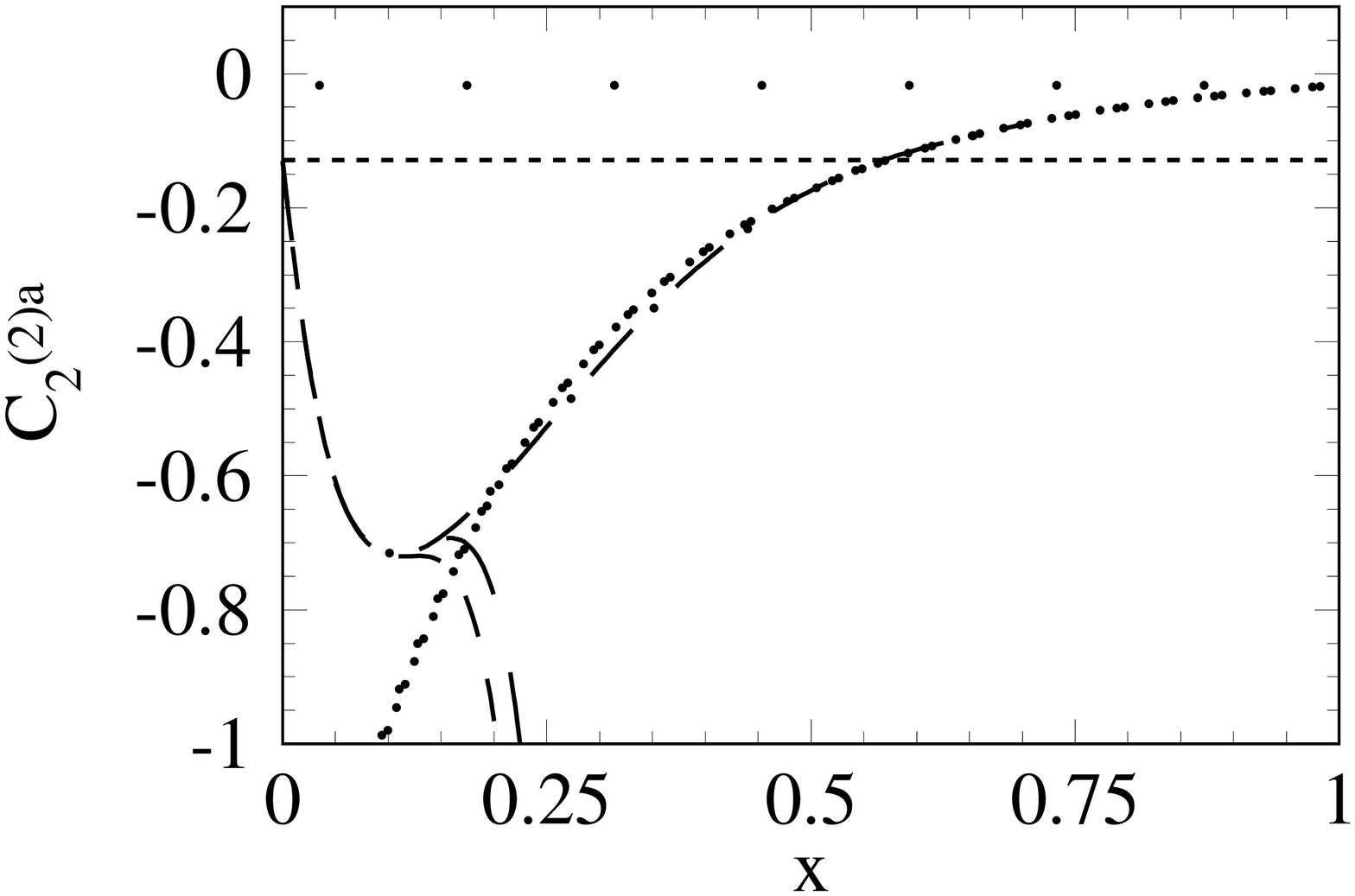}
    \\
    \includegraphics[width=0.4\linewidth]{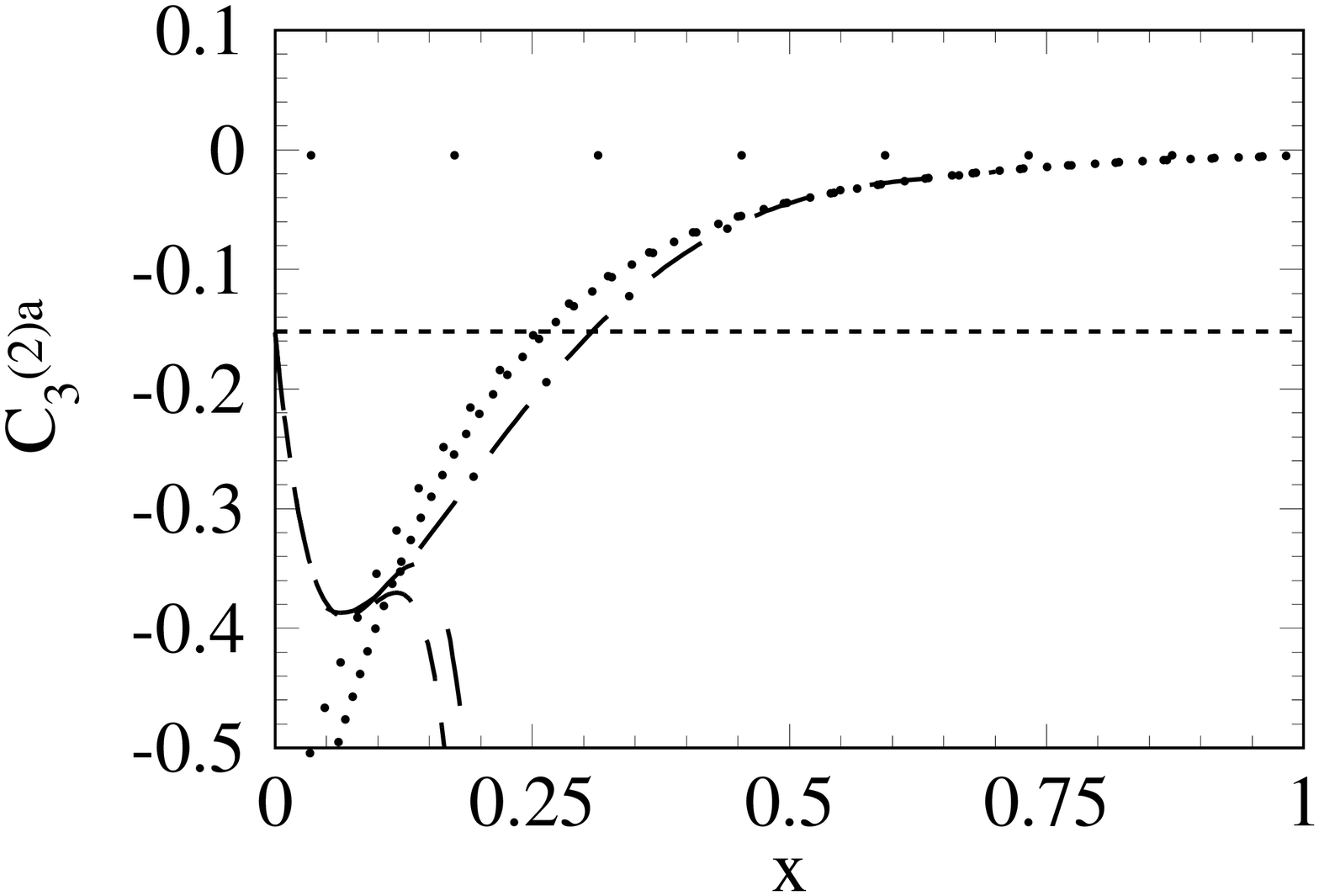}
    &
    \includegraphics[width=0.4\linewidth]{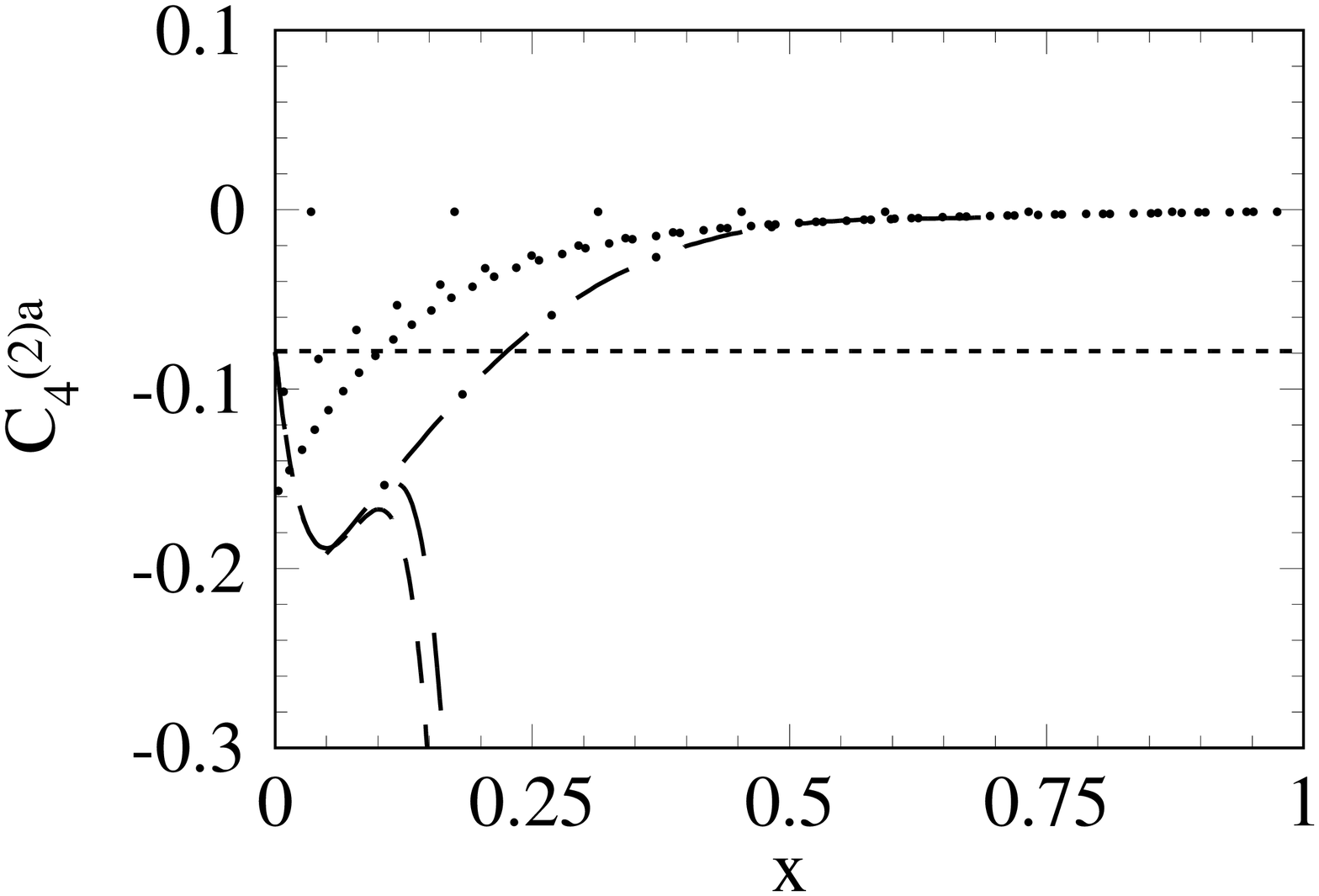}
  \end{tabular}
  \caption[]{\label{fig::c3aas2}Three-loop
    contribution to $\bar{C}_n^a$. The same notation as in
    Fig.~\ref{fig::c3pas2} is adopted.
  }
\end{figure}


\begin{figure}[t]
  \centering
  \begin{tabular}{cc}
    \includegraphics[width=0.4\linewidth]{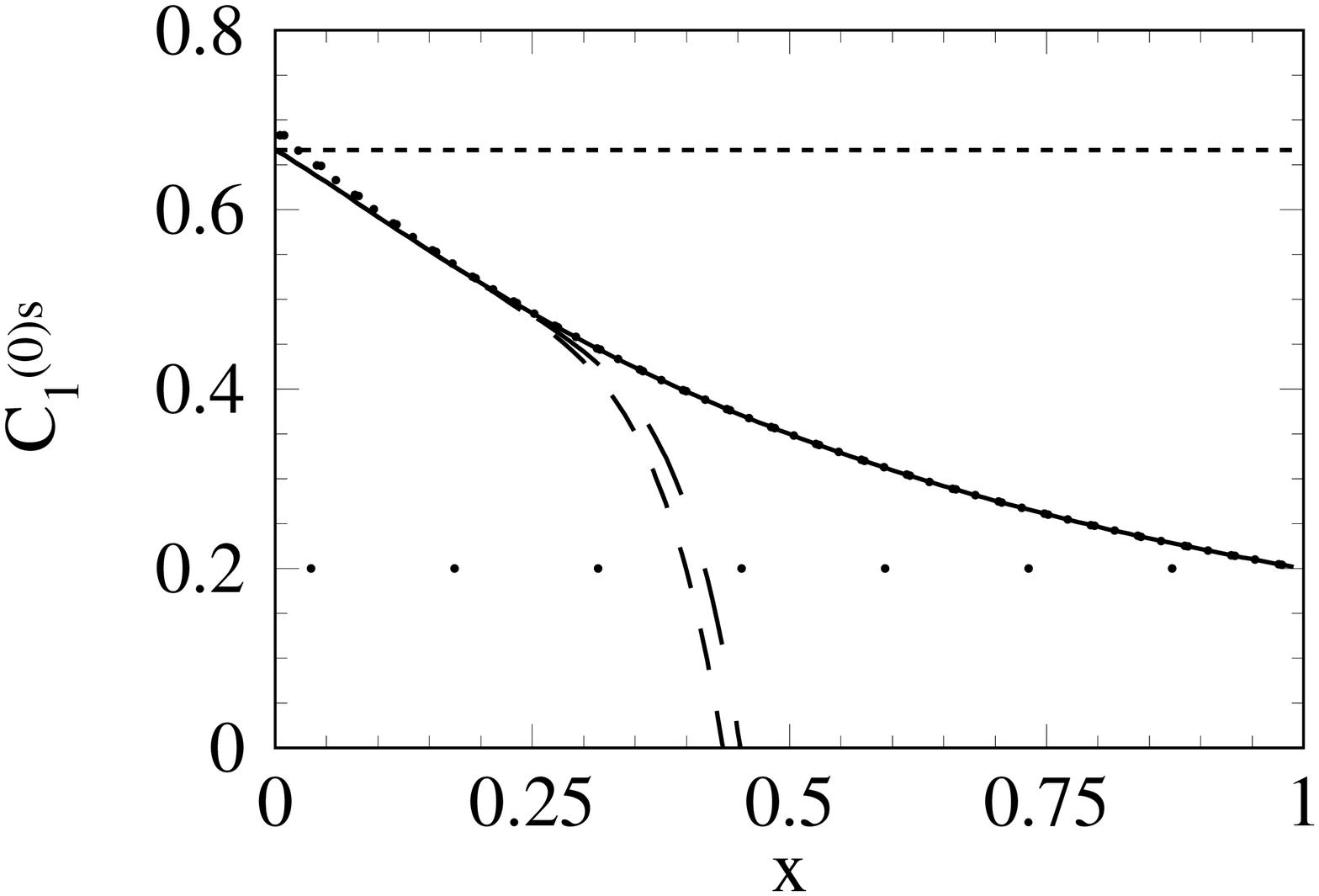}
    &
    \includegraphics[width=0.4\linewidth]{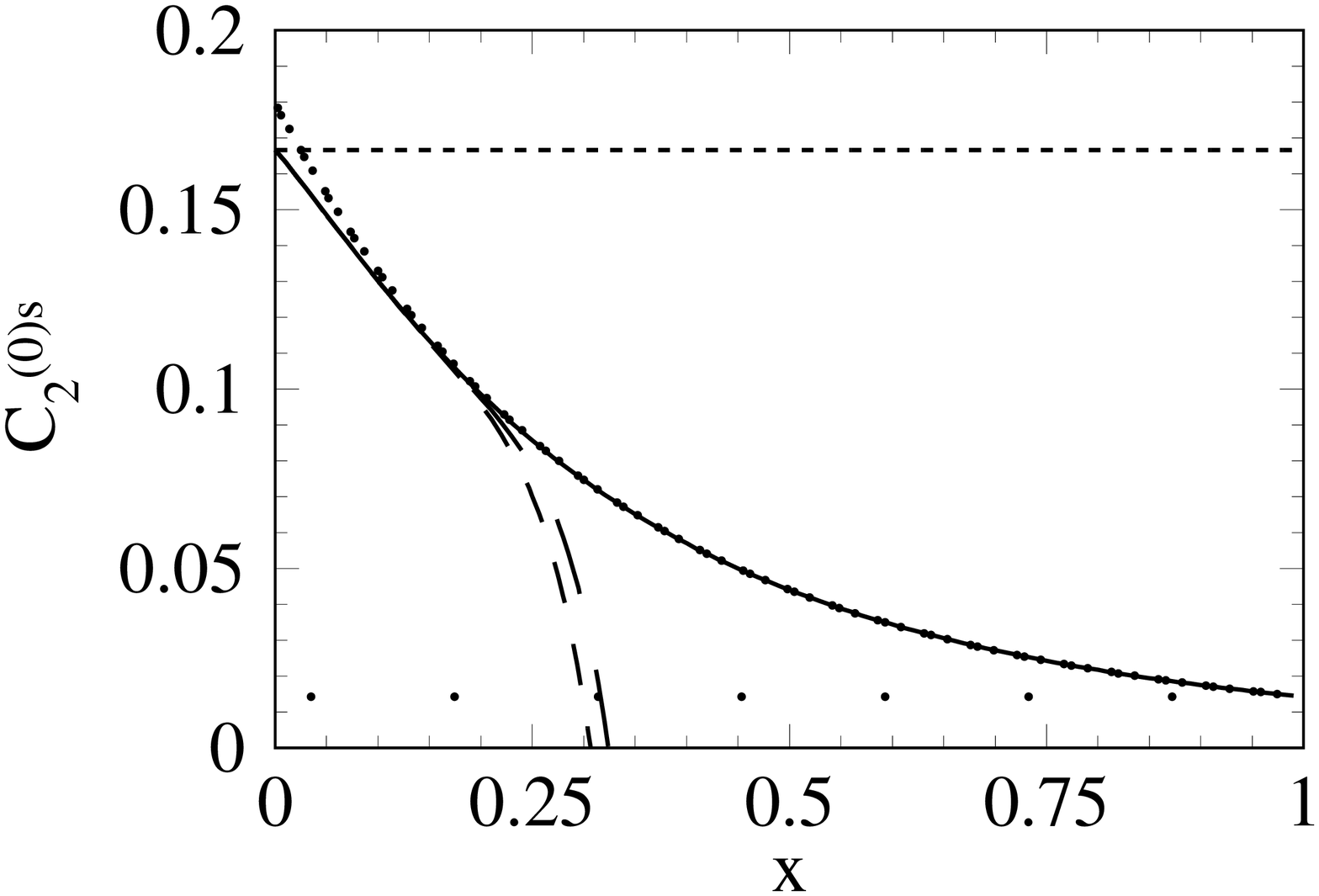}
    \\
    \includegraphics[width=0.4\linewidth]{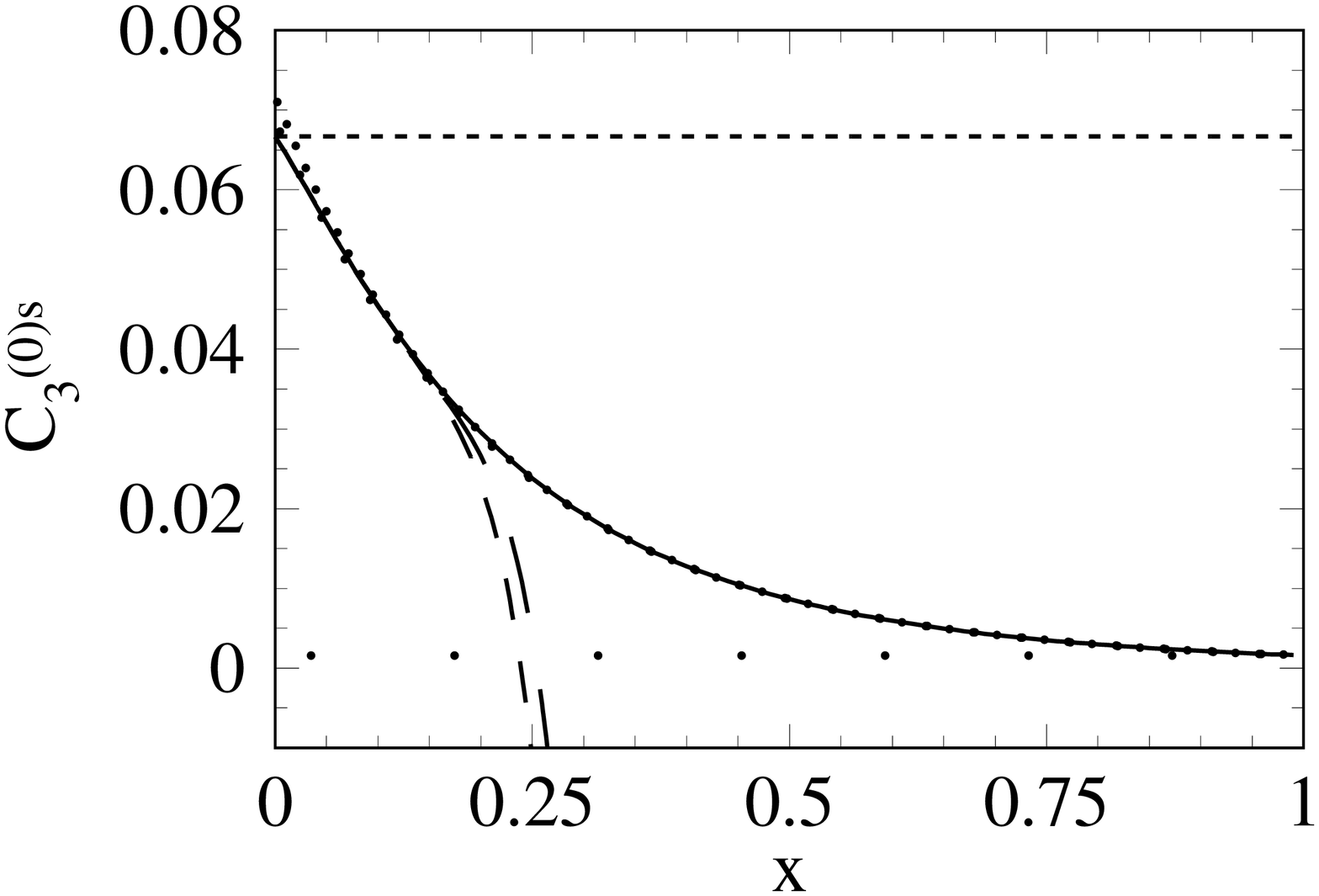}
    &
    \includegraphics[width=0.4\linewidth]{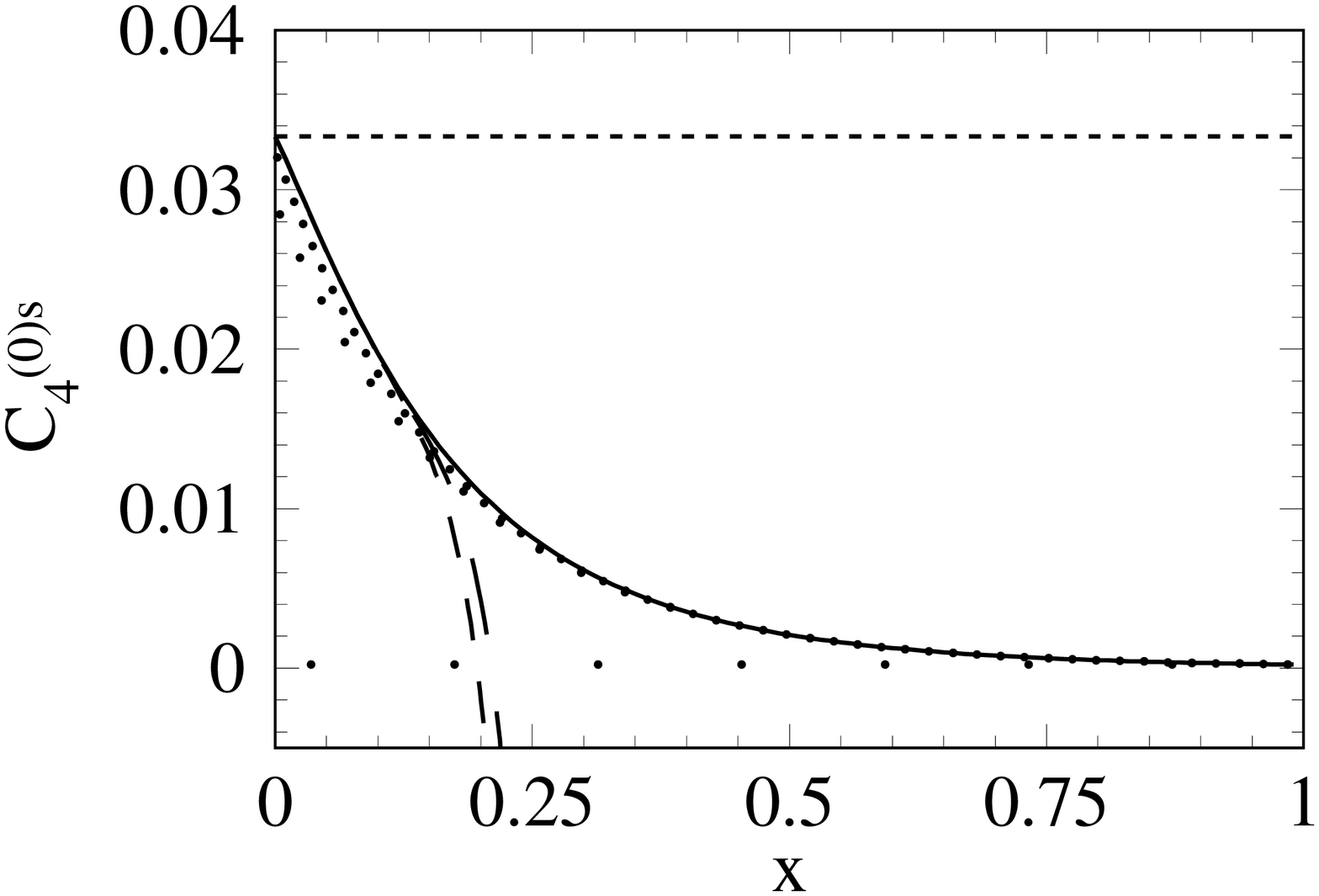}
  \end{tabular}
  \caption[]{\label{fig::c3sas0}One-loop
    contribution to $\bar{C}_n^s$. The same notation as in
    Fig.~\ref{fig::c3pas0} is adopted.
  }
\end{figure}

\begin{figure}[t]
  \centering
  \begin{tabular}{cc}
    \includegraphics[width=0.4\linewidth]{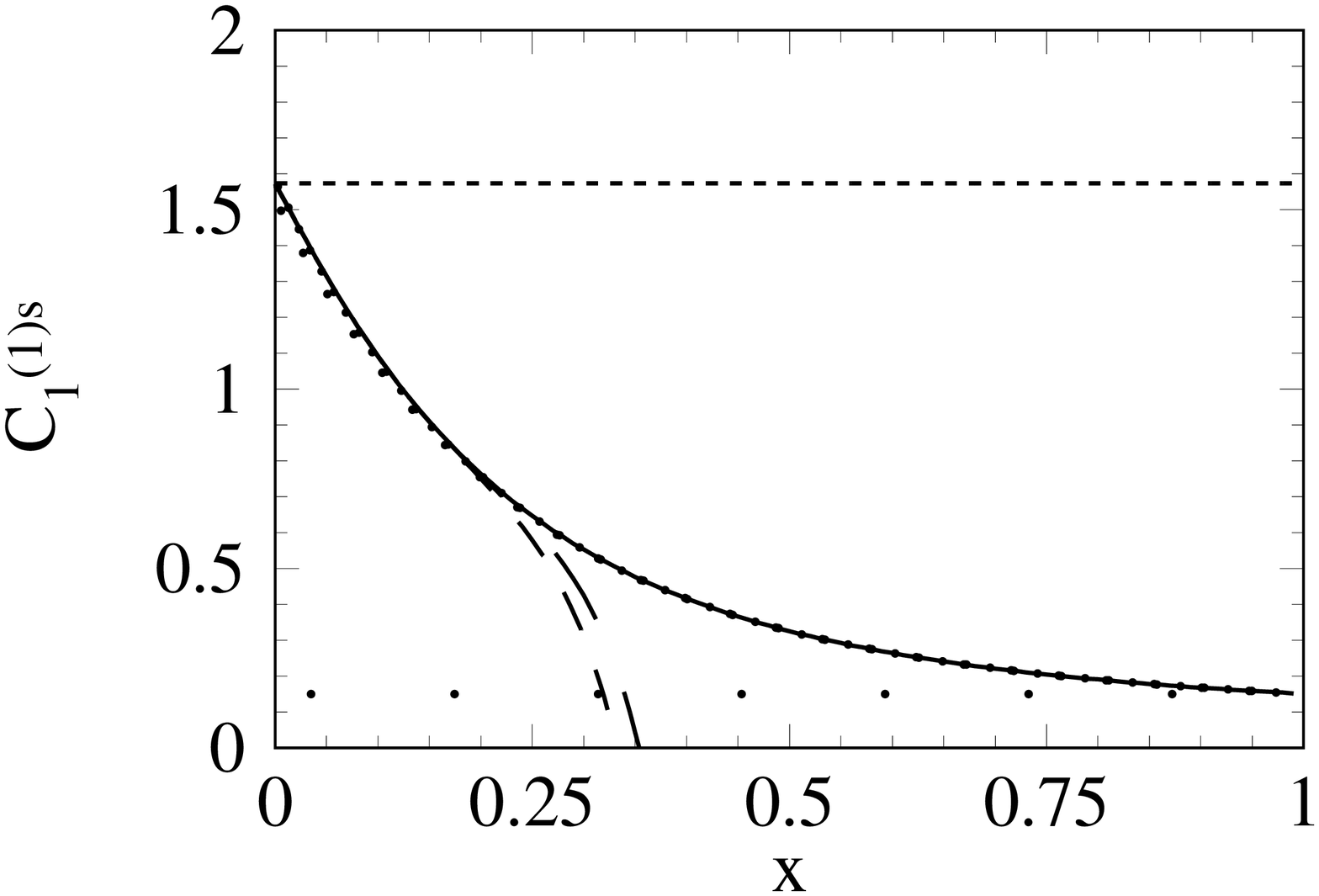}
    &
    \includegraphics[width=0.4\linewidth]{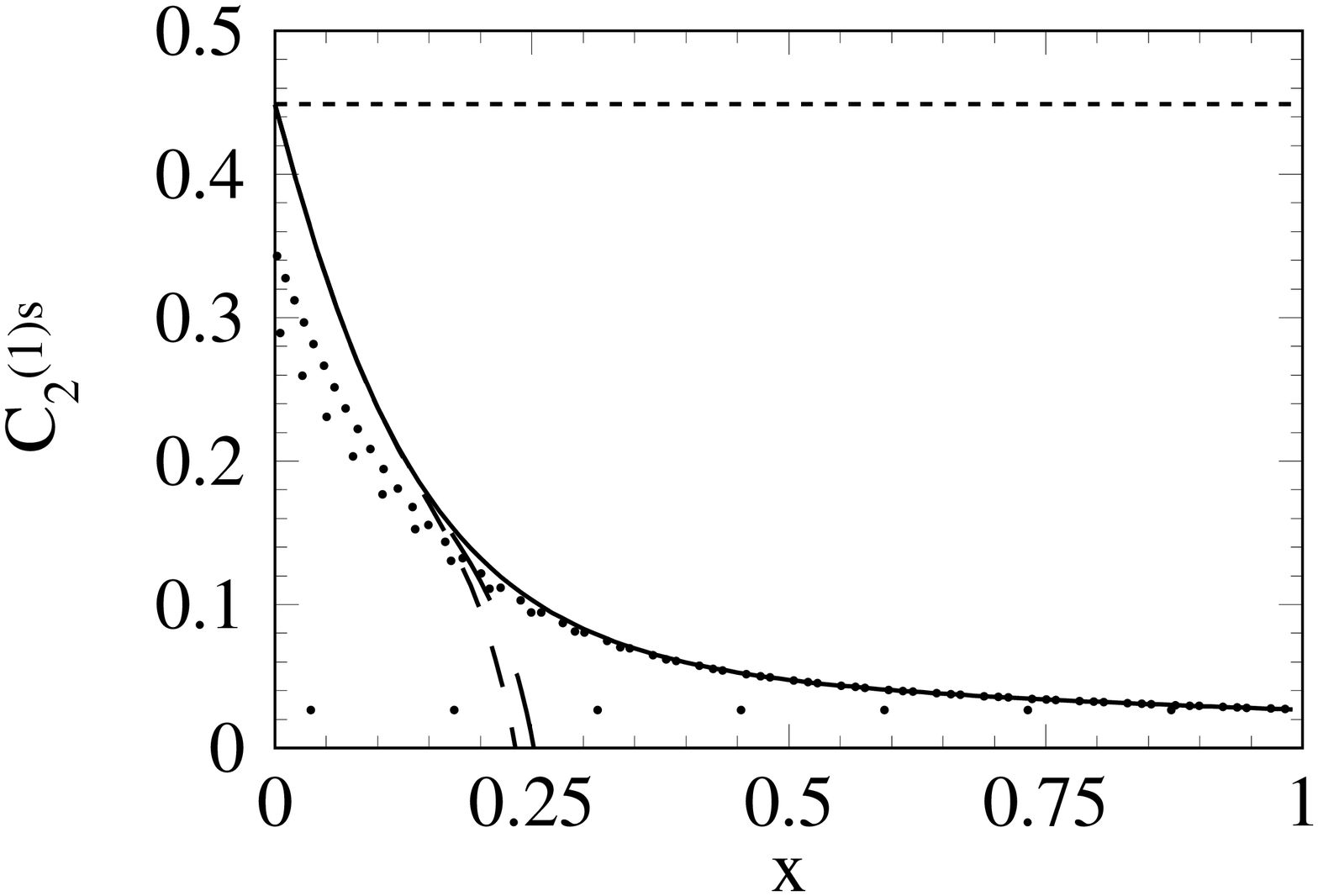}
    \\
    \includegraphics[width=0.4\linewidth]{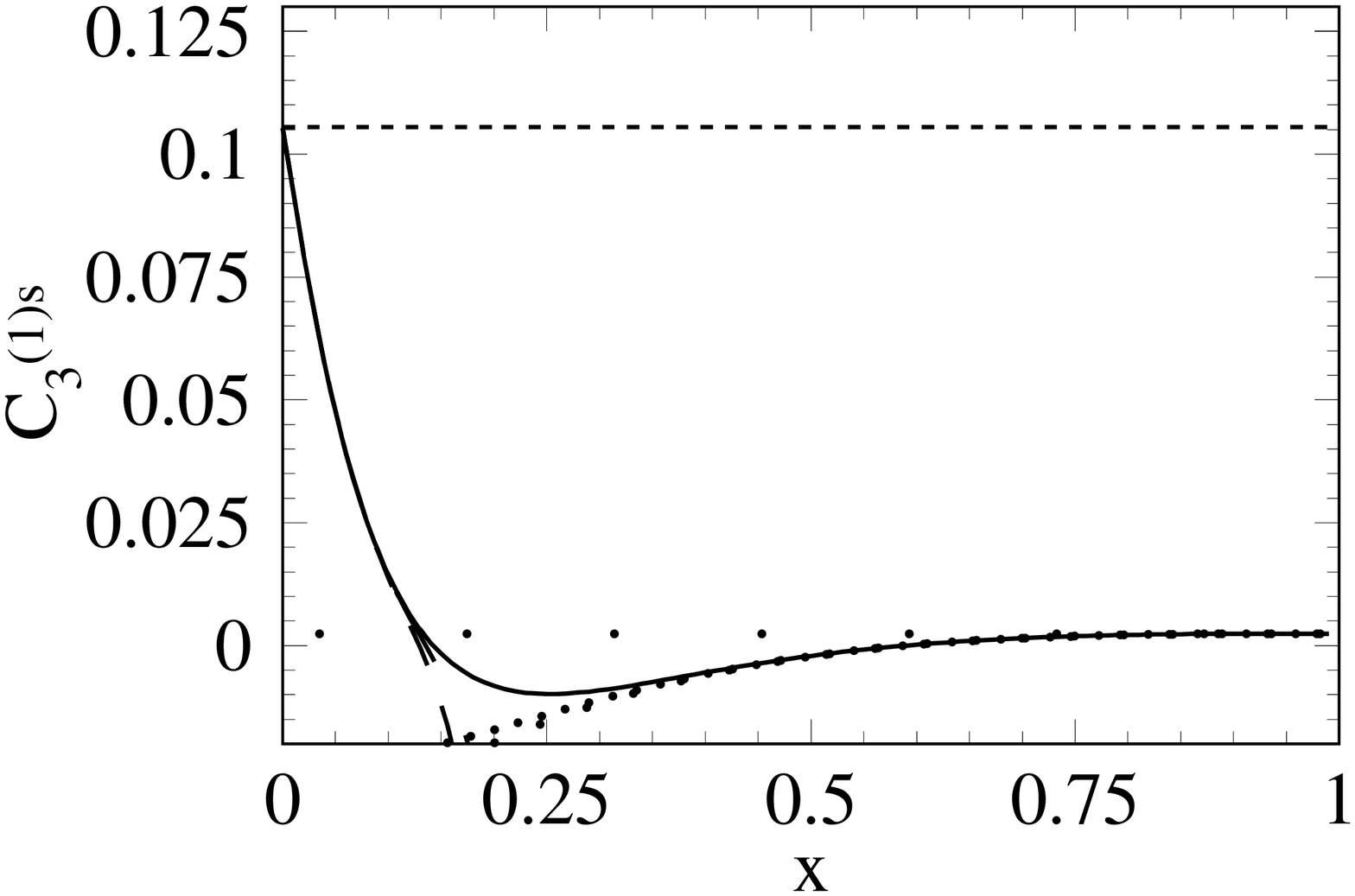}
    &
    \includegraphics[width=0.4\linewidth]{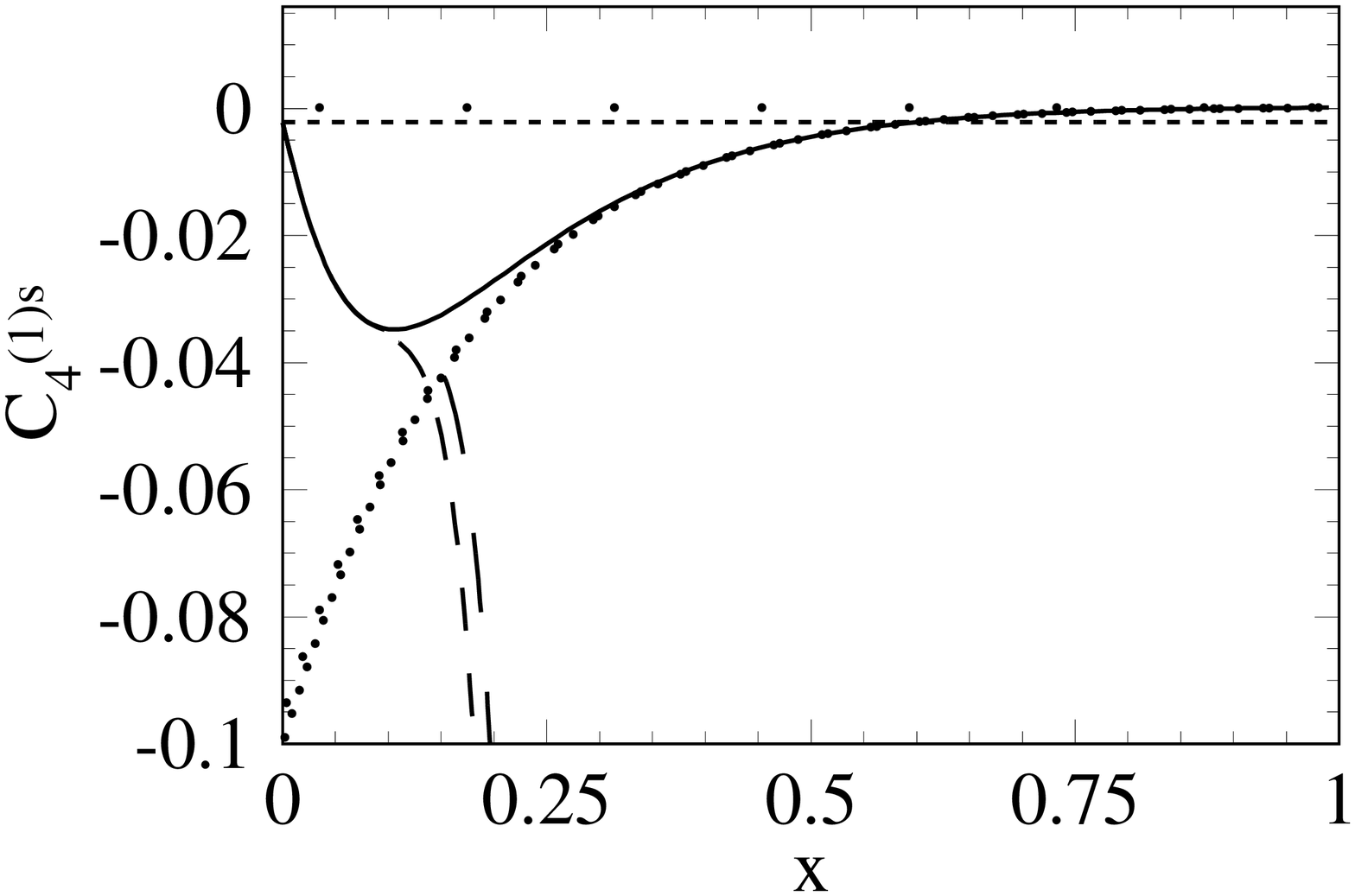}
  \end{tabular}
  \caption[]{\label{fig::c3sas1}Two-loop
    contribution to $\bar{C}_n^s$. The same notation as in
    Fig.~\ref{fig::c3pas0} is adopted.
  }
\end{figure}

\begin{figure}[t]
  \centering
  \begin{tabular}{cc}
    \includegraphics[width=0.4\linewidth]{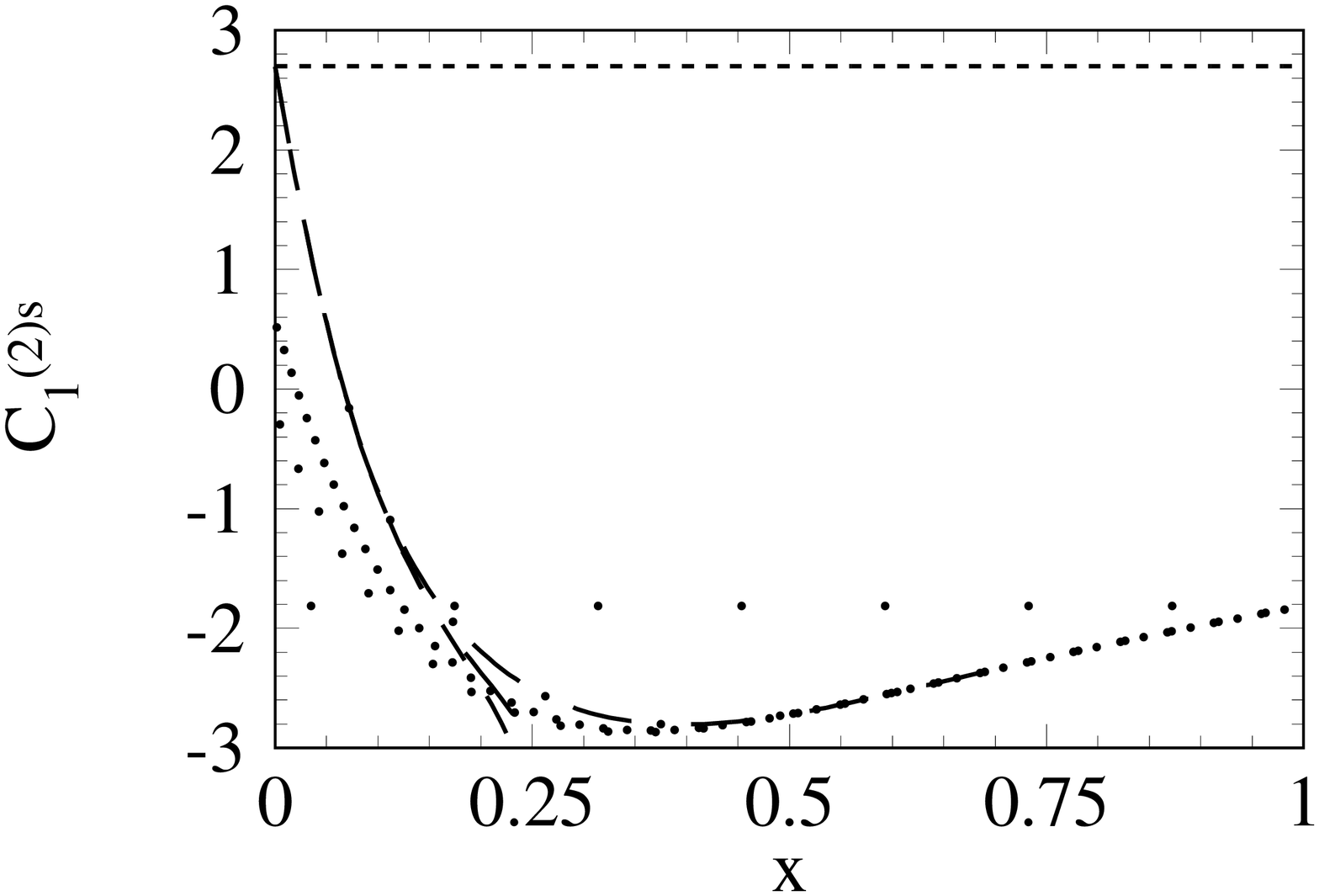}
    &
    \includegraphics[width=0.4\linewidth]{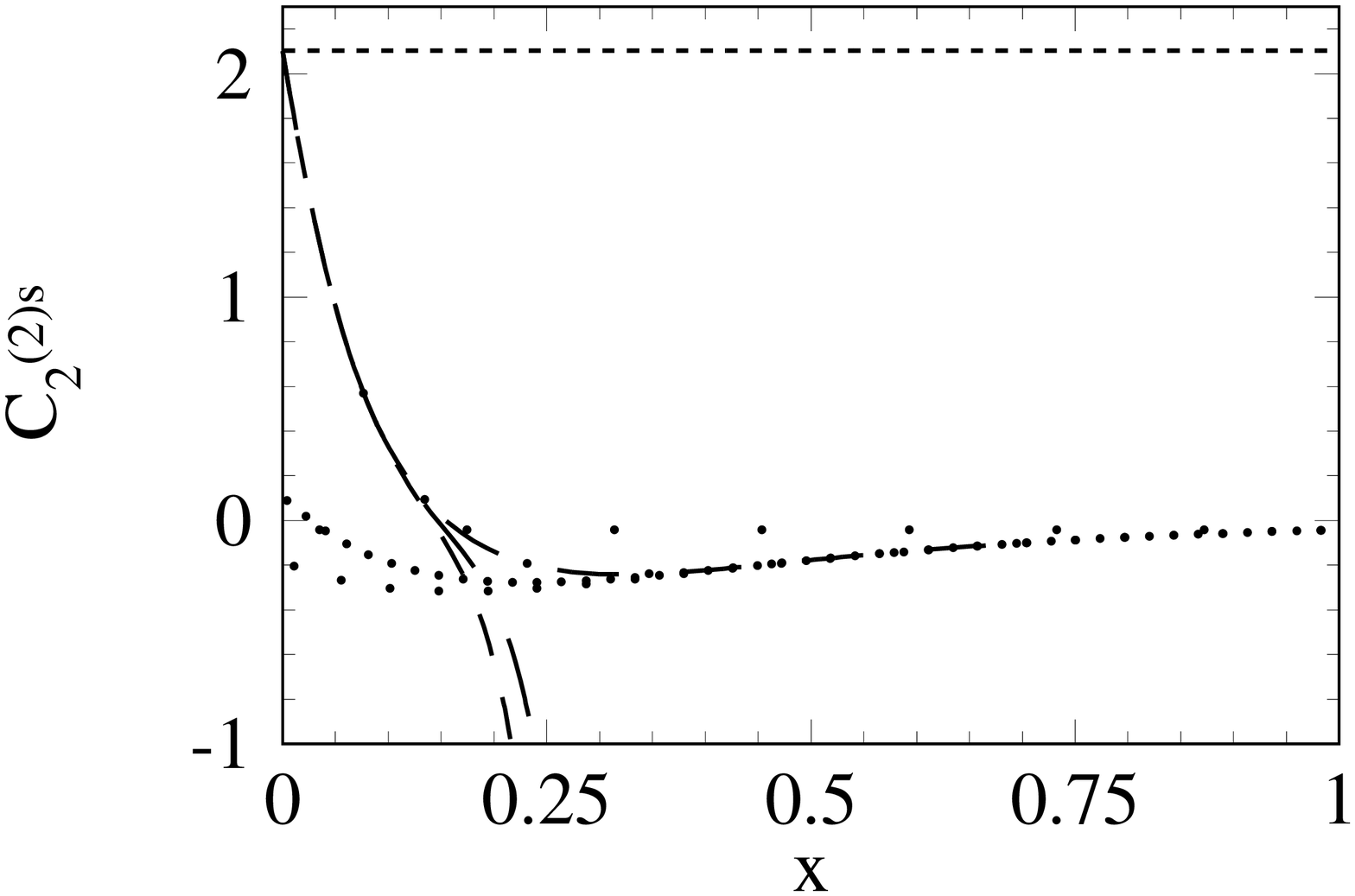}
    \\
    \includegraphics[width=0.4\linewidth]{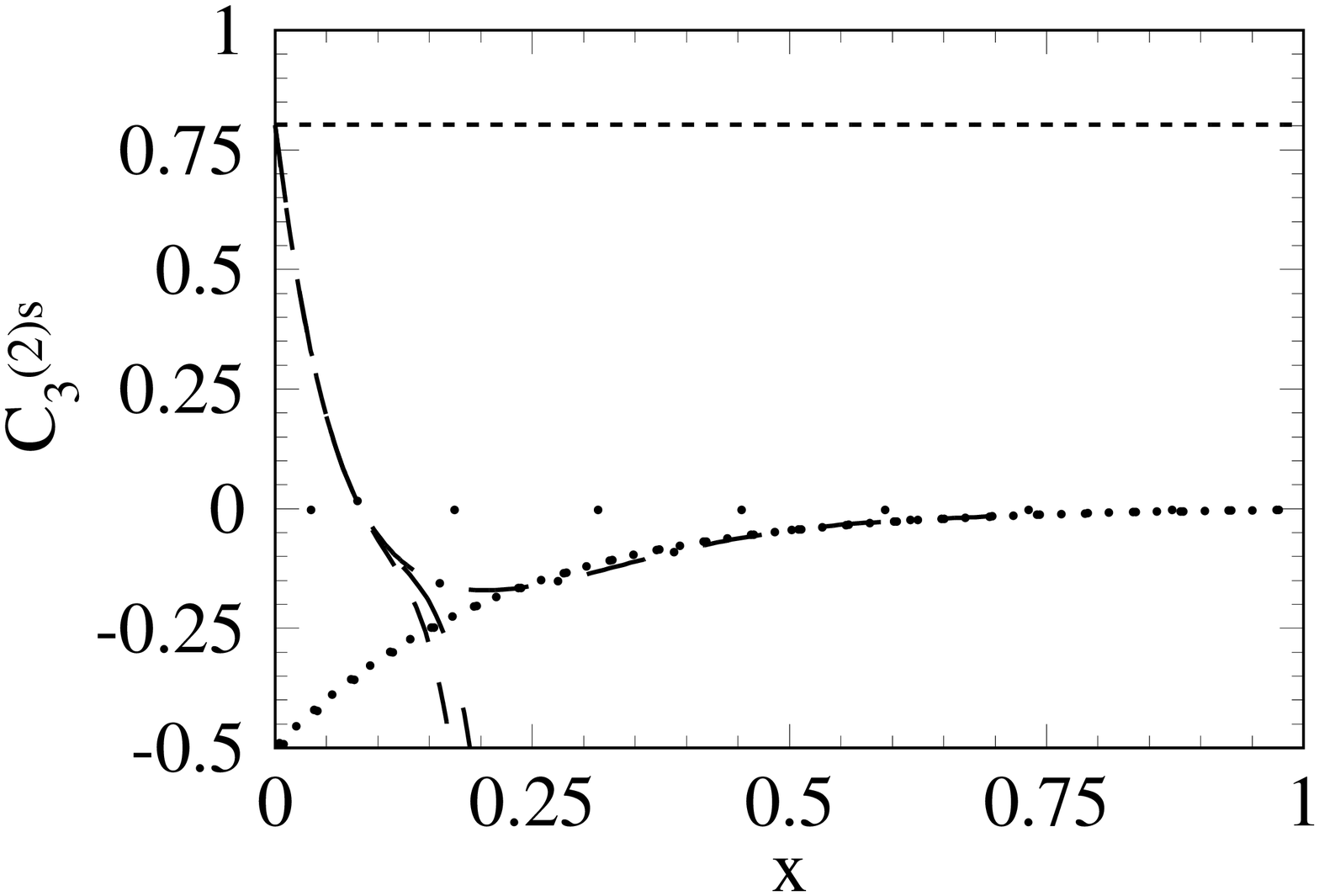}
    &
    \includegraphics[width=0.4\linewidth]{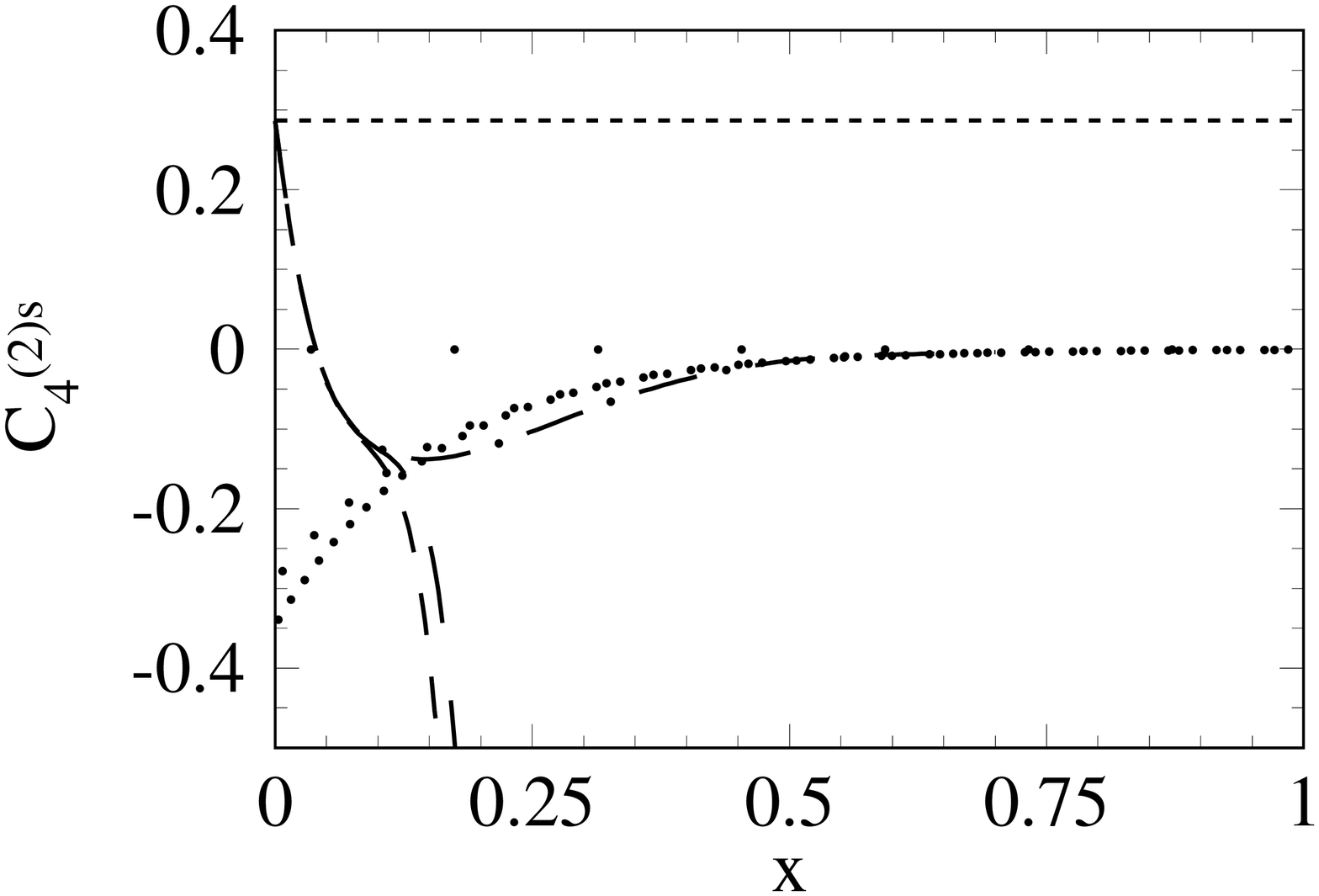}
  \end{tabular}
  \caption[]{\label{fig::c3sas2}Three-loop
    contribution to $\bar{C}_n^s$. The same notation as in
    Fig.~\ref{fig::c3pas2} is adopted.
  }
\end{figure}


\end{appendix}

\newpage



\end{document}